\documentclass[useAASTex,usenatbib,fleqn]{mn2e}

\usepackage[english]{babel}
\usepackage{graphicx}
\usepackage{fleqn}
\usepackage{amsfonts}
\usepackage{amssymb}
\usepackage{amsmath}
\usepackage{booktabs}
\usepackage{times}
\begin{document}
\title{Relativistic dynamics of stars near a supermassive black hole}
\author[Adrian S. Hamers, Simon F. Portegies Zwart and David Merritt]{Adrian S. Hamers$^{1}$, Simon F. Portegies Zwart$^{1}$ and David Merritt$^{2}$ \\
$^{1}$Leiden Observatory, Leiden University, PO Box 9513, NL-2300 RA Leiden, The Netherlands \\
$^{2}$School of Physics and Astronomy and Center for Computational Relativity and Gravitation, Rochester Institute of Technology, Rochester, \\ New York 14623, USA}


\date{Accepted for publication in MNRAS 2014 June 4.} 

\maketitle

\begin{abstract}
General relativistic precession limits the ability of gravitational encounters to increase the eccentricity $e$ of orbits near a supermassive black hole (SBH). This ``Schwarzschild barrier'' (SB) has been shown to play an important role in the orbital evolution of stars like the galactic center S-stars. However, the evolution of orbits below the SB, $e>e_\mathrm{SB}$, is not well understood; the main current limitation is the computational complexity of detailed simulations. Here we present an $N$-body algorithm that allows us to efficiently integrate orbits of test stars around a SBH including general relativistic corrections to the equations of motion and interactions with a large ($\gtrsim 10^3$) number of field stars. We apply our algorithm to the S-stars and extract diffusion coefficients describing the evolution in angular momentum $L$. We identify three angular momentum regimes, in which the diffusion coefficients depend in functionally different ways on $L$. Regimes of lowest and highest $L$ are well-described in terms of non-resonant relaxation (NRR) and resonant relaxation (RR), respectively. In addition, we find a new regime of ``anomalous relaxation'' (AR). We present analytic expressions, in terms of physical parameters, that describe the diffusion coefficients in all three regimes, and propose a new, empirical criterion for the location of the SB in terms of the $L$-dependence of the diffusion coefficients. Subsequently we apply our results to obtain the steady-state distribution of angular momentum for orbits near a SBH.
\end{abstract}

\section{Introduction}
\label{sect:introduction}
Near a supermassive black hole (SBH), evolution of stellar orbits due to gravitational encounters is influenced by three factors. (1) Orbits are nearly Keplerian. (2) The number, $N_\star(r)$, of stars contained within radius $r$ is likely to be small. (3) Relativistic corrections to the equations of motion can be important. Considerations (1) and (2) are the basis of ``resonant relaxation'' (RR) \citep{rauchtr96}, which identifies changes in orbital angular momenta  with torques due to the nearly-stationary mass rings corresponding to the Keplerian orbits. General relativity (GR) appears in this theory as one of several mechanisms capable of inducing orbital precession, hence setting the ``coherence time'' over which the torques can act \citep{rauchtr96}. But recent work reveals that GR can play a much more essential role, particularly in the case of orbits that are highly eccentric. Such orbits precess due to GR at a higher rate than most other orbits at the same radii. This rapid precession tends to quench the effects of the torques \citep{ha06}, but it also leads to a less obvious, and more striking, phenomenon: a ``barrier''  in angular momentum that ``reflects'' stars that strike it from above (i.e. from orbits of higher angular  momentum)  (\citealt{ma11}, hereafter MAMW11). Following MAMW11, we refer to the locus in (energy, angular momentum) space where these phenomena occur as the ``Schwarzschild barrier'' (SB), in recognition of the fact that the precession that underlies the phenomenon is due to the spinless, or Schwarzschild, part of the SBH metric. Compact objects are expected to dominate the stellar population at these small radii, and the existence of the SB is expected to  mediate their capture by the SBH (\citealt{ma11,bas14}; however, for spinning SBHs highly eccentric orbits may not suffer a blockade, \citealt{amaro-seoane13}). Capture events, or EMRIs (extreme-mass-ratio inspirals) (\citealt{sigurdrees97}), would otherwise be expected to be a potentially observable source of low-frequency gravitational waves \citep{amaro-seoane12}.

Many processes exist that can deposit stars onto highly eccentric orbits around a SBH. These processes include close encounters between stars \citep{goodman83}, encounters between stars and massive perturbers \citep{perets07} or a stellar disk \citep{chen14}, and the tidal disruption of stellar binaries that approach the SBH on nearly radial orbits \citep{hills88}. These ideas are relevant to models that attempt to explain the presence of young stars very near to the SBH in the Galactic center (GC). Some of these stars, the so-called S-stars, have orbits of high enough eccentricity that they must lie below the predicted location of the SB \citep{antmerr13}. If the S-stars were deposited initially onto orbits with even higher eccentricities than observed today (which would be the case, for instance, in the binary disruption model), then the fraction of S-stars initially below the SB was even higher in the past. The evolution of such highly eccentric orbits over Myr time scales is not well described by existing theory of resonant or non-resonant relaxation; it depends in critical ways on the barrier phenomena described above \citep{antmerr13}.

Progress in understanding the relativistic dynamics of nuclear star clusters has been driven in large part by the recent development of extremely accurate and efficient computer codes for solving the (small-) $N$-body problem \citep{mikkolaaarseth93,mikkolaaarseth02,mikkolataniwaka99,mikkola08}. But the new results summarized above also imply that the number of stars in a real galaxy that are subject to GR phenomena is probably much larger than can be handled efficiently by these codes. For instance, in the Milky Way, the number of stars and stellar remnants inside $r=a_\mathrm{SB,max}$, the largest semimajor axis for which the SB exists, is probably of order $10^3-10^4$. Efficient, Monte-Carlo algorithms for evolving test-orbits near the SB were developed in MAMW11 and applied to the S-star problem by \citet{antmerr13}, but these algorithms are based on an extremely simple model for the torquing potential and its time dependence.

A major goal of this paper is to develop an alternate algorithm that represents the field-star forces much more accurately than the Monte-Carlo routines in MAMW11, but which nevertheless is efficient enough to be used for realistically large $N$-values. Our code, called \textsc{Test Particle Integrator} (\textsc{TPI}), explicitly follows the motion of the field stars along their precessing, Keplerian orbits, but ignores interactions between them. The motion of the test stars is then followed by direct integration in the time-varying potential produced by the $N$ field stars. Relativistic terms are included in the equations of motion of both test and field stars via the post-Newtonian approximation. This algorithm contains all of the dynamics which are believed to be important for the evolution of orbits due to RR in the presence of relativity, excluding only the changes in the field-star distribution that would be due to the RR torques themselves, or to perturbations from the test stars.

In \S\,\ref{sect:methods} we describe \textsc{TPI} and perform a number of basic tests. In \S\,\ref{sect:belowSB} the orbital evolution below the SB is studied using simulations similar to those performed by MAMW11. By restricting to a small number of particles we can compare our results to results obtained from $N$-body codes in which the simplifying assumptions adopted in \textsc{TPI} are relaxed. In \S\,\ref{sect:S-stars} we apply our code to the S-star cluster; similar simulations with the other $N$-body codes used in \S\,\ref{sect:belowSB} are currently not feasible. Assuming that the S-stars are formed in highly eccentric orbits, which is consistent with the binary disruption model, and adopting a cusp of stellar black holes, we study the orbital evolution of the S-stars after their formation. 

The models explored here were designed to represent the Galactic center, but it is useful to ask how our results would generalize to other nuclei. To this end, in \S\,\ref{sect:belowSB:diffusion} and \ref{sect:S-stars:dfc} we extract angular-momentum diffusion coefficients from the simulations and compare them with existing theory. We argue in \S\,\ref{sect:timescales} that diffusion in angular momentum should be well described by NRR at very low $L$ ($e \gg e_\mathrm{SB}$), and by RR at high $L$ ($e\ll e_\mathrm{SB}$). But in the angular momentum regime near and ``below'' the SB (i.e. $e \gtrsim e_\mathrm{SB}$), neither RR nor NRR is applicable (MAMW11). By computing angular momentum diffusion coefficients from the simulations, we are able, for the first time, to demonstrate the existence of the three regimes and to quantify their $L$-dependence. This allows us, in \S\,\ref{sect:steady-state}, to estimate the steady-state angular momentum distribution implied by the Fokker-Planck equation. In \S\,\ref{sect:discussion} we discuss the implications of our results and we conclude in \S\,\ref{sect:conclusions}.

\section{Timescales}
\label{sect:timescales}
The focus in this paper is on orbits near a SBH that are very eccentric compared with the typical eccentricity expected in, say, a ``thermal'' distribution, $\langle e\rangle = 2/3$. The time scale over which such eccentric orbits evolve due to gravitational encounters with other stars can depend strongly on $e$. We begin by summarizing what is known about that dependence. As we will see, in regimes near or below the SB, i.e. $e\gtrsim e_\mathrm{SB}$, the eccentricity dependence is still poorly understood and that is one motivation for carrying out the simulations described below. 

\begin{figure}
\center
\includegraphics[scale = 0.4, trim = 0mm 0mm 0mm 0mm]{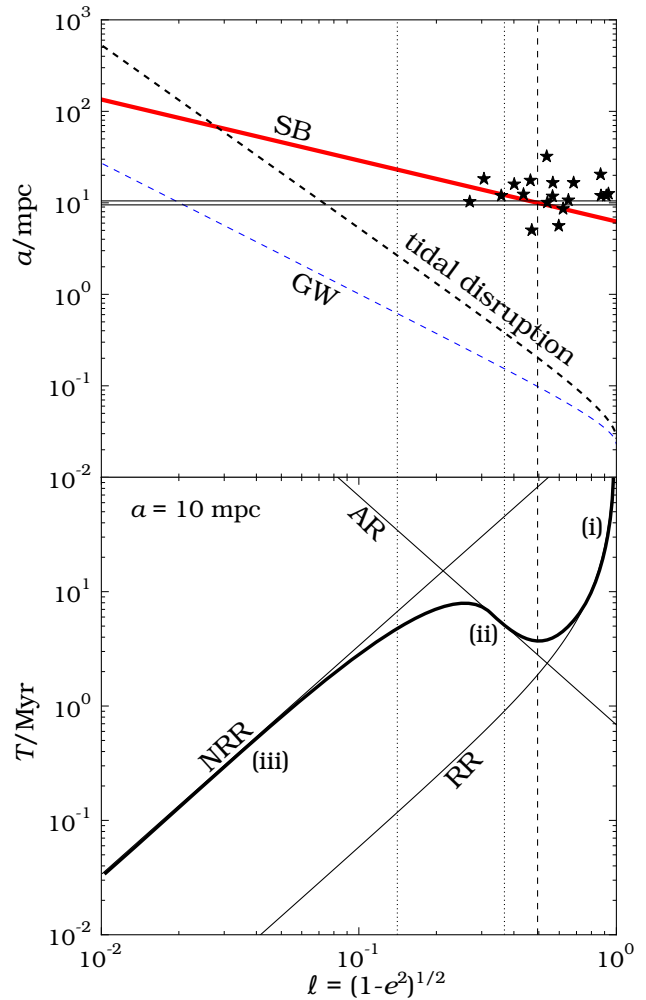}
\caption{\small Top panel: characteristic curves in the $(a,\ell)$ (semimajor axis, normalized angular momentum) plane for stars orbiting near the MW SBH. Curves are defined in the text; the assumed stellar mass and radius are $10 \, \mathrm{M_\odot}$ and $8 \, \mathrm{R_\odot}$, respectively. Shown as stars are the $(a,\ell)$ values of the 19 S-stars with well-determined orbits \citep{gil09}. Bottom panel: approximate time scales for $\ell$ to change by of order itself due to gravitational interactions with field stars, at a ``slice'' in semimajor axis at 10 mpc. RR = resonant relaxation; AR = anomalous relaxation; NRR = non-resonant relaxation. Three angular momentum regimes are indicated; see text. The vertical dotted lines indicate the range in $\ell$ expected for the tidal disruption of a stellar binary by the SBH (\citealt{hills88}; see also \S\,\ref{sect:S-stars}). The vertical dashed line marks the value of $\ell_\mathrm{SB}$ at $a = 10$ mpc, the semimajor axis in the bottom panel. }
\label{fig:introduction_figure}
\end{figure}

The top panel of Figure~\ref{fig:introduction_figure} plots several curves in the $(a,\ell)$ (semimajor axis, normalized angular momentum) plane that are relevant to stars orbiting near a SBH. This figure adopts an SBH mass $M_\bullet =4\times 10^6 \, \mathrm{M}_\odot$, the value in the Milky Way. 

The red curve labelled ``SB'' is given by
\begin{align}
\left(1-e^2\right)^{1/2} \equiv \ell = 
\frac{r_g}{a}\frac{M_\bullet}{M_\star(a)}\sqrt{N_\star(a)}.
\label{eq:ell_SB}
\end{align}
Here $r_g\equiv GM_\bullet/c^2$ is the gravitational radius of the SBH, $M_\star(r)$ is the mass in stars within radius $r$, and $N_\star(r)$ is the number of stars within $r$; $M_\star=m_\star N_\star$. The quantity $\ell$ is a normalized angular momentum: $\ell = L/L_c$ with $L_c(a) =\sqrt{GM_\bullet a}$ the angular momentum of a circular orbit of semimajor axis $a$. Equation~(\ref{eq:ell_SB}) is the approximate locus in the $(a,\ell)$ plane where the change in the angular momentum $L$ of a test star due to torques from the other stars, in one relativistic precession cycle (equation~\ref{eq:1PNprecession}), is of order $L$. In their (small-) $N$-body simulations, MAMW11 found that equation~(\ref{eq:ell_SB}) predicts very well the maximum eccentricity reached by orbits as they evolve due to gravitational encounters. In Figure~\ref{fig:introduction_figure}, we adopted for $M_\star(r)$ and $N_\star(r)$ the expressions given in  \S\,\ref{sect:S-stars:initial_cond}.

The curve labelled ``tidal disruption'' is the locus of orbits having periapses at the tidal disruption radius defined in \S\,\ref{sect:S-stars:initial_cond}. The curve labelled ``GW'' is an estimate of where changes in orbital eccentricity due to gravitational-wave enery loss occur at the same rate as changes due to two-body relaxation (MAMW11, Eq. 62).

The lower panel of Figure~\ref{fig:introduction_figure} shows how time scales for changes in orbital angular momentum are believed to depend on $\ell$. Here, ``time scale'' is defined as the time for $L$ to change by of order itself (and not, for instance, for $L$ to change by of order $L_c$). We identify three regimes. (i) Above the SB, resonant relaxation (RR) is effective \citep{rauchtr96}. On time intervals longer than the ``coherence time''-- the time for an orbit of typical eccentricity to precess -- orbital angular momenta are expected to undergo a random walk due to torques from the $\sqrt{N}$ asymmetry in the mass distribution. The associated time scale is $\sim\ell^2$ times $t_\mathrm{RR}$; the latter is given by equation~(\ref{eq:dfc_lRR2}). (ii) At $\ell\lesssim\ell_\mathrm{SB}$, GR precession strongly reduces the ability of the $\sqrt{N}$ torques to change $\ell$. While no adequate theory yet exists for diffusive evolution in this regime, approximate arguments (MAMW11) suggest that the evolution time scale should increase rapidly with decreasing $\ell$. The curve labelled AR (``anomalous relaxation'') shows, qualitatively, how the evolution time scale might depend on $\ell$ in this regime. (iii) At sufficiently small $\ell$, the diffusion time due to AR is expected to become very long due to the rapid GR precession. But non-resonant relaxation (NRR) is not affected by the precession, and because the time scale for NRR to change $\ell$ is proportial to $\ell^2$, at sufficiently small $\ell$, this time must become shorter than the AR time scale.

As we discuss below, the value of $\ell$ at the transition between regimes (ii) and (iii) depends on various quantities, including the number of stars (for a given $M_\star$, say). This result is likely to be important when calculating rates of ``barrier penetration,'' since the dominant mechanism driving diffusion past the SB will be different in different nuclear models. We will argue that in the simulations of MAMW11, the particle number was small enough that the NRR regime extended all the way to the SB for some values of the semimajor axis; while in more realistic nuclear models, diffusion below the SB would need to contend with AR.

The time scales plotted in Figure~\ref{fig:introduction_figure} do not tell the whole story. For instance, there is a separate time scale associated with drift in angular momentum (due to the first-order diffusion coefficients) and that time scale is relevant to the ``bounce'' phenomenon that occurs near the SB, as described below. At sufficiently small $L$, orbits around a spinning SBH will also be affected by spin-orbit (Lense-Thirring) torques \citep{mv12}, a phenomenon whose consequences for the angular momentum evolution will not be explored here.

\section{Method}
\label{sect:methods}
In \textsc{TPI} we exploit the property that well within the sphere of influence of a SBH the motion of the stars is dominated by the SBH, i.e. the stellar motion is well described in terms of perturbed Keplerian orbits. Torques acting on these stars give rise to exchange of angular momentum between stars. This process is known as resonant relaxation (RR) and affects the eccentricities of the orbits. Furthermore, two-body (non-resonant) interactions affect the orbital energies in addition to their angular momenta. When considering a large ensemble of stars, however, these processes should not strongly affect the {\it mean} angular momenta and energies provided that the system is dynamically relaxed. On the other hand, energy exchange and RR are important when considering individual stars. This consideration motivates a split between dynamically relaxed field stars and test stars that evolve dynamically in time as a consequence of both angular momentum and energy exchanges with field stars. We define a test star as a particle with zero mass, i.e. a particle that does not affect the field stars and other test stars. 

The field stars are assumed to follow uniformly-precessing Kepler orbits with constant semimajor axis $a$, eccentricity $e$, inclination $i$ and longitude of the ascending node $\Omega$. The argument of periapsis $\omega$ is advanced linearly in time according to the rate prescribed by analytical formulae that include precession due to general relativity (Schwarzschild precession) and Newtonian precession due to the distributed mass in stars (mass precession). The advance per orbital period $P$ due to Schwarzschild precession, to first post-Newtonian (PN) order, is given by \citep{weinberg72}:
\begin{align}
\Delta \omega_{\mathrm{1PN},P} = 6 \pi \frac{G (m_\star + M_\bullet)}{a \left(1-e^2\right)c^2}.
\label{eq:1PNprecession}
\end{align}
Here $m_\star$ is the field star mass, $M_\bullet$ is the SBH mass, $G$ is the gravitational constant and $c$ is the speed of light. Periapsis advance due to mass precession depends on the detailed distribution of the mass. In all the models considered here, we assume a spherical field-star distribution with density $\rho_\star(r) \propto r^{-2}$. In this case, the apsidal advance due to mass precession per orbital period is \citep[][Eq. (4.87)]{bookmerritt13}:
\begin{align}
\Delta \omega_{\mathrm{MP},P} = - 2 \pi \frac{M_\star(a)}{M_\bullet} \frac{\sqrt{1-e^2}}{1+\sqrt{1-e^2}}.
\label{eq:MP}
\end{align}
Here $M_\star(a)$ is the total field star mass within radius $r=a$.

In \textsc{TPI} the motion of the field stars is calculated with a Kepler solver that advances the positions and velocities for a given time interval assuming unperturbed Keplerian ellipses. The resulting positions and velocities $\boldsymbol{r}$ and $\boldsymbol{v}$ are subsequently rotated in the orbital plane to account for the in-plane precession resulting from both  Schwarzschild and mass precession:
\begin{subequations}\label{eq:rot}
\begin{align}
\boldsymbol{r} \rightarrow \cos(\Delta \omega) \, \boldsymbol{r} + \sin(\Delta \omega) \, \boldsymbol{\hat{\ell}}\times \boldsymbol{r}, \\
\boldsymbol{v} \rightarrow \cos(\Delta \omega) \, \boldsymbol{v} + \sin(\Delta \omega) \, \boldsymbol{\hat{\ell}}\times \boldsymbol{v}.
\end{align}
\end{subequations}
Here $\Delta \omega = \Delta \omega_{\mathrm{1PN},\Delta t} + \Delta \omega_{\mathrm{MP},\Delta t}$ is the total precession angle in time interval $\Delta t$, and $\boldsymbol{\hat{\ell}} = \boldsymbol{r} \times \boldsymbol{v}/|| \boldsymbol{r} \times \boldsymbol{v}||$ is the unit specific angular momentum vector. By treating the motion of the field stars in this way the interactions between field stars are modeled in an approximate method that neglects two-body encounters and resonant torques. This makes it computationally feasible to include a large ($\gtrsim 10^3$) number of field stars. 

The test stars are integrated using a direct-summation $N$-body code. It is advantageous to employ Kustaanheimo-Stiefel regularization \citep{ks65} for their motion around the SBH. Tests have shown that in the absence of field stars this method reduces the required number of integration steps while at the same time it increases the accuracy. In \textsc{TPI} each test star forms a regularized and independent two-body system with the SBH. The perturbing acceleration $\boldsymbol{a}$ is given by: 
\begin{align}
\boldsymbol{a} = \boldsymbol{a}_\mathrm{SBH, \, PN} + \boldsymbol{a}_\mathrm{field, \, 0PN}.
\label{eq:a_pert}
\end{align}
Here $\boldsymbol{a}_\mathrm{SBH, \, PN}$ is the PN acceleration from the SBH. We have implemented 1PN, 2PN and 2.5PN terms for a non-spinning SBH \citep{damder81}, and 1.5PN and 2.0PN terms that arise from spin of the SBH \citep{kidder95}. In most of the simulations presented here we restrict to including only the 1PN terms. The quantity $\boldsymbol{a}_\mathrm{field, \, 0PN}$ is the Newtonian acceleration from the field stars. To integrate the regularized equations of motion we use a standard 4th order Hermite predict, evaluate and correct integration scheme \citep{makino91}. 

The implementation of the PN terms in our algorithm can be compared to other algorithms based on geodesic solvers \citep{bas14}. In the latter algorithms the Schwarzschild metric is used to obtain relativistic corrections to the equations of motion, thereby assuming that the metric is determined solely by the SBH, i.e. that the mass $m$ of the particle orbiting the SBH can be neglected compared to $M_\bullet$. This is similar to our algorithm, in which $m$ in the PN terms is set to zero (i.e. the symmetric mass ratio $\nu=mM_\bullet/(m+M_\bullet)^2=0$). A major difference is that in our algorithm the corrections are included to finite order of $v/c$, whereas a geodesic solver is in principle accurate to arbitrary order, provided that $m$ is sufficiently small and that therefore there is no dissipation due to gravitational waves. 

Test stars have individual block time steps $\Delta t_\mathrm{block}$ that are determined dynamically using time symmetrization \citep{fun96}. At the end of each integration step a new time-symmetric time step $\Delta t$ is calculated from:
\begin{align}
\nonumber & \Delta t = \frac{1}{2} ||\boldsymbol{u}||^2 \left [ f\left (\boldsymbol{u}^{(2)}_b, \boldsymbol{u}^{(3)}_b,\boldsymbol{u}^{(4)}_b \right) + f\left (\boldsymbol{u}^{(2)}_e, \boldsymbol{u}^{(3)}_e,\boldsymbol{u}^{(4)}_e \right) \right]; \\
& f\left (\boldsymbol{u}^{(2)}, \boldsymbol{u}^{(3)},\boldsymbol{u}^{(4)} \right) = \eta \times \mathrm{min} \left [\frac{ ||\boldsymbol{u}^{(2)}|| }{ ||\boldsymbol{u}^{(3)}||}, \left ( \frac{ ||\boldsymbol{u}^{(2)}|| }{ ||\boldsymbol{u}^{(4)}||} \right )^{1/2} \right ].
\label{eq:timestepfunction}
\end{align}
Here $\boldsymbol{u}$ is the regularized position vector, $(i)$ indicates the $i^\mathrm{th}$ derivative with respect to the regularized time, $\eta$ is a time step parameter and the indices $b$ and $e$ indicate the beginning and end of the current step, respectively. Subsequently the block time step is computed from $\Delta t_\mathrm{block} = 2^k \times \Delta t_\mathrm{min}$, where $k$ is the largest positive integer such that $2^k \times \Delta t_\mathrm{min} < \Delta t$ and $\Delta t_\mathrm{min} = 1 \times 10^{-14} \, \mathrm{yr}$ is the minimum time step that we allow in the simulations. The time between iterations is given by the minimum of the test star block time steps. At each iteration the positions and velocities of the field stars are shifted in their Kepler orbits and rotated according to equation~(\ref{eq:rot}). 

In the case of a large number of test and/or field stars (typically if either number is $\gtrsim 10^3$) the evaluation of $\mathbf{a}_\mathrm{field, \, 0PN}$ for all the test stars is the most computationally expensive part of the integration of the equations of motion. For this reason we have implemented parallel computation of $\boldsymbol{a}_\mathrm{field, \, 0PN}$ on CPUs using \textsc{OpenMP} as well as GPU-accelerated evaluation of $\boldsymbol{a}_\mathrm{field, \, 0PN}$ using the \textsc{Sapporo} library \citep{gab09}.

In \textsc{TPI} the detection of captures of test stars by the SBH is implemented. We assume that during the integration step the test star moves in a straight line $\boldsymbol{r}(s) = \boldsymbol{r}_b + s (\boldsymbol{r}_e - \boldsymbol{r}_b)$ where $\boldsymbol{r}_b$ and $\boldsymbol{r}_e$ are the (non-regularized) position vectors at the beginning and the end of the integration step, and $s \in [0,1]$ is a parameter. We check if any of the points on this trajectory satisfies $\boldsymbol{r}(s)^2 = r_\mathrm{capt}^2$ for $s \in [0,1]$, where $r_\mathrm{capt}$ is the capture radius. If this is the case then the test star has either just grazed or penetrated the capture sphere and we register a capture event. After a test star has been captured it is recorded and removed from the simulation.

To validate \textsc{TPI} we have performed several simple tests of interactions between test stars and the SBH and between test and field stars. These tests are described in Appendix \ref{app:simple_tests}. In this paper the time step parameter is set to $\eta = 0.02$; this choice is motivated in the latter appendix. Tests of \textsc{TPI} in the regime below the SB, which is the main focus of this paper, are described in detail in \S\,\ref{sect:belowSB}, where we also compare our results with those from other, slower, $N$-body codes in which the simplifying assumptions adopted in \textsc{TPI} are relaxed. 

\begin{figure*}
\center
\includegraphics[scale = 0.46, trim = 0mm 0mm 0mm 0mm]{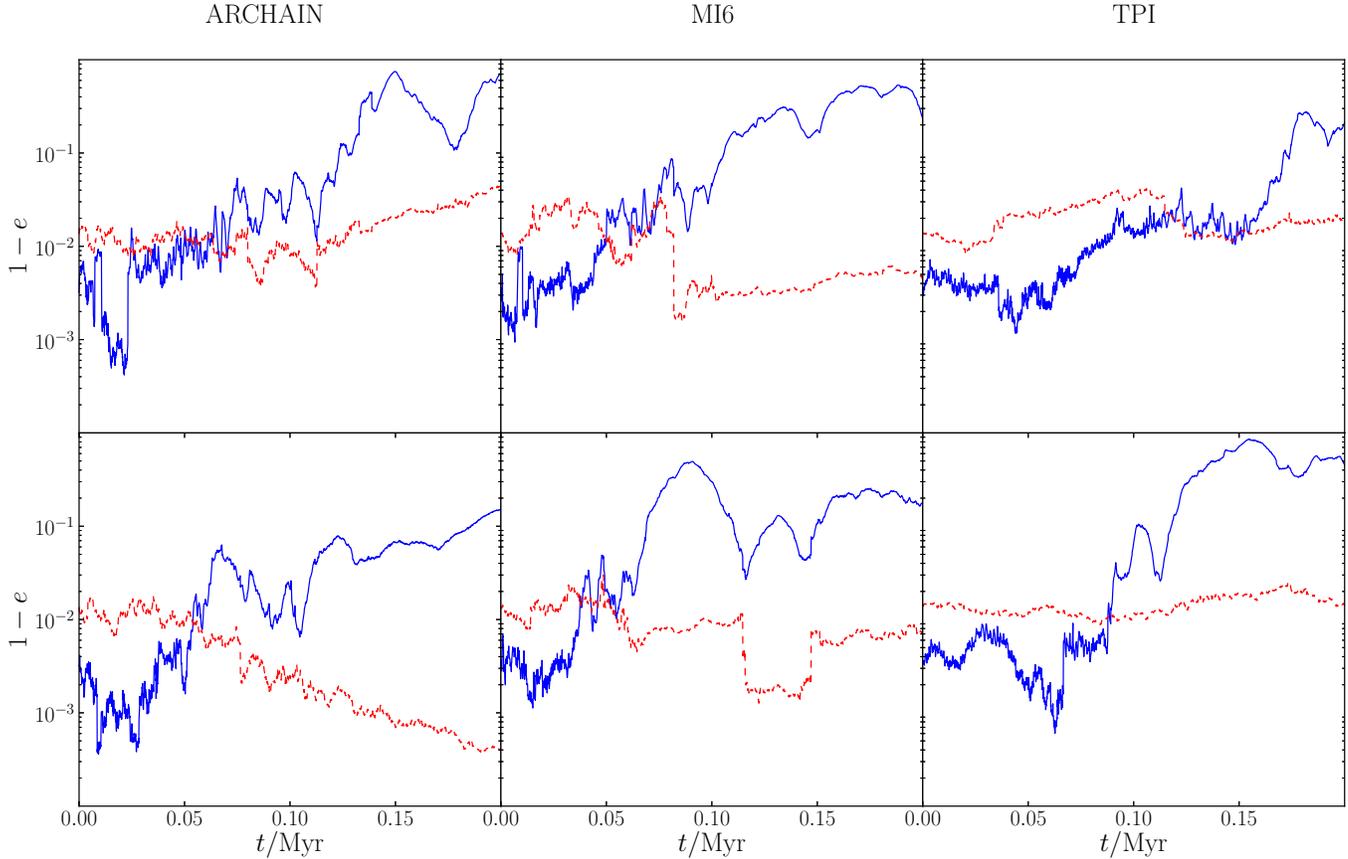}
\caption{\small Eccentricity evolution (blue solid lines) and the SB (equation~(\ref{eq:ell_SB}); red dashed lines) as function of time for test stars with initially $a = 2 \, \mathrm{mpc}$ and $\log_{10}(1-e) = -2.5$ as computed with \textsc{ARCHAIN} (first column) \textsc{MI6} (second column) and \textsc{TPI} (third column). We show cases in which the test star is not captured and moves to above the SB. The initial conditions differ in each panel, hence the panels should not be compared directly. }
\label{fig:comparison_et_all}
\end{figure*}

Before describing the results of the code comparisons, we note that even the more accurate algorithms discussed below contain potentially important approximations. These codes include the Newtonian terms from the SBH and the $N$ bodies, plus the 1PN terms from the SBH alone. The latter terms are proportional to $G^2M_\bullet^2/r^3c^2$, or to $(GM_\bullet/r)(v^2/c^2)$, with $r$ the distance from the SBH. At 1PN order, one can potentially do better, since the full $N$-body Hamiltonian is known, the so-called EIH (Einstein-Infeld-Hoffmann) Hamiltonian \citep{eih38}. The EIH equations of motion also include terms of order $M_\bullet m_\star$ and $m_\star^2$ \citep{will13}. Given the small values of $m_\star/M_\bullet$ considered here, only the former, or ``cross'', terms are likely to matter. In the context of apsidal precession, one expects the cross terms to induce changes of order 
\begin{equation}
(\Delta\omega)_\mathrm{cross} \approx (\Delta\omega)_\mathrm{M} \times
(\Delta\omega)_\mathrm{PN}, \nonumber
\end{equation}
that is, the product of the shifts due to mass precession and to Schwarzschild precession considered individually. Over sufficiently long times, the effects of the cross terms will of course accumulate, and it is an open question whether this might significantly impact the evolution of orbits near the SB.

\section{Orbital evolution below the SB; small-$N$ simulations}
\label{sect:belowSB}
\subsection{Initial conditions}
\label{sect:belowSB:init}
As mentioned in \S\,\ref{sect:introduction} there exist several processes that can deposit stars below the SB on time scales of the order the Kepler period $P(a)$. Here we study the evolution of orbits after deposition below the SB using simulations with \textsc{TPI}. We also include simulations performed with two direct-summation $N$-body codes, \textsc{MI6} \citep{nitmak08,iwasawa11} and \textsc{ARCHAIN} \citep{mikkola08}. \textsc{MI6} uses a mixed fourth-order and sixth-order Hermite integration scheme. The SBH is kept fixed at the origin, simplifying the equations of motion. In particular, this allows for PN accelerations to be calculated for star-SBH interactions only, avoiding the calculation of PN accelerations for star-star interactions. The latter are assumed to be negligible compared to the former. In \textsc{MI6} 1PN and 2.5PN accelerations are included. The \textsc{ARCHAIN} code is an essentially exact $N$-body code owing to chain regularization and it includes 1PN, 2PN and 2.5PN terms.

The initial conditions of our simulations were similar to those of the $N$-body simulations performed by MAMW11 and \citet{bas14}. We sampled field stars of mass $m_\star = 50 \, \mathrm{M}_\odot$ in Kepler orbits around a SBH of $M_\bullet = 1.0 \times 10^6 \, \mathrm{M_\odot}$ with the following orbital distributions: semimajor axes $a$ were sampled randomly between $a_\mathrm{min} = 0.1 \, \mathrm{mpc}$ and $a_\mathrm{max} = 10 \, \mathrm{mpc}$ corresponding to a stellar density distribution $\rho_\star(r) \propto r^{-2}$; a thermal eccentricity distribution was assumed and orbital angles were sampled randomly. The total number of field stars was $N_\mathrm{max} \equiv N_\star(a_\mathrm{max}) = 50$. We also carried out simulations with larger $N_\star(a_\mathrm{max})$ with \textsc{TPI}; the latter are discussed in \S\,\ref{sect:belowSB:diffusion}.

In the case of \textsc{ARCHAIN} and \textsc{MI6} we placed five of the field stars below the SB in the $(a,e)$ parameter space. We define above and below the SB as $\ell \equiv \sqrt{1-e^2} > \ell_\mathrm{SB}$ and $\ell < \ell_\mathrm{SB}$, respectively, where $\ell_\mathrm{SB} = \ell_\mathrm{SB}(a)$ is defined in equation (\ref{eq:ell_SB}); for the $N$-body simulations $M_\star(a) = m_\star N_\star(a) \approx m_\star N_\mathrm{max} \, (a/a_\mathrm{max})$. We will refer to these five stars as test stars, but we note that in the case of \textsc{ARCHAIN} and \textsc{MI6} these stars are not massless and have the same mass as the field stars. In the case of \textsc{TPI} we initiated five test stars below the SB at the same values of $a$ and $e$ as those of the five test stars in \textsc{ARCHAIN} and \textsc{MI6}. In each simulation the five test stars shared a common value of $a$ and $e$ but were initiated with different (random) orbital angles and phases. We carried out a series of simulations with the following combinations of the initial values of $a$ and $e$:
\begin{align}
\begin{array}{ll}
a = 2 \, \mathrm{mpc}; & \log_{10}(1-e) \in \{-3.0,-2.5,-2.0\}; \\
a = 4 \, \mathrm{mpc}; & \log_{10}(1-e) \in \{-3.3, -2.9\}; \\
a = 8 \, \mathrm{mpc}; & \log_{10}(1-e) = -3.8.
\end{array}
\label{eq:test_star_ae}
\end{align}
For each combination of $a$ and $e$ (i.e. each simulation with five test stars below the SB) we ran simulations with five different random realizations, obtaining 25 time series for each $(a,e)$ pair. 

The integration time per simulation was set to 1 Myr. The capture radius was $r_\mathrm{capt} = 8 \, r_g \approx 3.8 \times 10^{-4} \, \mathrm{mpc}$, consistent with the capture radius of a compact object onto a non-spinning SBH \citep{will12}. In all simulations we included 1PN terms; we also carried out integrations in which the 2.5PN terms were included (in case of \textsc{ARCHAIN}, 2PN terms are included as well). However, because the 2.5PN terms cannot be included self-consistently in \textsc{TPI} we present in \S\,\ref{sect:belowSB} only results in which the 2.5PN terms were excluded, with the exception of \S\,\ref{sect:belowSB:capture}.

\subsection{Qualitative behavior}
\label{sect:belowSB:qualitative}
We show in Figure \ref{fig:comparison_et_all} the eccentricity evolution for a test star with initially $a = 2 \, \mathrm{mpc}$ and $\log_{10}(1-e) = -2.5$ as computed with each of the three codes, without the 2.5PN terms. We select two cases (corresponding to the two rows) in which the test star crosses the SB from below to above. Note that the initial conditions differ in each panel of Figure \ref{fig:comparison_et_all}, hence the panels should not be compared directly. Based on these and similar plots, we make the following qualitative observations.
\begin{enumerate}
\item Below the SB the eccentricity varies in an approximately periodic fashion, on a (short) time scale consistent with the Schwarzschild precession time. There is also a component of its evolution that can be described as a random walk. (The latter was referred to as ``anomalous relaxation'' in \S\ref{sect:timescales}.)
\item Above the SB the eccentricity variations are much larger, extending to $e\approx 0$, and have a longer associated time scale. These features can be explained qualitatively in terms of RR, 
which is not quenched above the barrier.
\item Stars above the SB tend to remain there, since their trajectories ``bounce'' on striking the barrier from above.
\item As a consequence of items (ii) and (iii), the SB acts as a diode or a one-way membrane: stars can only easily cross it in one direction, from below (high $e$) to above (low $e$).
\end{enumerate}
In \S\,\ref{sect:belowSB:e_osc} and \S\,\ref{sect:belowSB:diffusion} we explore some of these properties more quantitatively, and we also use them as a means of comparing the different codes. 

\begin{table}
\begin{tabular}{cccccccc}
\toprule
$a/\mathrm{mpc}$ & $\log_{10}(1-e)$ & \multicolumn{6}{c}{$N_\mathrm{capt}$} \\
& & \multicolumn{2}{c}{ARCHAIN} & \multicolumn{2}{c}{MI6} & \multicolumn{2}{c}{TPI} \\
& & W & WO & W & WO & W & WO \\
\midrule
2 & -2.0 & 4  & 8  & 7  & 6  & 1  & 4 \\
2 & -2.5 & 9  & 14 & 11 & 9  & 13 & 8 \\ 
2 & -3.0 & 13 & 15 & 18 & 16 & 15 & 17 \\
4 & -2.9 & 11 & 13 & 9  & 7  & 13 & 11 \\
4 & -3.3 & 18 & 17 & 18 & 17 & 11 & 12 \\
8 & -3.8 & 18 & 13 & 22 & 15 & 21 & 21 \\
\bottomrule
\end{tabular}
\caption{Number of captured test stars at $t = 1 \, \mathrm{Myr}$ for the three different codes. A distinction is made between simulations with (W) and without (WO) 2.5PN terms. }
\label{table:test_star_captures}
\end{table}

\subsection{Capture rates}
\label{sect:belowSB:capture}
Before turning to our observations from \S\,\ref{sect:belowSB:qualitative} we present in Table \ref{table:test_star_captures} the number of captured stars at the end of the simulation for the three codes \textsc{ARCHAIN}, \textsc{MI6} and \textsc{TPI}. We include both simulations with (W) and without (WO) 2.5PN terms. Although the exact number of captured stars $N_\mathrm{capt}$ differs between the three codes, in all three cases there is a similar trend of increasing $N_\mathrm{capt}$ with both $a$ and $e$. For example, for each of the codes without the 2.5PN terms $N_\mathrm{capt}$ increases by a factor $\sim 3 - 4$ from $a=2\,\mathrm{mpc}$ and $\log_{10}(1-e) = -2.0$ to $a=8\,\mathrm{mpc}$ and $\log_{10}(1-e) = -3.8$. 

\begin{figure}
\center
\includegraphics[scale = 0.435, trim = 0mm 0mm 0mm 0mm]{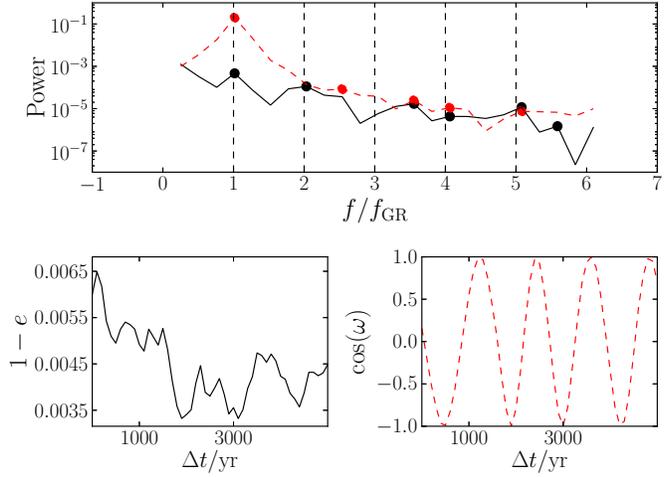}
\caption{\small Top panel: power spectra of the eccentricity (black solid line) and the cosine of the argument of periapsis (red dashed line) for a simulation with initially $a = 2 \, \mathrm{mpc}$ and $\log_{10}(1-e) = -2.5$ as computed with \textsc{TPI}. Local maxima are indicated with bullets. The time series of the eccentricity and argument of periapsis used for the power spectra are shown in the bottom left and bottom right panels (time is expressed with respect to the beginning of the sampling interval). Refer to the text for the method used to select the sampling interval. Most of the power is contained at $f=f_\mathrm{GR}$.}
\label{fig:power_spectrum_example}
\end{figure}

\begin{figure}
\center
\includegraphics[scale = 0.435, trim = 0mm 0mm 0mm 0mm]{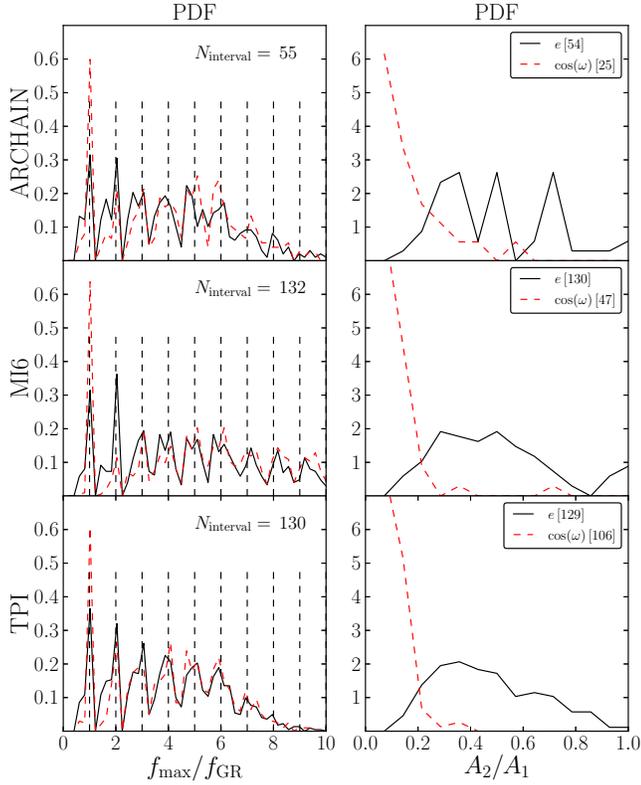}
\caption{\small Left column: distributions of the local frequency maxima determined from the power spectra for the three codes. Black solid lines apply to the eccentricity and red dashed lines apply to the argument of periapsis. Right column: distributions of the ratios of the amplitude at $f_\mathrm{max} \approx 2 f_\mathrm{GR}$ to the amplitude at $f_\mathrm{max} \approx f_\mathrm{GR}$. The numbers in brackets are the number of data points in the frequency bin. }
\label{fig:power_spectrum_all}
\end{figure}

\subsection{Eccentricity oscillations below the SB}
\label{sect:belowSB:e_osc}
\subsubsection{Frequency of oscillations}
\label{sect:belowSB:e_osc:freq}
We obtained power spectra of the eccentricity and argument of periapsis from the simulations below the SB using the following method. For each time in the simulation $t_0$ we computed the time scale $2\,t_\mathrm{GR}$ for Schwarzschild precession to change $\omega$ by $2\pi$, $2\,t_\mathrm{GR} = 2\pi P/\Delta \omega_{\mathrm{1PN},P} = (1/3) (a/r_g) (1-e^2)P$ (cf. equation~(\ref{eq:1PNprecession})). Subsequently we recalculated $2\,t_\mathrm{GR}$ based on the mean values of $a$ and $e$ in the interval $t_0 < t < t_0 + \Delta t$, where $\Delta t = 8 \, t_\mathrm{GR}$\footnote{The factor 8 in $\Delta t = 8 \, t_\mathrm{GR}$ is a compromise between a long sampling interval (leading to much noise induced by two-body encounters) and a short sampling interval (leading to too few data points).}. This procedure was repeated until convergence with respect to $t_\mathrm{GR}$ had occurred. The interval was rejected if for any of the points within it the star was above the SB, the number of points was less than 50 or the fractional changes in $a$ and $e$ satisfied $|\Delta a/a| > 0.04$ and $|\Delta e/e| > 0.04$, respectively. The latter criteria serve to minimize noise in the power spectra induced by sudden changes in $a$ and $e$ due to NRR. Power spectra of the eccentricity and argument of periapsis were subsequently computed for the accepted intervals. The starting search time for the subsequent interval was $t_0 + \Delta t$. 
 
We show in Figure \ref{fig:power_spectrum_example} an example of power spectra obtained using the above method in a simulation with initially $a = 2 \, \mathrm{mpc}$ and $\log_{10}(1-e) = -2.5$, as computed with \textsc{TPI}. There is a peak in both power spectra at $f = f_\mathrm{GR}$, where $f_\mathrm{GR} \equiv 1/(2\,t_\mathrm{GR})$. This is consistent with our observation in \S\,\ref{sect:belowSB:qualitative} that below the SB the eccentricity oscillations occur on the Schwarzschild precession time scale. The peak in the power spectrum at $f\approx f_\mathrm{GR}$ is higher for the argument of periapsis compared to the eccentricity because Schwarzschild precession affects the argument of periapsis directly, whereas the effect on the eccentricity is indirect, i.e., through the $\sqrt{N}$ torques. 

We applied the above method to all simulated $(a,e)$ pairs of the test stars (cf. equation~(\ref{eq:test_star_ae})). For the obtained power spectra we determined the local maxima (shown for one example in the top panel of Figure \ref{fig:power_spectrum_example} with bullets) and we recorded the corresponding frequencies $f_\mathrm{max}$ and amplitudes $A$, where $A$ is the square root of the power. We show in the first column of Figure \ref{fig:power_spectrum_all} the resulting distributions of $f_\mathrm{max}$ for the three codes. There is a clear peak in the eccentricity spectra at $f_\mathrm{max} \approx f_\mathrm{GR}$. This peak can be interpreted as implying that the torquing potential (due to the $\mathcal{O}(N^{-1/2})$ asymmetry in the field star distribution) is basically lopsided, or $m=1$, in character (MAMW11). Higher-order terms in the multipole expansion of the field star potential would give rise to eccentricity oscillations at higher integer frequencies of $f_\mathrm{GR}$. The results shown in the first column of Figure \ref{fig:power_spectrum_all} indicate that these higher-order contributions are important, though typically not dominant.

We determined the amplitudes $A_n$ of the peaks at higher integer frequencies $f_\mathrm{max} \approx n f_\mathrm{GR}$ and we normalized these to $A_1$, the amplitude at $f_\mathrm{max} \approx f_\mathrm{GR}$. Frequencies for each $n$ were selected from data satisfying $n-0.3 < f_\mathrm{max}/f_\mathrm{GR} < n+0.3$, where the limits are motivated by the distributions shown in first column of Figure \ref{fig:power_spectrum_all}. We show the resulting distributions of $A_2/A_1$ for the three codes in the second column of Figure \ref{fig:power_spectrum_all}. These distributions have peaks at roughly similar locations for each of the three codes, $A_2/A_1 \approx 0.4$. In the case of \textsc{ARCHAIN} there appear to be several peaks, which may be due to a lack of data. We show in Table \ref{table:amplitude_ratios} the median and median absolute deviations of $A_n/A_1$ for $n \leq 8$. These values are consistent between the three codes. 

It is conceivable that precession of the field stars affects the ratios $A_n/A_1$ in the above analysis. We have verified that for the duration of the sampling intervals the field stars do not precess by a large amount, i.e. in the case of \textsc{MI6} on average only 3 (out of 45) field star orbits precess over an angle $\Delta \omega > \pi/2$ during $\Delta t$. In addition, we have carried out the analysis for \textsc{TPI} without field star precession (i.e. by setting $\Delta \omega = 0$ in equation~(\ref{eq:rot})) and we found no substantially different results, e.g. $A_2/A_1 \approx 0.52 \pm 0.21$, consistent with $A_2/A_1 \approx 0.42 \pm 0.16$ in the case where precession of the field stars is included (cf. Table \ref{table:amplitude_ratios}). 

\begin{table}
\begin{tabular}{cccc}
\toprule
$n$ & \multicolumn{3}{c}{$A_n/A_1$} \\
    & \textsc{ARCHAIN} & \textsc{MI6} & \textsc{TPI} \\
\midrule
2  &  0.46  $\pm$  0.19  [54] &  0.55  $\pm$  0.29  [130] &  0.42  $\pm$  0.16  [129] \\
3  &  0.22  $\pm$  0.09  [53] &  0.23  $\pm$  0.09  [131] &  0.25  $\pm$  0.10  [126] \\
4  &  0.18  $\pm$  0.06  [46] &  0.14  $\pm$  0.06  [113] &  0.18  $\pm$  0.06  [123] \\
5  &  0.13  $\pm$  0.05  [43] &  0.11  $\pm$  0.06  [111] &  0.13  $\pm$  0.06  [113] \\
6  &  0.12  $\pm$  0.06  [38] &  0.12  $\pm$  0.05  [106] &  0.12  $\pm$  0.05  [93] \\
7  &  0.08  $\pm$  0.05  [21] &  0.09  $\pm$  0.05  [82] &  0.07  $\pm$  0.02  [46] \\
8  &  0.09  $\pm$  0.05  [15] &  0.08  $\pm$  0.05  [75] &  0.06  $\pm$  0.03  [18] \\
\bottomrule
\end{tabular}
\caption{Amplitudes $A_n$ for the eccentricity power spectra at frequencies $f_\mathrm{max} \approx n f_\mathrm{GR}$ normalized to $A_1$. The values are the median values and the median absolute deviations. The numbers in brackets are the number of data points in the frequency bin.}
\label{table:amplitude_ratios}
\end{table}

\begin{figure}
\center
\includegraphics[scale = 0.435, trim = 0mm 0mm 0mm 0mm]{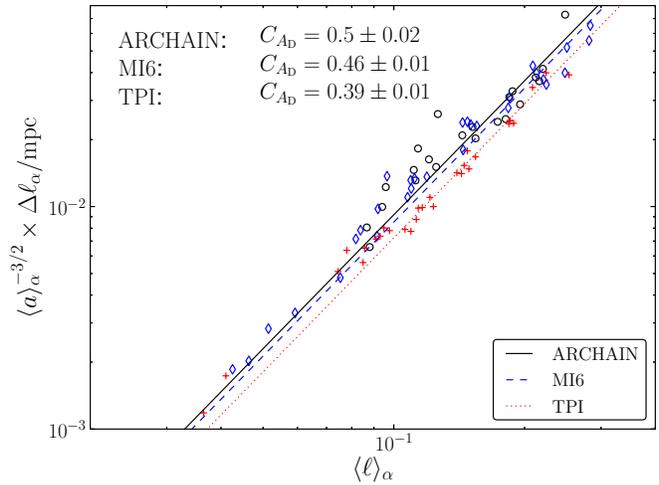}
\caption{\small Amplitude of eccentricity oscillations binned in mean value of $a$ and $\langle \ell \rangle$ as function of $\langle \ell \rangle$, averaged over all orientations $\alpha$. Black circles: \textsc{ARCHAIN}; blue diamonds: \textsc{MI6}; red plusses: \textsc{TPI}. The solid, dashed and dotted lines show least squares fits to the data for the three codes, respectively, weighted by the bin size corresponding to each data point. The fitted values of $C_{A_\mathrm{D}}$ are shown in the top left. }
\label{fig:e_amplitude}
\end{figure}

\subsubsection{Amplitude of oscillations}
\label{sect:belowSB:e_osc:amplitude}
Here a method is presented to obtain the amplitude of eccentricity oscillations below the SB and the results are compared to theoretical predictions. We expect the amplitude of the latter oscillations to depend on the (dimensionless) angular momentum $\ell = \sqrt{1-e^2}$ and hence the distance in angular momentum to the SB. This is due to more rapid Schwarzschild precession for lower $\ell$ and therefore more efficient quenching of the effects of the $\sqrt{N}$ torques that would otherwise drive RR. 

We adopt the Hamiltonian model presented in MAMW11 that includes Schwarzschild precession, mass precession and the effects of a lopsided mass distribution, assumed to be oriented with respect to the orbit with an angle $\alpha$. Let $\Delta \ell \equiv \ell_\mathrm{max} - \ell_\mathrm{min}$ be the amplitude of oscillations in $\ell$, where $\ell_\mathrm{min}$ and $\ell_\mathrm{max}$ are the minimum and maximum angular momenta during one oscillation of duration $2\,t_\mathrm{GR}$, respectively. Then if $\ell$ is sufficiently small, i.e. if the second and third terms of MAMW11 Eq. 41a can be neglected with respect to the first term, $\Delta \ell$ depends on the average angular momentum $\langle \ell \rangle \equiv (1/2)(\ell_\mathrm{min} + \ell_\mathrm{max})$ via the relation (MAMW11 Eq. 46):
\begin{align}
\Delta \ell \approx 2 \langle \ell \rangle^2 A_\mathrm{D} \sin(\alpha).
\label{eq:deltaell}
\end{align}
Here $A_\mathrm{D}$ is a dimensionless parameter that specifies the strength of the lopsided component of the distributed mass in the Hamiltonian model. In terms of the model parameters $A_\mathrm{D}$ is expressed by (MAMW11 Eq. 43b\footnote{The factor $(1/3)$ in MAMW11 Eq. 43b should be replaced by $(1/2)$ (David Merritt, private communication).}):
\begin{align}
A_\mathrm{D} = \frac{C_{A_\mathrm{D}}}{2} \frac{S}{G M_\bullet/a^2} \frac{a}{r_g} = \frac{C_{A_\mathrm{D}}}{2\sqrt{N_\star(a)}} \frac{ M_\star(a)}{M_\bullet} \frac{a}{r_g}.
\label{eq:AD}
\end{align}
Here $S$ is the amplitude of the lopsided distortion, and $M_\star(a)$ and $N_\star(a) = M_\star(a)/m_\star$ are the enclosed stellar mass and the number of enclosed stars, respectively. The parameter $C_{A_\mathrm{D}}$ captures unspecified uncertainties in this model.

We obtained $\Delta \ell$ from the simulations with a method similar to that used to obtain power spectra. In order to minimize the effect of directed changes in $\ell$ over time scales longer than $t_\mathrm{GR}$ the sampling interval was shortened to $\Delta t = 4 \, t_\mathrm{GR}$ and the number of required points per sampling interval was reduced to 20. For the resulting sampling intervals we recorded the minimum and maximum values of $\ell$ and the mean value of $a$, $\langle a \rangle$. We binned the data into 100 bins of $\langle a \rangle$  with $1 < \langle a \rangle /\mathrm{mpc} < 10$ and 10 bins of $\langle \ell \rangle $ with $0 < \langle \ell \rangle < 0.3$. For each bin we computed the mean values of $\langle \ell \rangle$ and $\Delta \ell$, which amounts to averaging these quantities over the angle $\alpha$. We rejected bins if the bin size was less than or equal to 5.

In order to compare results from the simulations to the prediction of equation~(\ref{eq:deltaell}) we average this equation over the unit sphere and substitute $A_\mathrm{D}$ using equation~(\ref{eq:AD}) with $N_\star(a) \approx (a/a_\mathrm{max}) N_\mathrm{max}$. Subsequently we obtain:
\begin{align}
\langle a \rangle_\alpha^{-3/2} \times \Delta \ell_\alpha \approx C_{A_\mathrm{D}} \frac{\pi}{4} \frac{m_\star}{M_\bullet} \sqrt{\frac{N_\mathrm{max}}{a_\mathrm{max}}} r_g^{-1} \langle \ell \rangle_\alpha^2.
\label{eq:deltaell_av_alpha}
\end{align}
Here the subscript $\alpha$ indicates the average over the unit sphere. We show in Figure \ref{fig:e_amplitude} the resulting amplitudes for the three codes and linear least squares fits to the data according to equation~(\ref{eq:deltaell_av_alpha}), where we used the number of points in each bin as relative weights. The data from the simulations is consistent with the prediction $\Delta \ell \propto \langle \ell \rangle^2$, for each of the three codes. The resulting values of $C_{A_\mathrm{D}}$ that we obtain from the fits are shown in the top left of Figure \ref{fig:e_amplitude}.

\begin{figure}
\center
\includegraphics[scale = 0.435, trim = 0mm 0mm 0mm 0mm]{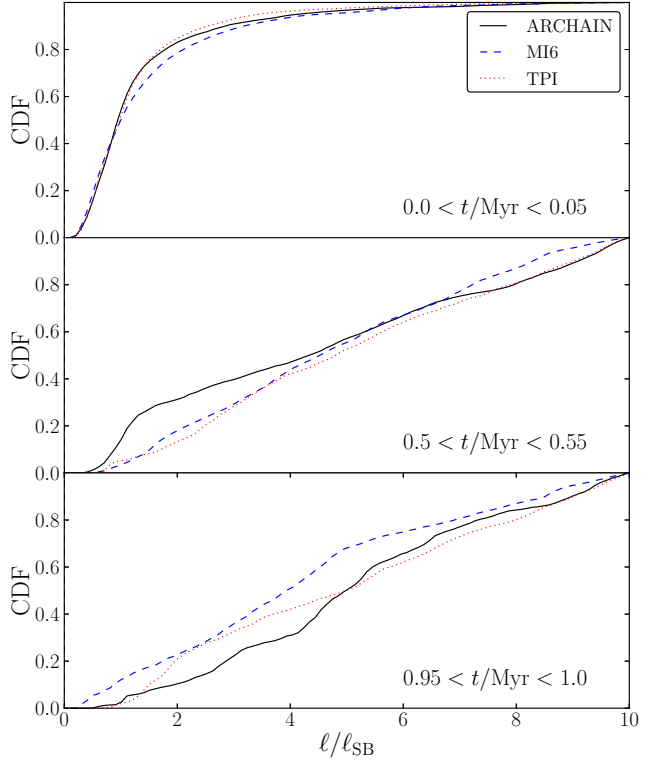}
\caption{Cumulative distributions of the dimensionless angular momentum $\ell$ normalised to the angular momentum associated with the SB, $\ell_\mathrm{SB}$ (cf. equation~(\ref{eq:ell_SB})). Three time intervals are shown for the three codes. Black solid lines: \textsc{ARCHAIN}; blue dashed lines: \textsc{MI6}; red dotted lines: \textsc{TPI}. }
\label{fig:distance_evolution}
\end{figure}

\subsection{Diffusion in angular momentum above and below the SB}
\label{sect:belowSB:diffusion}
In \S\,\ref{sect:belowSB:e_osc} we described the eccentricity oscillations that occur below the SB. However, in our simulations, not all orbits remain below the SB indefinitely. To illustrate this we show in Figure \ref{fig:distance_evolution} the cumulative distributions of $\ell/\ell_\mathrm{SB}$, with $\ell_\mathrm{SB}$ given by equation~(\ref{eq:ell_SB}), at three time intervals $0<t/\mathrm{Myr}<0.05$, $0.50<t/\mathrm{Myr}<0.55$ and $0.95<t/\mathrm{Myr}<1.0$. At the earliest time in the simulations the majority of orbits are below the SB ($\ell/\ell_\mathrm{SB}<1$). As time progresses the latter quantity gradually increases and by the end of the simulation nearly all orbits ($\gtrsim 90\%$) have diffused above the SB (i.e. $\ell \gg \ell_\mathrm{SB}$). The deviations in the cumulative distributions of $\ell/\ell_\mathrm{SB}$ between the three codes appear to increase with time. This may be due to various reasons, including exponential divergence in the gravitational $N$-body problem and the increase in the amplitude of eccentricity oscillations above the SB (cf. Figure \ref{fig:comparison_et_all}), therefore reducing the number of data points for larger $\ell$. Nevertheless, there does not appear to be a systematic difference between the distributions for the three codes.

In this section we carry out a quantitative analysis of the angular momentum diffusion. We obtained from the simulations the first-order ($n=1$) and second-order ($n=2$) diffusion coefficients, $\langle (\Delta \ell)^n \rangle$, describing changes in $\ell$.  Each diffusion coefficient was computed for a given initial value of $\ell$ and for a time interval, $\Delta t$, normalized to the orbital period $P$: $\tau=\Delta t/P$. These quantities were binned in linear bins of size 200 with $0 < \ell < 1$ and size 5 with $10^3 < \tau < 2\times 10^3$, respectively. For each time $t_i$ in the simulation we selected times $t_j>t_i$ with associated time lags $\tau_{ij}  = (t_j-t_i)/P_i$ in the range $\tau < \tau_{ij} < \tau + \Delta \tau$ with $\Delta \tau = 10$. We rejected any time $t_j$ if the absolute value of the change of the semimajor axis at time $t_j$ relative to $t_i$, $|(a_i-a_j)/a_i|$, exceeded 0.2, or if the test star was captured or unbound at $t_j$. For the remaining $t_j$ we computed the corresponding change of orbital angular momentum (normalized to the angular momentum of a circular orbit $L_c$), $\Delta \ell_{ij} = \ell_j - \ell_i$. Subsequently, we computed the first-order diffusion coeffient from $\langle \Delta \ell \rangle = \mathrm{mean} ( \Delta \ell_{ij} ) / \mathrm{mean} (\tau_{ij} P)$ and the second-order diffusion coefficient from $\langle (\Delta \ell)^2 \rangle = \mathrm{mean} [ (\Delta \ell_{ij})^2 ] / \mathrm{mean} (\tau_{ij} P)$, where the mean is taken over each bin of $\ell$ and $\tau$.

In the method described above the diffusion coefficients are functions of the time lag $\Delta t$. One expects that over some finite range in $\Delta t$, the results will not depend too strongly on $\Delta t$. According to \citet{VanKampen1992}, when evaluting diffusion coefficients in some quantity $x$, $\Delta t$ must be ``so small that $x$ cannot change very much during $\Delta t$, but large enough for the Markov assumption to apply". In our case, an additional condition applies: for $\ell < \ell_\mathrm{SB}$, a lower limit on $\Delta t$ is given by $t_\mathrm{GR}(\ell) \propto \ell^2$, since  we are interested in directed changes in the {\it mean} value of $\ell$ below the SB, averaged over the time scale of the angular momentum oscillations, which is $\sim t_\mathrm{GR}$ (cf. \S\,\ref{sect:belowSB:e_osc}). 

We argue in \S\, \ref{sect:S-stars:dfc} that for $\ell=\ell_\mathrm{SB}$, the characteristic time for $\ell$ to change by of order itself is the ``coherence time''  $t_\mathrm{coh}(a)$, defined as the time for a typical field star orbit, of semimajor axis $a$, to change its orientation.  We adopt $t_\mathrm{coh}^{-1} = \langle t_\mathrm{MP}\rangle^{-1} + \langle t_\mathrm{GR}\rangle^{-1}$, where $\langle t_\mathrm{MP} \rangle$ and $\langle t_\mathrm{GR} \rangle$ are the field star mass precession and Schwarzschild precession time scales averaged over a thermal distribution in eccentricity, respectively. As indicated in Figure~\ref{fig:introduction_figure}, the time scale for changes in $\ell$ increases away from the SB, both toward higher and lower $\ell$. Thus, in both the AR and RR angular momentum regimes defined in that figure, we expect that setting $\Delta t \lesssim t_\mathrm{coh}$ will ensure that $\ell$ ``does not change very much" during $\Delta t$.

The time scale $t_\mathrm{GR}(\ell)$ decreases rapidly as $\ell$ decreases from $\ell_\mathrm{SB}$; clearly, for $0<\ell<\ell_\mathrm{SB}$, $t_\mathrm{GR}(\ell)$ is maximal for $\ell=\ell_\mathrm{SB}$. In our simulations $t_\mathrm{GR}(\ell_\mathrm{SB})$ is typically comparable to or smaller than $t_\mathrm{coh}$. For example, for $a = 2 \, \mathrm{mpc}$, $t_\mathrm{GR}(\ell_\mathrm{SB}) \approx 1.3 \times 10^3 \, P$, whereas $t_\mathrm{coh} \approx 1.6 \times 10^3 \, P$. This demonstrates that, by choosing $\Delta t \sim t_\mathrm{coh}$, we satisfy both lower and upper limits of $\Delta t$ for $\ell<\ell_\mathrm{SB}$.

These arguments aside, the validity of an assumed value of $\Delta t$ can be checked by comparing the values of the diffusion coefficients derived for larger and smaller $\Delta t$. An example is given in Appendix \ref{sect:app:deptau}. 

\citet{eilonkupalex09} also carried out extensive $N$-body simulations to study the efficiency of RR in small-$N$ systems. Their pioneering work differed from ours in two important respects: their integrations were Newtonian, i.e., the effects of relativistic precession were not included, and they did not investigate the $L$-dependence of the diffusion rate, choosing instead to focus on the time dependence of the mean angular momentum changes induced by the torques. We can, however, compare our results to theirs in the high-$L$ regime where the effects of relativity are unimportant, as shown below.

\begin{figure*}
\center
\includegraphics[scale = 0.46, trim = 0mm 0mm 0mm 0mm]{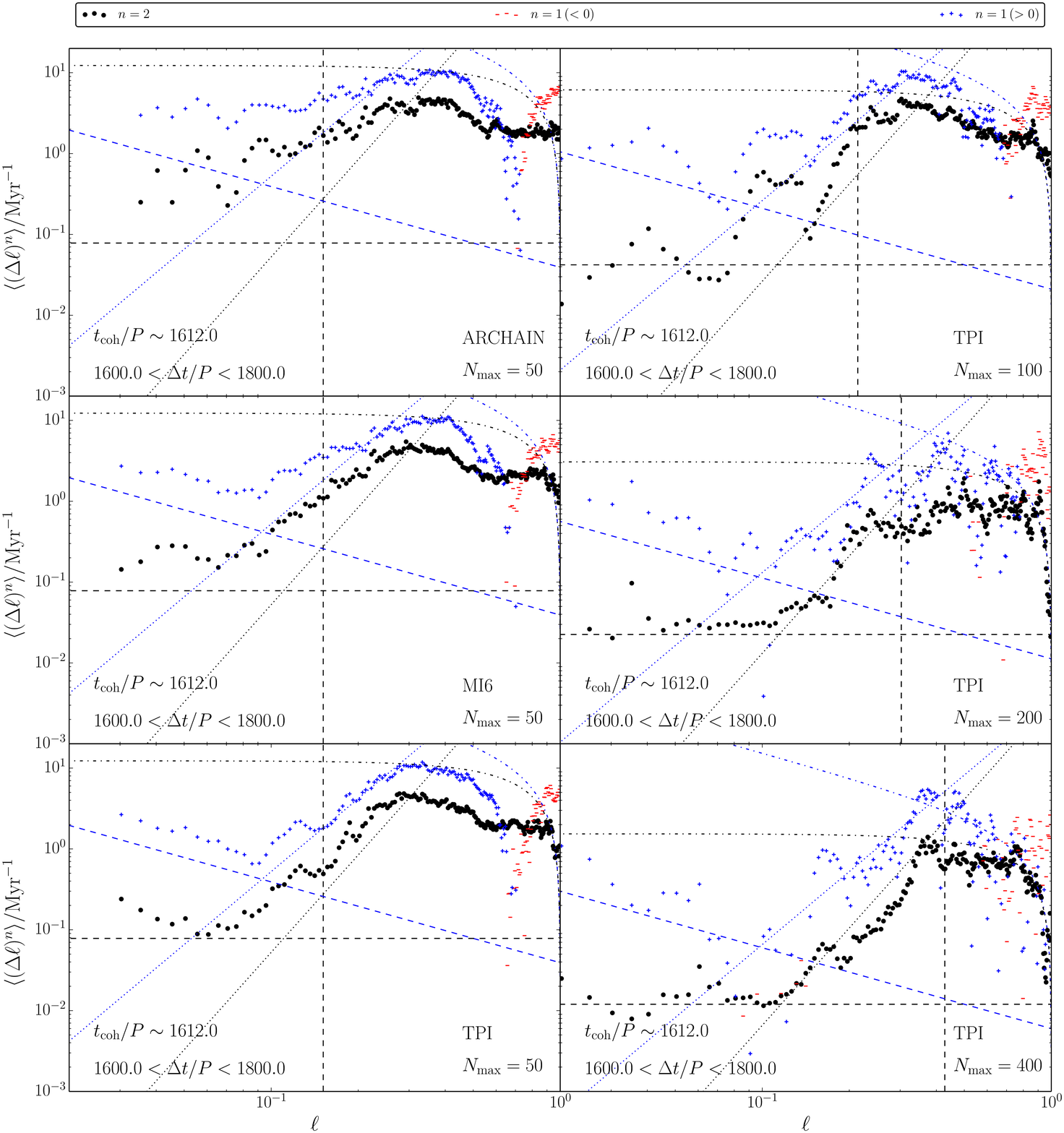}
\caption{\small First-order and second-order diffusion coefficients as function of $\ell \equiv L/L_c$. Left column: based on the $N_\mathrm{max} = 50$ simulations, distinguishing between the three codes (initially $a = 2 \, \mathrm{mpc}$ and $\log_{10}(1-e) = -2.0$, $-2.5$ or $-3.0$). Right column: based on simulations with \textsc{TPI} with $N_\mathrm{max} = 100,200$ and 400 (initially $a=2\,\mathrm{mpc}$ and $\log_{10}(1-e) = -2.5$). Positive (negative) first-order diffusion coefficients are shown in blue (red); second-order diffusion coefficients are shown in black. Minuses, plusses and bullets: quantities obtained from the simulations. Dashed lines: the predicted NRR diffusion coefficient in the limit $\ell \rightarrow 0$, equation~(\ref{eq:dfc_NRR_l0}). Black dot-dashed lines: the second-order incoherent RR prediction, equation~(\ref{eq:dfc_lRR2}), with $\beta_s = 1.6 \, \sqrt{1-\ell^2}$ \citep{gurhop07}. The blue dot-dashed lines show an {\it ad hoc} relation for the first-order RR coefficient, equation~(\ref{eq:dfc_lRR1}). Dotted lines: predictions for $\ell \lesssim \ell_\mathrm{SB}$ according to the model presented in \S\,\ref{sect:S-stars:dfc}. The vertical black dashed line shows the predicted value of $\ell$ at the SB, equation~(\ref{eq:ell_SB}). In each panel the time lags shown are comparable to the coherence time (see text). }
\label{fig:diffusion_coefficients}
\end{figure*}

We show in Figure \ref{fig:diffusion_coefficients} our derived diffusion coefficients $\langle \Delta \ell \rangle$ (blue plusses and red minuses for positive and negative values, respectively) and $\langle (\Delta \ell)^2 \rangle$ (black bullets) as function of $\ell$. In the left column results are shown for the three codes and the simulations with $N_\mathrm{max} = 50$, combining data from the test stars for the runs with initially $a = 2 \, \mathrm{mpc}$ and $\log_{10}(1-e) = -2.0$, $-2.5$ and $-3.0$. In each panel of Figure \ref{fig:diffusion_coefficients} the value of $\ell$ associated with the SB, $\ell_\mathrm{SB}$ (cf. equation~(\ref{eq:ell_SB})), is indicated with the vertical black dashed line. The coherence times and the adopted time lag bins, expressed in units of orbital period, are indicated in the bottom left of each panel. There appear to be no systematic differences in the diffusion coefficients between the different codes shown in the first column of Figure \ref{fig:diffusion_coefficients}. 

As mentioned in \S\,\ref{sect:introduction}, the simulations with $N_\mathrm{max} = 50$ field stars likely do not give a good description of the environment close to a SBH because the number of field stars within the initial volume of the simulation is too low (50, whereas $10^3 - 10^4$ would be more realistic). For this reason we carried out additional simulations with \textsc{TPI} with larger numbers of field particles, i.e. $N_\mathrm{max} = 100, 200$ and $400$ (simulations with $N_\mathrm{max} = 4800$ are discussed in \S\,\ref{sect:S-stars}). The field star mass $m_\star$ was adjusted to keep the enclosed stellar mass within any radius constant with respect to the $N_\mathrm{max} = 50$ simulations. The adopted values are $m_\star = 25, 12.5$ and $6.25 \, \mathrm{M}_\odot$ for $N_\mathrm{max} = 100, 200$ and $400$, respectively. The initial orbital elements of the test stars were $a = 2.0 \, \mathrm{mpc}$ and $\log_{10}(1-e) = -2.5$. Other parameters were identical to those in the $N_\mathrm{max}=50$ simulations (cf. \S\,\ref{sect:belowSB:init}). The diffusion coefficients derived from these simulations with larger $N_\mathrm{max}$ are shown in the right column of Figure \ref{fig:diffusion_coefficients}.

Some theoretical predictions exist for the dependence of the diffusion coefficients on $\ell$, and we can compare these predictions with our results. We refer the reader to Figure \ref{fig:introduction_figure} which identifies the three regimes in angular momentum.

As discussed in \S\ref{sect:timescales}, we expect that non-resonant relaxation (NRR) will dominate diffusion in angular momentum in the limit $\ell \rightarrow 0$. Our argument was that -- by definition -- NRR is unaffected by coherence-time arguments, and hence that the rapid GR precession that occurs in this low-$\ell$ regime has no consequence for the rate of non-resonant diffusion in angular momentum.

The orbit-averaged, NRR diffusion coefficients in the limit $\ell \rightarrow 0$ for test masses near a SBH are \citep{cohnk78,cohn79}: 
\begin{subequations}\label{eq:dfc_NRR_l0}
\begin{align}
\langle \Delta \ell \rangle_\mathrm{NRR} &\rightarrow \frac{1}{4\ell} A(E); \\
\left \langle \left (\Delta \ell \right )^2 \right \rangle_\mathrm{NRR} &\rightarrow \frac{1}{2} A(E).
\end{align}
\end{subequations}
Here $A(E)$ is a function of orbital energy, or, equivalently, of semimajor axis. It is given by (e.g., Appendix B of MAMW11):
\begin{align}
A(E)^{-1} = \frac{C_\mathrm{NRR}(\gamma)}{\log(\Lambda)} \left ( \frac{M_\bullet}{M_\star(a)} \right )^2 N_\star(a) P(a).
\label{eq:AE}
\end{align}
Here $C_\mathrm{NRR}(\gamma)$ is a dimensionless quantity that depends on the field star density slope $\gamma$. It can be evaluated using the procedure outlined in Appendix B of MAMW11. Explicit expressions for $C_\mathrm{NRR}(\gamma)$ as function of $\gamma$ are included in Appendix \ref{app:C_gamma}. The value that applies to the simulations presented here is $C_\mathrm{NRR}(2) = (9/7)\{1/[12 \log(2)-1]\}\approx 0.18$. For the Coulomb logarithm $\Lambda$ we adopt $\Lambda = 2 M_\bullet/m_\star$ (MAMW11). The diffusion coefficients described by equation~(\ref{eq:dfc_NRR_l0}) are plotted in Figure \ref{fig:diffusion_coefficients} with the dashed blue and black lines for the first-order and second-order coefficients, respectively. 

For the simulations with $N_\mathrm{max}=50$, it can be seen in Figure \ref{fig:diffusion_coefficients} that the first- and second-order diffusion coefficients gradually approach the NRR predictions for $\ell \ll \ell_\mathrm{SB}$. As $N_\mathrm{max}$ is increased, the correspondence between measured and predicted diffusion coefficients becomes quite good in this regime. This reinforces the hypothesis that NRR is indeed the mechanism that is primarily responsible for changes in $\ell$ as $\ell \rightarrow 0$. 

The other limiting case is $\ell \gg \ell_\mathrm{SB}$. In this high-angular-momentum regime, we expect that the dominant diffusion mechanism is (incoherent) resonant relaxation (RR) \citep[][p. 274]{bookmerritt13}. Only a limited set of predictions are available for the dependence of the RR diffusion coefficients on $\ell$, and as far as we are aware, no attempt has ever been made to compute the first-order coefficient in the incoherent RR regime.

The second-order coefficient can be written in the form: 
\begin{align}
\left \langle \left (\Delta \ell \right )^2 \right \rangle_\mathrm{RR}^{-1} = \beta_s^{-2} \left [ \frac{M_\bullet}{M_\star(a)} \right ]^2 N_\star(a) \frac{P(a)^2}{t_\mathrm{coh}} .
\label{eq:dfc_lRR2}
\end{align}
Here $t_\mathrm{coh}$ is the ``coherence time'' as introduced above, and $\beta_s$ is a parameter describing the efficiency of RR in the coherent regime, i.e. for time intervals $\Delta t\lesssim t_\mathrm{coh}$ during which $\ell$ increases approximately linearly with time. Our adopted coherence time is given by $t_\mathrm{coh}^{-1} = \langle t_\mathrm{MP}\rangle^{-1} + \langle t_\mathrm{GR}\rangle^{-1}$, where $\langle t_\mathrm{MP} \rangle$ and $\langle t_\mathrm{GR} \rangle$ are the field star mass precession and Schwarzschild precession time scales for $\omega$ to change by $\pi$ radians, averaged over a thermal distribution in eccentricity, respectively. These quantities are given explicitly by $\langle t_\mathrm{GR} \rangle = (1/12) (a/r_g) P(a)$ and $\langle t_\mathrm{MP} \rangle = (3/2)[M_\bullet/M_\star(a)] P(a)$ for $\gamma=2$ (cf. equations~(\ref{eq:1PNprecession}) and (\ref{eq:MP})). 

In Figure \ref{fig:diffusion_coefficients} we show equation~(\ref{eq:dfc_lRR2}) with the black dot-dashed lines assuming $\beta_s = 1.6 \, \sqrt{1-\ell^2}$, which is the Newtonian result obtained by \citet{gurhop07}. For the $N_\mathrm{max}=50$ case, there are some systematic differences between the observed and predicted diffusion coefficients, which are similar in all three codes. The measured values are systematically smaller, and  there is also a local minimum in $\langle (\Delta \ell)^2 \rangle$ at $\ell \approx 0.8$, which is not predicted. However, as $N_\mathrm{max}$ is increased, the local minimum gradually disappears, and the second-order coefficient is increasingly better described by the RR prediction. We can not claim to have a good explanation for the larger disagreement for smaller $N_\mathrm{max}$, but one possibility may be the increased importance of non-resonant relaxation when $N_\mathrm{max}$ is small. We include in this category processes like strong encounters and multi-body effects, which, although not well described by equations like (\ref{eq:dfc_NRR_l0}), are likewise unaffected by coherence-time arguments and which become increasingly important in stellar systems as $N$ is decreased. We note that the simulations with even larger $N_\mathrm{max}$ (cf. \S\,\ref{sect:S-stars:dfc}) also show good agreement with theory in this regime.

We remark that the result $\beta_s = 1.6 \, \sqrt{1-\ell^2}$ in the high-$L$ regime can be compared to the work of \citet{eilonkupalex09} (cf. section 4.3 of the latter paper). \citet{eilonkupalex09} determined a value of $\beta_{s,\mathrm{EKA09}} = 1.05$ averaged over their simulations (cf. their table 1), in which a thermal distribution of eccentricities was assumed. Averaging $\beta_s = 1.6 \, \sqrt{1-\ell^2} = 1.6 e$ over a thermal eccentricity distribution one finds $\langle \beta_s\rangle \approx 1.07$, which is in excellent agreement with the result of \citet{eilonkupalex09}.

As noted above, there does not appear to be any discussion in the literature about the expected form of the first-order RR diffusion coefficient. Figure \ref{fig:diffusion_coefficients} plots the {\it ad hoc} expression:
\begin{align}
 \left \langle \Delta \ell \right \rangle_\mathrm{RR} \approx \ell^{-1} \left \langle \left ( \Delta \ell \right )^2 \right \rangle_\mathrm{RR}.
\label{eq:dfc_lRR1}
\end{align}
This expression must be very approximate; it is clear that it cannot be valid for $\ell \approx 1$ because the first-order diffusion coefficient is expected (and is observed) to become negative as $\ell\rightarrow 1$. Nevertheless, as Figure \ref{fig:diffusion_coefficients} suggests, it is a reasonable approximation for $\ell_\mathrm{SB}\lesssim \ell \ll 1$ (in Figure \ref{fig:diffusion_coefficients} equation~(\ref{eq:dfc_lRR1}) is plotted without modifying the normalization). 

Finally, we consider the diffusion coefficients in the third of the three angular-momentum regimes defined in Figure \ref{fig:introduction_figure}: $\ell\lesssim \ell_\mathrm{SB}$, called ``anomalous relaxation'' (AR) in that figure. Figure \ref{fig:diffusion_coefficients} suggests that this regime becomes increasingly well-defined in the simulations as $N_\mathrm{max}$ increases: a distinct ``knee'' appears at $\ell\approx \ell_\mathrm{SB}$, below which $\langle\Delta\ell\rangle$ and $\langle(\Delta\ell)^2\rangle$ both drop rapidly toward smaller $\ell$, before flattening off in the NRR regime. We interpret this behavior as a manifestation of the rapid quenching of RR below the SB; indeed the location of this knee might be taken as an empirical definition of the location of the barrier. (The location of the knee is consistent with the value of $\ell_\mathrm{SB}$ as predicted by equation~(\ref{eq:ell_SB}), even though the nuclear model in Figure \ref{fig:diffusion_coefficients} is rather different than the one considered in MAMW11.) As shown below, the ``knee'' becomes even better defined in simulations with still larger values of $N_\mathrm{max}$; we will argue that this is due to a greater separation between the AR and NRR regimes.

A mechanism that would drive angular momentum diffusion in the $\ell\lesssim\ell_\mathrm{SB}$ region was discussed in MAMW11. Here we note -- following the discussion in that paper -- that diffusion in this regime is not expected to be well described either in terms of resonant nor non-resonant relaxation. In \S\,\ref{sect:S-stars:dfc} we return to the behavior of diffusion in angular momentum for $\ell \lesssim \ell_\mathrm{SB}$ and present a theoretical model for diffusion in this regime.

\section{Dynamical evolution of the S-stars}
\label{sect:S-stars}
\subsection{Initial conditions}
\label{sect:S-stars:initial_cond}

\begin{figure}
\center
\includegraphics[scale = 0.43, trim = 0mm 0mm 0mm 0mm]{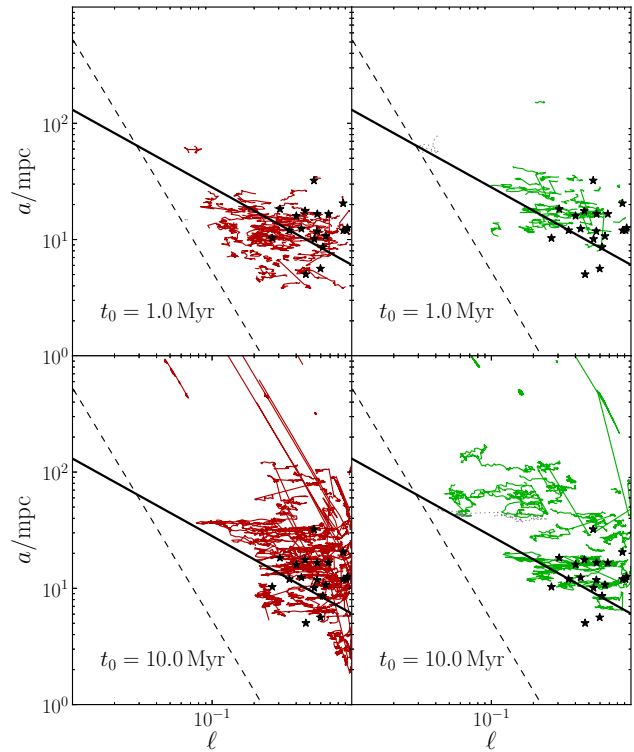}
\caption{\small Planes of semimajor axis versus dimensionless angular momentum $\ell$ for all 10 realizations of the 19 S-stars in our simulations with \textsc{TPI}. Tracks are shown for two time ranges $t_0$, with $0.8 \, t_0 < t < t_0$. Red (green) tracks apply to orbits that are initially below (above) the SB; dotted grey tracks apply to stars that become captured or unbound during the interval shown. The predicted position of the SB (equation~(\ref{eq:lSBSstar})) is indicated with the black solid line. The black dashed line shows the pericenter distance corresponding to the tidal disruption radius. Black stars indicate the orbital elements of the observed S-stars with $a<32.2 \, \mathrm{mpc}$ from \citet{gil09}.}
\label{fig:S_stars_ae_plane}
\end{figure}

We have demonstrated the validity of the results of \textsc{TPI} in \S\,\ref{sect:belowSB} using comparisons to more accurate, but slower, $N$-body codes. Here we proceed with simulations of the S-star cluster in which the number of field particles is larger by a factor of $\sim 10^2$; such simulations are currently not feasible with the other $N$-body codes discussed in \S\,\ref{sect:belowSB}.

The S-star cluster consists of main-sequence (MS) B-type stars at projected distances $r_p \lesssim 0.''8 \approx 32 \, \mathrm{mpc}$ \citep{genzel03,eis05,ghez08,gil09} from the central SBH, Sgr A* (assuming a distance to the GC of 8.3 kpc, \citealt{gil09}). The strong tidal field of the SBH at these radii makes it unlikely that the S-stars formed {\it in situ} \citep{morris93}, hence various formation scenarios have been proposed in which the S-stars formed elsewhere and migrated to their current locations (see \citealt{alex05} and \citealt{genzeleisen10} for reviews). \citet{antmerr13} (hereafter AM13) used Monte-Carlo simulations to show that binary disruption best matches the observed eccentricity distribution of the S-stars. In this process a stellar binary is tidally disrupted by the SBH, unbinding one of the stars from the SBH and leaving the other star in a tight and highly eccentric ($0.93 \lesssim e \lesssim 0.99$) orbit around the SBH \citep{hills88}. For the observed semimajor axes of the S-stars the high initial eccentricities predicted by this process imply that some of the S-stars were deposited below the SB (cf. Figure 1 of AM13). The number of stars that were deposited below the SB in this case depends on the assumed distribution of the field stars near the SBH. The diffusion processes discussed in \S\,\ref{sect:timescales} and \S\,\ref{sect:belowSB} could therefore be important for the dynamical evolution of some of the S-stars in the first few Myr after being deposited in the GC.

In the Monte-Carlo simulations of AM13 the orbital evolution of the S-stars was described using equations of motion derived from an orbit-averaged Hamiltonian, and two-body relaxation effects were not taken into account. Here we do take into account diffusion driven by two-body relaxation using \textsc{TPI}. (As discussed in more detail below, we confirm that the neglect of NRR by those authors was a reasonable approximation, at least as far as the eccentricity distribution is concerned.) We adopted from AM13 a field star number distribution $N_\star(a) = N_\mathrm{max} (a/a_\mathrm{max})^{3-\gamma}$ with $N_\mathrm{max} = 4.8 \times 10^3$, $a_\mathrm{max} = 0.2 \, \mathrm{pc}$ and  $\gamma = 2$; the field star mass was set to $m_\star = 10 \, \mathrm{M}_\odot$. This distribution is consistent with steady-state models of the GC of a cusp of stellar remnants \citep{ha06}. We simulated 19 S-stars, adopting the semimajor axes with $a<32.2 \, \mathrm{mpc}$ from the sample of S-stars for which orbital fits were obtained by \citet{gil09}. In our simulations the S-stars were treated as test stars; for each S-star there were 10 random realizations, each with an initial eccentricity sampled from a thermal distribution with $0.93 < e < 0.99$ and a random orientation, consistent with binary disruption. The probability for $e>e_\mathrm{SB}$ in this model is $\approx 0.72$. The capture radius was set to an approximation of the tidal disruption radius, $r_\mathrm{capt} = 2 R \, (M_\bullet/m)^{1/3}$, where $R = 8 \, \mathrm{R}_\odot$ and $m=10\, \mathrm{M}_\odot$ \citep{antlomb11}. We included only 1PN terms in the simulations and therefore we assumed a non-spinning SBH. The integration time was constrained by computational limitations and was set to $20 \, \mathrm{Myr}$. 

In these simulations, equation (\ref{eq:ell_SB}) predicts: 
\begin{align}
\ell_\mathrm{SB}(a)  &= \frac{r_g}{a}\frac{M_\bullet}{M_\star(a)}\sqrt{N_\star(a)} \nonumber \\
&= \ell_\mathrm{SB,10} \left(\frac{a}{10\;\mathrm{mpc}}\right)^{-3/2}, \nonumber \\
\ell_\mathrm{SB,10} &\approx 0.49.
\label{eq:lSBSstar}
\end{align}

\subsection{Orbital evolution}
\label{sect:S-stars:orb_ev}
We show in Figure \ref{fig:S_stars_ae_plane} the $(a,1-e)$-plane for all 10 realizations of the 19 S-stars in our simulations for $t<10\, \mathrm{Myr}$. Tracks are shown for two times $t_0$ in the simulations. Red (green) tracks apply to orbits that are initially below (above) the SB; dotted grey tracks apply to stars that become captured or unbound during the interval shown (cf. \S\,\ref{sect:S-stars:ej_dis}). A fraction $\sim 0.72$ of the stars start below the SB (red solid line in Figure \ref{fig:S_stars_ae_plane}). The orbits of the majority of these rapidly diffuse to larger semimajor axis and/or smaller eccentricity: by 10 Myr, most of them have evolved to locations above the SB. The orbits that are initially above the SB, on the other hand, tend to remain in this region. Note that some penetration to regions below the SB does occur, however, and that some stars remain below the SB even after $10 \, \mathrm{Myr}$. In what follows, we discuss this evolution in more detail.

\begin{figure*}
\center
\includegraphics[scale = 0.46, trim = 0mm 0mm 0mm 0mm]{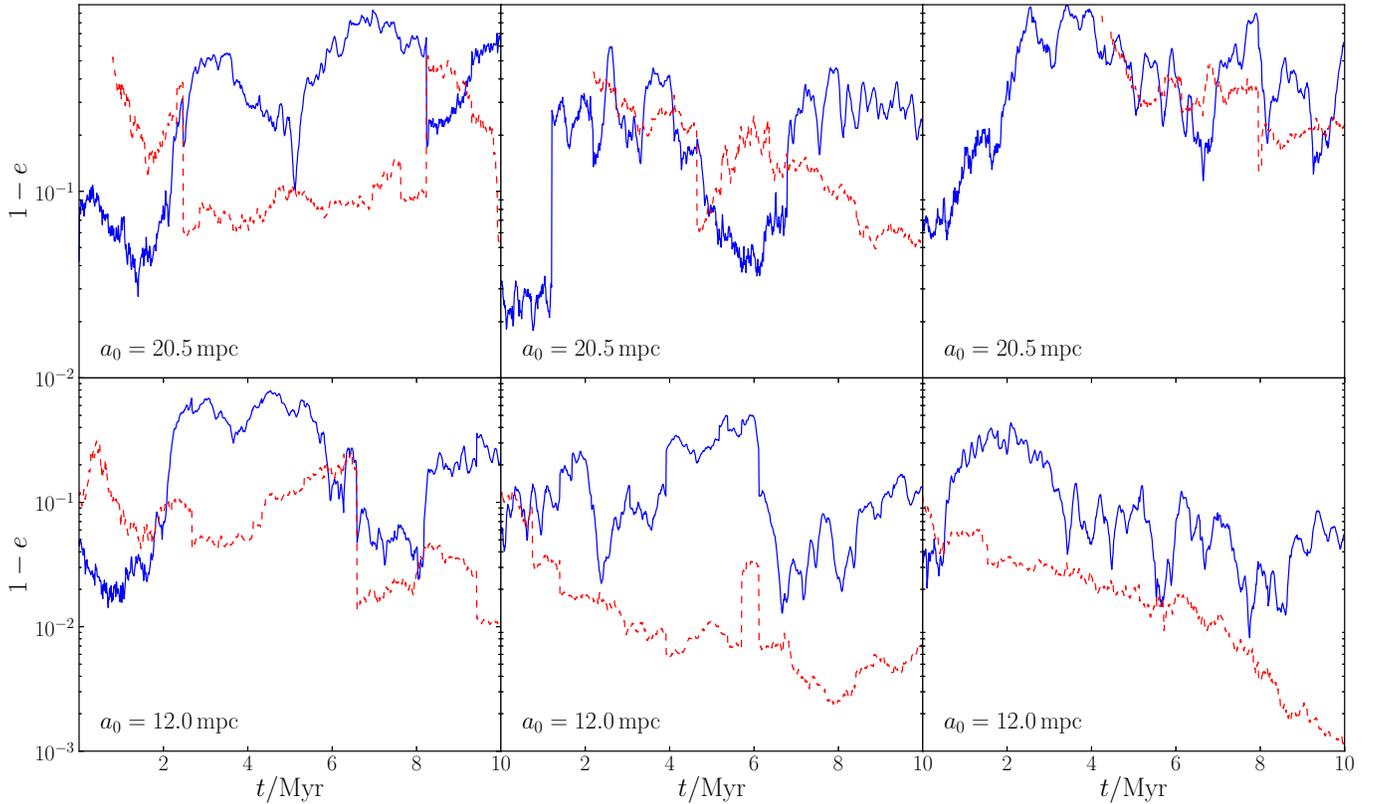}
\caption{\small The eccentricity evolution (blue solid lines) for six realizations of the S-stars in the simulations with initial eccentricity $e_0 > e_\mathrm{SB}$. Red dashed lines show the predicted value of $e_\mathrm{SB}$ (cf. equation~(\ref{eq:lSBSstar})). Tracks are selected for which no capture or unbinding event occurs within the time interval shown. The initial semimajor axis is indicated in the bottom left of each panel. }
\label{fig:S_stars_ome_time}
\end{figure*}

In Figure \ref{fig:S_stars_ome_time} we show the eccentricity evolution for six realizations of S-stars in the simulations with initial eccentricity $e_0 > e_\mathrm{SB}$. By 10 Myr these orbits have diffused to locations above the SB. At several instances the orbit, after having diffused to $e<e_\mathrm{SB}$, becomes more eccentric again and reaches $e \approx e_\mathrm{SB}$. The orbit is then ``reflected,'' however, to lower eccentricity. This behaviour is consistent with that seen in the $N$-body simulations of MAMW11. 

It is significant that the relation proposed by MAMW11 for the location of the SB, and which is plotted as the red dashed line in Figure \ref{fig:S_stars_ome_time}, appears to predict remarkably well the value of the eccentricity at which RR ``turns on'' in these simulations. This, in spite of the fact that the number of stars in the new simulations is a factor $\sim 10^2$ higher than in those of MAMW11. We interpret this success as confirming, to a greater degree than was possible in MAMW11, the general validity of the relation (\ref{eq:ell_SB}).

\begin{figure}
\center
\includegraphics[scale = 0.435, trim = 0mm 0mm 0mm 0mm]{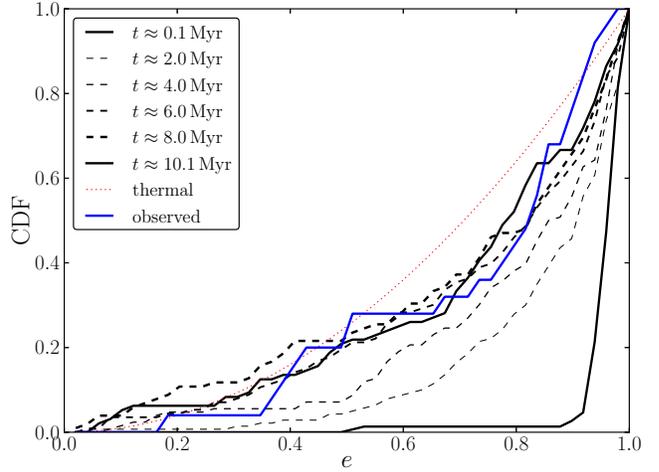}
\caption{\small The cumulative eccentricity distribution of all realizations of the S-stars in our simulations at various times between $t\approx0$ and $10\,\mathrm{Myr}$, assuming burst formation. The initial and final distributions are shown with black solid lines. Intermediate times are shown with black dashed lines; the thickness increases with time. The blue solid line shows the observed distribution of the S-stars \citep{gil09}. The red dotted line shows a thermal distribution $N(e) = e^2$. }
\label{fig:S_stars_e_hist}
\end{figure}

We show in Figure \ref{fig:S_stars_e_hist} the evolution of the cumulative eccentricity distribution for all realizations of the S-stars in our simulations. This distribution evolves rapidly from a near $\delta$-function at $e\sim 1$ that reflects the initial conditions, to a much more uniform distribution. The distribution does not appear to converge to a ``thermal'' form, $N(e) = e^2$, but on average remains more eccentric. Interestingly, the distribution appears to converge to a form that is closer to the observed, ``super-thermal'' distribution of the S-stars \citep{gil09}, shown in Figure \ref{fig:S_stars_e_hist} with the blue solid line. 

\begin{figure}
\center
\includegraphics[scale = 0.435, trim = 0mm 0mm 0mm 0mm]{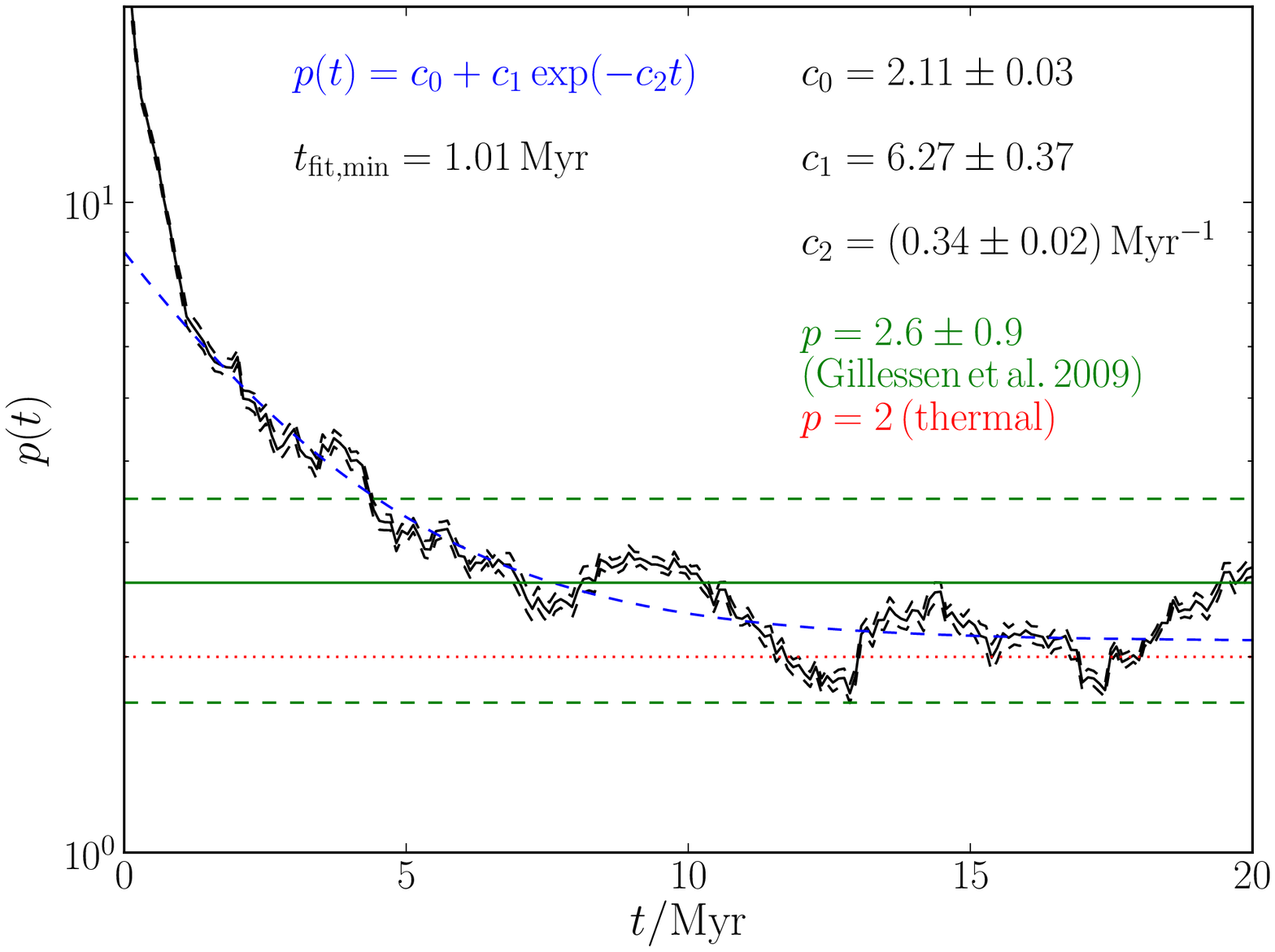}
\caption{\small The evolution of the slope $p$ with time for all realizations of the S-stars in our simulations (black solid line), in the case of formation in a burst. We exclude four S-stars, for which the computation had not advanced to 20 Myr by the time of writing. The uncertainty in $p$ is indicated with black dashed lines. The blue dashed line shows a least-squares fit of the form $p_\mathrm{fit}(t) = c_0 + c_1 \exp(-c_2 t)$; we exclude data for $t<1.01 \, \mathrm{Myr}$. The green solid and dashed lines indicate the observed value $p_\mathrm{obs}=2.6 \pm 0.9$  \citep{gil09}. The red dashed line shows $p=2$ (thermal distribution). }
\label{fig:S_stars_p_evolution}
\end{figure}

To investigate this apparent correspondence with observations more quantitatively, we fitted the cumulative eccentricity distribution in our simulations to a power law, $N(e) = e^p$, and we show the time evolution of $p$ in Figure \ref{fig:S_stars_p_evolution} with solid and dashed black lines. There is an initial rapid decrease of $p$ from $\sim 22$ to $\sim 7$ over the course of $\sim 1\, \mathrm{Myr}$. The subsequent evolution is slower, with $p$ decreasing to $\sim 3$ after $4 \, \mathrm{Myr}$. The form of $p(t)$ for $1 \lesssim t/\mathrm{Myr} \lesssim 7$ is well-fitted by a decaying exponential function, $p_\mathrm{fit}(t) = c_0 + c_1 \exp(-c_2 t)$; we find best-fit values $c_0 \approx 2.11$, $c_1 \approx 6.27$ and $c_2 \approx 0.34 \, \mathrm{Myr}^{-1}$ (the fitted curve is shown with the blue dashed line in Figure \ref{fig:S_stars_p_evolution}). After $\sim 7 \, \mathrm{Myr}$, the detailed evolution of $p(t)$ deviates slightly from a decaying exponential function. The overall evolution is still consistent with a decaying exponential, however. Interestingly, in our simulations $p(t)$ appears to oscillate roughly between $p\approx 2$, consistent with a thermal distribution, and $p \approx 2.6$, consistent with observations ($p_\mathrm{obs} = 2.6 \pm 0.9$ \citealt{gil09}).

In the results presented above it was assumed that all 19 S-stars are deposited in the GC in a single burst at $t=0$. We used these results as a template to estimate the evolution of $p(t)$ in the case of continuous formation of S-stars. The details are presented in Appendix \ref{app:cont}. The effect of continuous formation is to slow the evolution of $p$ as function of time. We find that in the case of continuous formation the time for $p$ to decrease to $p=2.6$ is increased by a factor of $\sim 3.6$ from $\sim 7 \, \mathrm{Myr}$ to $\sim 25 \, \mathrm{Myr}$. We discuss implications of the evolution of the eccentricity distribution in \S\,\ref{sect:discussion:S_star_implications}.

\begin{figure}
\center
\includegraphics[scale = 0.435, trim = 0mm 0mm 0mm 0mm]{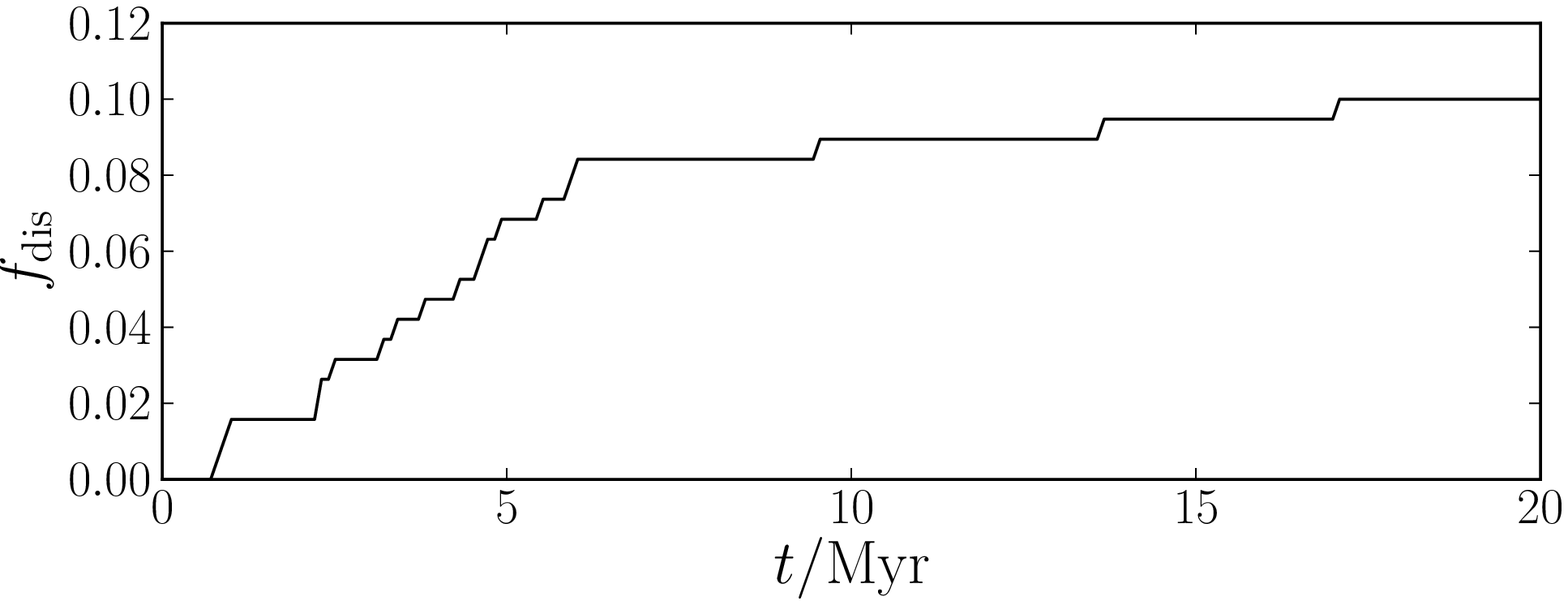}
\includegraphics[scale = 0.435, trim = 0mm 0mm 0mm 0mm]{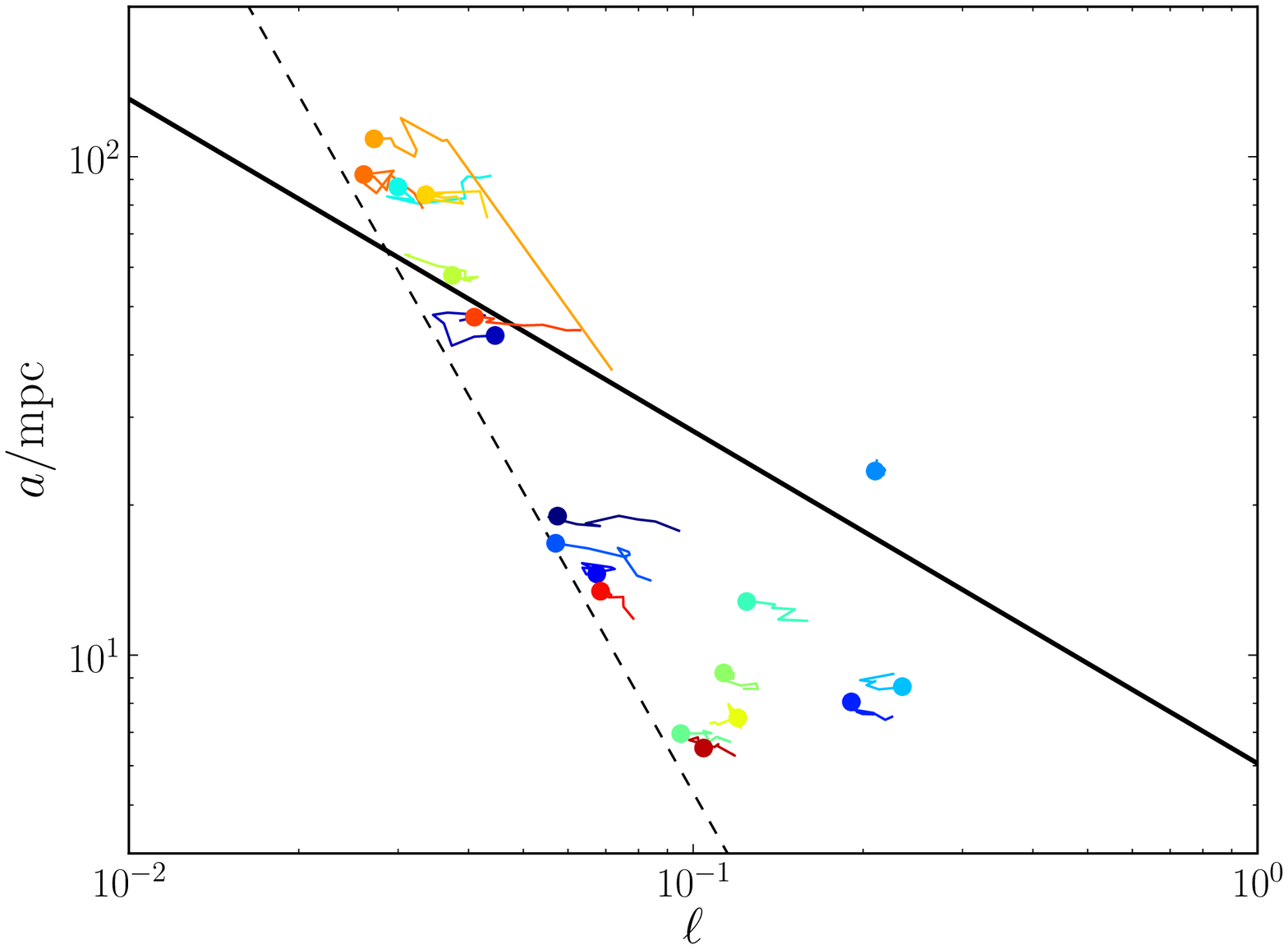}
\caption{\small Top: the cumulative fraction of tidally disrupted S-stars (for all 10 realizations) as function of time. Bottom: orbital tracks prior to disruption. The orbital elements prior to the disruption event (determined at apocenter) are shown with bullets; in addition, tracks of 10 orbital periods prior to disruption are shown. The black solid line shows the SB according to equation~(\ref{eq:lSBSstar}). }
\label{fig:S_stars_captures}
\end{figure}

\begin{figure}
\center
\includegraphics[scale = 0.435, trim = 0mm 0mm 0mm 0mm]{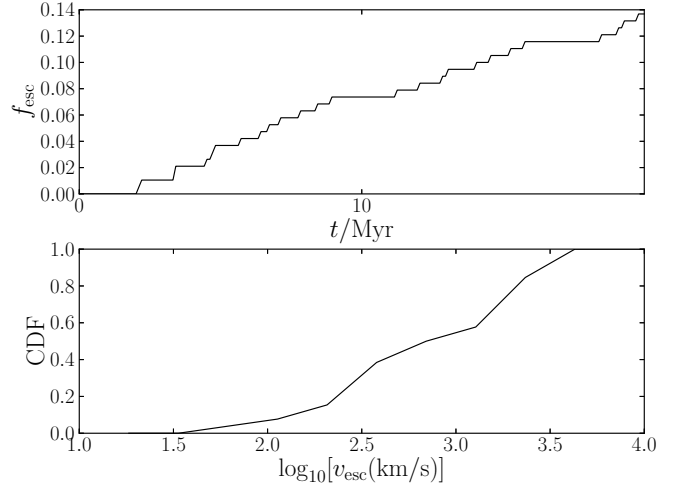}
\caption{\small Top: the cumulative fraction of unbound S-stars (for all 10 realizations) as function of time. Bottom: the cumulative distribution of the escape velocity $v_\mathrm{esc} = \sqrt{2E}$ from the SBH for the unbound stars, not taking into account the deceleration from the Galactic bulge. }
\label{fig:S_stars_escapers}
\end{figure}

\subsection{Tidally disrupted and ejected stars}
\label{sect:S-stars:ej_dis}
We show in the top panel of Figure \ref{fig:S_stars_captures} the cumulative fraction of S-stars that are tidally disrupted, i.e. the stars that at some time in the simulation approach the SBH within the assumed tidal disruption radius $r_\mathrm{capt} = 2 R \, (M_\bullet/m)^{1/3}$ with $R = 8 \, \mathrm{R}_\odot$ and $m=10\, \mathrm{M}_\odot$. The majority of disruptions occurs at $t < 10 \, \mathrm{Myr}$: initially the orbits are highly eccentric, making stars susceptible to disruption. As the eccentricity decreases and the orbits reach the SB the probability for capture decreases. This is borne out by the bottom panel of Figure \ref{fig:S_stars_captures} in which we show orbital tracks in the $(a,\ell)$-plane prior to disruption. Most of the orbits are close to the disruption boundary prior to disruption and most of the latter orbits are below the SB. We note that the eccentricity oscillations described in \S\,\ref{sect:belowSB} potentially enhance disruptions because during the oscillations the eccentricity can reach a higher value than the mean eccentricity. Only few (2 out of 19) disruptions occur above the SB and with relatively high angular momentum ($\ell>10^{-1}$), in which case a strong two-body encounter is required to produce the required small pericenter distance (i.e. an interaction typically associated with the full loss-cone). The cumulative fraction of disrupted stars is $\sim 0.10$ after $20 \, \mathrm{Myr}$, which is an order of magnitude larger than the fraction of $\lesssim 0.01$ found by AM13. This may suggest that NRR, which was not taken into account in the calculations of AM13, is important for determining the rate of tidal disruptions. 

Furthermore we show in the top panel of Figure \ref{fig:S_stars_escapers} the cumulative fraction of S-stars that become unbound from the SBH (i.e. stars with orbital energy $E>0$).  Unlike the fraction of tidally disrupted stars, the fraction of unbound stars continues to increase steadily after $t\approx10 \, \mathrm{Myr}$. This likely reflects the property that strong two-body encounters leading to ejection can in principle occur at any eccentricity and semimajor axis, whereas two-body encounters leading to tidal disruption are more likely if the eccentricity is high, in which case a small perturbation to the orbit is required for disruption. After 20 Myr the cumulative fraction of unbound stars is $\approx 0.14$. The distribution of the escape velocity $v_\mathrm{esc} = \sqrt{2E}$ from the SBH (not taking into account deceleration from the Galactic bulge) is plotted in the bottom panel of Figure \ref{fig:S_stars_escapers}. This distribution is peaked near $v_\mathrm{esc} \sim 10^3 \, \mathrm{km/s}$, which is comparable to the escape velocities of hypervelocity stars \citep{hills88}. 

\subsection{Diffusion coefficients}
\label{sect:S-stars:dfc}
\begin{figure*}
\center
\includegraphics[scale = 0.46, trim = 0mm 0mm 0mm 0mm]{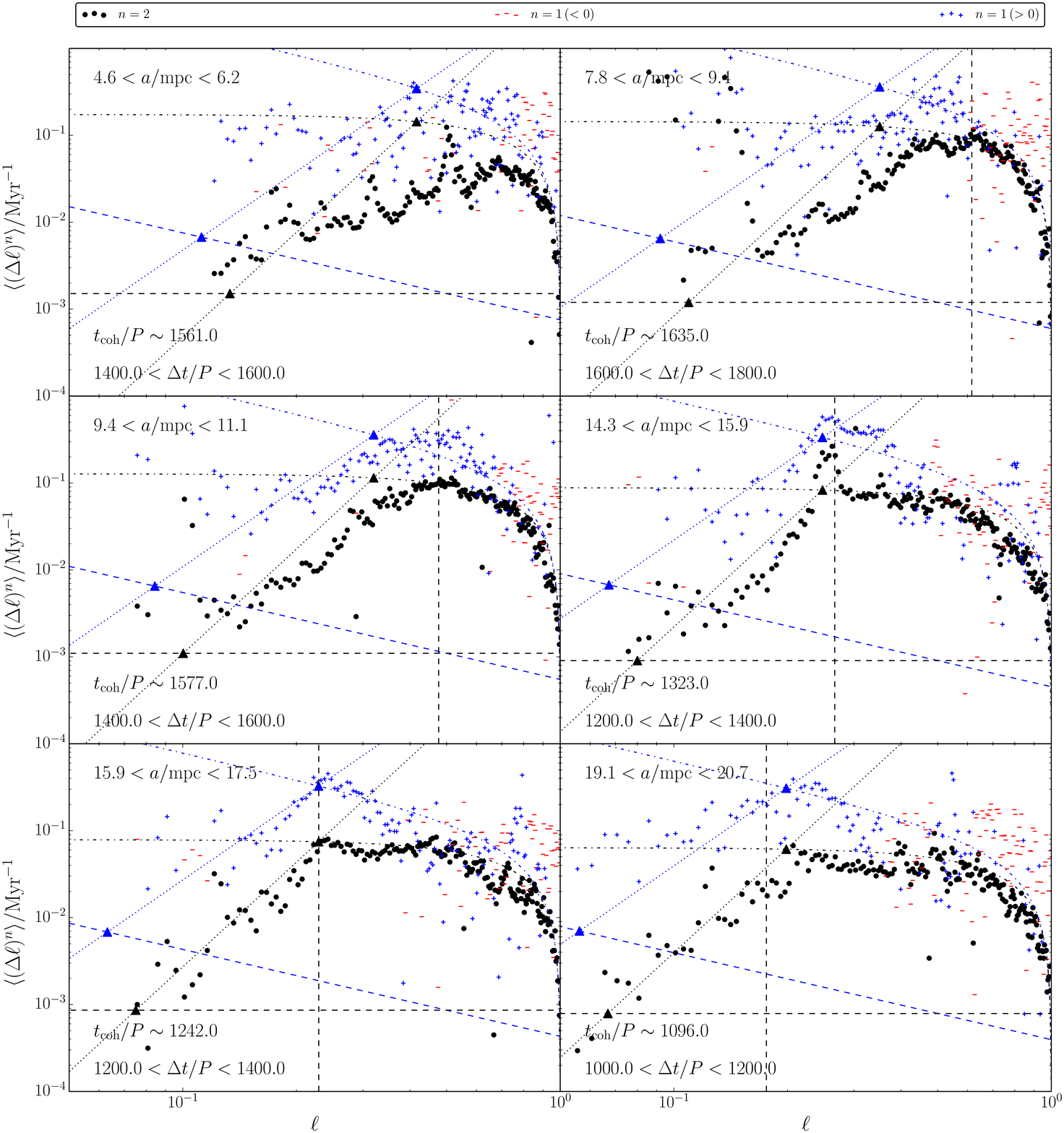}
\caption{\small First-order and second-order diffusion coefficients as function of $\ell \equiv L/L_c$ obtained from the S-star simulations. Positive (negative) first-order diffusion coefficients are shown in blue (red); second-order diffusion coefficients are shown in black. Minuses, plusses and bullets: quantities obtained from the simulations. Dashed lines: the predicted NRR diffusion coefficient in the limit $\ell \rightarrow 0$, equation~(\ref{eq:dfc_NRR_l0}). Black dot-dashed lines: the second-order incoherent RR prediction, equation~(\ref{eq:dfc_lRR2}), with $\beta_s = 1.6 \, \sqrt{1-\ell^2}$ \citep{gurhop07}. The blue dot-dashed lines show an {\it ad hoc} relation for the first-order RR coefficient, equation~(\ref{eq:dfc_lRR1}). Dotted lines: predictions according to the model presented in \S\,\ref{sect:S-stars:dfc}; we have set $C_{A_\mathrm{D}} = 0.5$ and $C_1=C_2=2.6$. The vertical black dashed line shows the predicted value of $\ell$ at the SB, equation~(\ref{eq:lSBSstar}). In each panel the time lags shown are comparable to the coherence time (see text). The triangles indicate the quantities $\ell_{\mathrm{a},n}$ and $\ell_{\mathrm{b},n}>\ell_{\mathrm{a},n}$ (cf. equation~(\ref{eq:labn})); blue: $n=1$; black: $n=2$. }
\label{fig:diffusion_coefficients_S_stars}
\end{figure*}

Diffusion coefficients in the S-star simulations were computed using the same technique as in \S\,\ref{sect:belowSB:diffusion}. In addition to binning the data with respect to the initial value of $\ell$ and the time lag $\tau$, here data was also binned with respect to semimajor axis. We show in Figure \ref{fig:diffusion_coefficients_S_stars} the resulting first-order and second-order diffusion coefficients for all realizations of the S-stars in our simulations for six ranges of the semimajor axis. As in Figure \ref{fig:diffusion_coefficients}, the time lags shown in Figure \ref{fig:diffusion_coefficients_S_stars} are chosen such that the coherence time lies within the time lag bin. We note that by setting the time lag to values that are substantially longer, the diffusion coefficient plots tend to change in appearance. This is illustrated and explained in Appendix \ref{sect:app:deptau}. 

In the regime $\ell \gg \ell_\mathrm{SB}$ the second-order diffusion coefficients from our simulations are consistent with the RR prediction, equation~(\ref{eq:dfc_lRR2}), with $\beta_s = 1.6 \, \sqrt{1-\ell^2}$ \citep{gurhop07}. As noted in \S\,\ref{sect:belowSB:diffusion}, this agreement is increasingly good with increasing $N_\mathrm{max}$, a trend that continues here. Furthermore, the ``knee'' feature of the diffusion coefficients near $\ell\approx \ell_\mathrm{SB}$, which was observed in \S\,\ref{sect:belowSB:diffusion} as $N_\mathrm{max}$ was increased, is also clearly present in Figure \ref{fig:diffusion_coefficients_S_stars}. 

As noted above, the form of the diffusion coefficients in the $\ell\lesssim\ell_\mathrm{SB}$ regime (``anomalous relaxation,'' AR) is not well understood theoretically. The rather abrupt decrease in the measured diffusion coefficients as $\ell$ decreases past $\sim \ell_\mathrm{SB}$ is expected, at least qualitatively, since the SB is defined as the value of $\ell$ for which the rapid GR precession quenches the effects of the $\sqrt{N}$ torques. We find from our simulations that the dependence of the diffusion {\it times}:
\begin{align}
T_n \equiv \left|\frac{\langle\left(\Delta\ell\right)^n\rangle}{\ell^n} \right|^{-1},
\ \ \ \ n = \{1,2\},
\end{align}
on $\ell$ in this regime is often well fit by a relation of the form
\begin{align}
T_{1,2}(a,\ell) = \mathrm{constant}(a) \times \ell^{-2}, \ \ \ \ \ell\lesssim \ell_\mathrm{SB} .
\end{align}
This is the dependence that was assumed in making the lower panel of Figure~\ref{fig:introduction_figure}.

In fact, an $\ell^{-2}$ dependence below the barrier is predicted by the simple Hamiltonian model described in Section VB of MAMW11, in which a random walk in $\ell$ results from assuming sudden, random changes in the direction of the $\sqrt{N}$ torquing potential each $\sim t_\mathrm{coh}$. We briefly summarize here the results of an analytic calculation based on that model (Merritt, D. 2013, unpublished). 

In the small-$\ell$ limit, the averaged Hamiltonian of Merritt et al. (2011) predicts, for times $\Delta t \lesssim t_\mathrm{coh}$,
\begin{subequations}\label{Equation:Hamilton}
\begin{align}
\ell^{-1}(\omega) &= \frac{1}{2\ell_1\ell_2} \left[\left(\ell_2-\ell_1\right)\sin(\omega) +
\left(\ell_1 + \ell_2\right)\right] \\
&= A_\mathrm{D}\left[\sin(\omega)+h\right] .
\end{align}
\end{subequations}
Here, $\{\ell_1,\ell_2\}$ are the extreme values of $\ell$ during a GR precession cycle, $\ell_\mathrm{av}=(1/2)(\ell_1+\ell_2)$, and $h=-H/A_\mathrm{D} =(\ell_1^{-1} + \ell_2^{-1})/(2A_\mathrm{D}) \approx 1/(A_\mathrm{D}\ell_\mathrm{av})$ is a normalized, averaged (secular) Hamiltonian $H$. (We have set $\sin i=\pi/2$ in Eq. (41) of Merritt et al. (2011), i.e., the torquing potential is assumed to be aligned with the $x$ axis.) Equation~(\ref{Equation:Hamilton}) describes changes in $\ell$ due to the $\sqrt{N}$ torques as the orbit precesses, at a (slightly) non-constant rate, due to GR. As noted above, the amplitude of the $\ell-$ oscillations in this regime scales as $\sim \ell_\mathrm{av}^2$. These oscillations, by themselves, do not imply any directed evolution in $\ell_\mathrm{av}$. But if the direction of the torquing potential is suddenly changed, after a time $\sim t_\mathrm{coh}$, the orbit will have been given a new value of $h$ and correspondingly different values of $\ell_1$ and $\ell_2$. Assuming that the changes in the direction of the torquing potential each $t_\mathrm{coh}$ are random, one finds for the first- and second-order diffusion coefficients of $h$ in this model:
\begin{align}
\langle\Delta h\rangle \approx -\frac{1}{t_\mathrm{coh}} \frac{1}{h} ,\ \ \ \ 
\langle\left(\Delta h\right)^2\rangle \approx \frac{1}{t_\mathrm{coh}}.
\end{align}
The corresponding time scales are:
\begin{align}
\left|\frac{\langle\Delta h\rangle}{h}\right|^{-1} \approx 
\left|\frac{\langle\left(\Delta h\right)^2\rangle}{h^2}\right|^{-1} \approx
h^2 t_\mathrm{coh} \approx
\frac{t_\mathrm{coh}}{A_\mathrm{D}^2\ell_\mathrm{av}^2} ,
\end{align}
consistent with the $\sim\ell^{-2}$ dependence observed in the simulations.
 
Accordingly, we suggest the following functional forms for the diffusion coefficients in the AR regime:
\begin{subequations}\label{eq:dfc_AR}
\begin{align}
\langle \Delta \ell \rangle &\approx \frac{C_1 }{\tau}\ell^3; \\
\langle (\Delta \ell)^2 \rangle &\approx \frac{C_2}{\tau} \ell^4,
\end{align}
\end{subequations}
with $\tau=t_\mathrm{coh}/A_\mathrm{D}^2$; furthermore, if the simple model presented above is valid, we expect $C_1\approx C_2 = \mathcal{O}(10^0)$.

\begin{figure}
\center
\includegraphics[scale = 0.435, trim = 0mm 0mm 0mm 0mm]{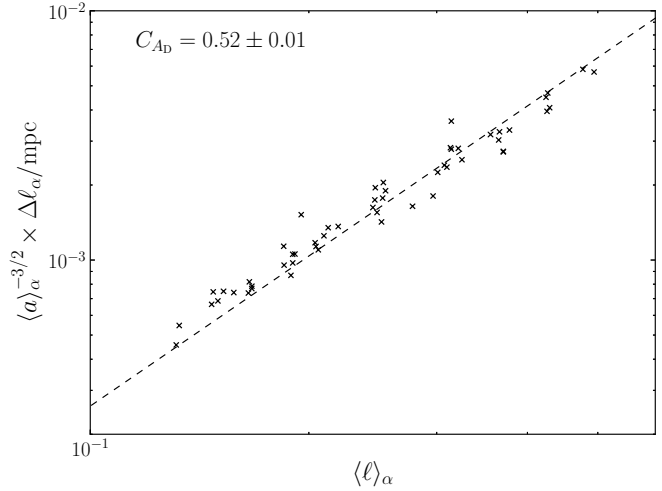}
\caption{\small Amplitude of eccentricity oscillations for the S-star simulations, binned in the mean value of $a$ and $\langle \ell \rangle$ as function of $\langle \ell \rangle$, and averaged over all orientations $\alpha$. The best-fit curve is shown with the black dashed line; the fitted value of $C_{A_\mathrm{D}}$ is shown in the top left. }
\label{fig:e_amplitude_S_stars}
\end{figure}

The quantities in equation~(\ref{eq:dfc_AR}) depend on the parameter $A_\mathrm{D}$ and the latter contains the fit parameter $C_{A_\mathrm{D}}$ (cf. equation~(\ref{eq:AD})). We used the same technique based on the amplitude of the eccentricity oscillations as in \S\,\ref{sect:belowSB:e_osc:amplitude} to determine this parameter for the S-star simulations; the results are shown in Figure \ref{fig:e_amplitude_S_stars}. Based on this result we adopt $C_{A_\mathrm{D}} = 0.5$ and we plot the predicted diffusion coefficients, equation~(\ref{eq:dfc_AR}), in Figure \ref{fig:diffusion_coefficients_S_stars} with the dotted lines. We find best agreement with the data for $C_1\approx C_2\approx2.6$. For reference have also included these predictions for the simulations that were discussed in \S\,\ref{sect:belowSB} in Figure \ref{fig:diffusion_coefficients}. 

Based on the results presented in Figure \ref{fig:diffusion_coefficients_S_stars}, we can approximate the first and- second-order diffusion coefficients as piecewise-continuous functions of $\ell$:
\begin{align}
\nonumber \left \langle \left (\Delta \ell \right )^n \right \rangle(\ell) & \approx \left \{ \begin{array}{ll}
\left \langle \left (\Delta \ell \right )^n \right \rangle_\mathrm{NRR}(\ell), & \ell<\ell_{\mathrm{a},n}; \\
\left \langle \left (\Delta \ell \right )^n \right \rangle_\mathrm{AR}(\ell), & \ell_{\mathrm{a},n} \leq \ell \leq \ell_{\mathrm{b},n}; \\
\left \langle \left (\Delta \ell \right )^n \right \rangle_\mathrm{RR}(\ell), & \ell>\ell_{\mathrm{b},n}.
\end{array} \right. \\
\label{eq:dif_coef_approx}
\end{align}
Here $\langle (\Delta \ell )^n \rangle_\mathrm{NRR}$, $\langle (\Delta \ell )^n \rangle_\mathrm{AR}$, $\langle \Delta \ell \rangle_\mathrm{RR}$ and $\langle (\Delta \ell )^2 \rangle_\mathrm{RR}$ are given explicity by equations (\ref{eq:dfc_NRR_l0}), (\ref{eq:dfc_AR}), (\ref{eq:dfc_lRR1}) and (\ref{eq:dfc_lRR2}), respectively (in the latter equation we adopt $\beta_s = \alpha_s \, \sqrt{1-\ell^2}$ with $\alpha_s=1.6$). We emphasize that equation~(\ref{eq:dfc_lRR1}) is {\it ad hoc} and theoretically not well motivated, as discussed in \S\,\ref{sect:belowSB:diffusion}. Moreover, it fails to describe the simulations for $\ell \gg \ell_{\mathrm{b},1}$. In \S\,\ref{sect:steady-state} we present a modified (but still not theoretically motivated) analytic prescription for $\langle \Delta \ell \rangle_\mathrm{RR}$ that better describes the data for $\ell \gg \ell_{\mathrm{b},1}$. 

The quantities $\ell_{\mathrm{a},n}$ and $\ell_{\mathrm{b},n}$ are defined such that $\left \langle \left (\Delta \ell \right )^n \right \rangle(\ell)$ is a continuous function of $\ell$. From equations (\ref{eq:dfc_NRR_l0}), (\ref{eq:dfc_AR}), (\ref{eq:dfc_lRR1}) and (\ref{eq:dfc_lRR2}) it follows that:
\begin{subequations}\label{eq:labn}
\begin{align}
\ell_{\mathrm{a},n} &= \left [\frac{n \, \log(\Lambda)}{C_\mathrm{NRR}(\gamma) C_n C_{A_\mathrm{D}}^2} \left ( \frac{r_g}{a} \right )^2 \frac{t_\mathrm{coh}(a)}{P(a)} \right ]^{1/4}; \\
 \ell_{\mathrm{b},n} &= \frac{1}{\sqrt{2}} \left [-C_{\mathrm{b},n} + \left ( C_{\mathrm{b},n}^2 + 4C_{\mathrm{b},n} \right )^{1/2} \right ]^{1/2}; \\
C_{\mathrm{b},n} &\equiv \frac{4 \alpha_s^2}{C_n C_{A_\mathrm{D}}^2} \left ( \frac{r_g}{a} \right )^2 \left[ \frac{t_\mathrm{coh}(a)}{P(a)} \right]^2. 
\end{align}
\end{subequations}
We note that in the nuclear models considered here, $C_{\mathrm{b},n} \ll 1$, hence $\ell_{\mathrm{b},n} \approx C_{\mathrm{b},n}^{1/4}$. In Figure, \ref{fig:diffusion_coefficients_S_stars} $\ell_{\mathrm{a},n}$ and $\ell_{\mathrm{b},n} > \ell_{\mathrm{a},n}$ are indicated with the two blue (black) triangles for $n=1$ ($n=2$). We also note that if $C_1=C_2$, which we observe is approximately the case in our simulations, then $\ell_{\mathrm{a},2} = 2^{1/4} \ell_{\mathrm{a},1}$ and $\ell_{\mathrm{b},1} = \ell_{\mathrm{b},2}$.

\begin{figure}
\center
\includegraphics[scale = 0.435, trim = 0mm 0mm 0mm 0mm]{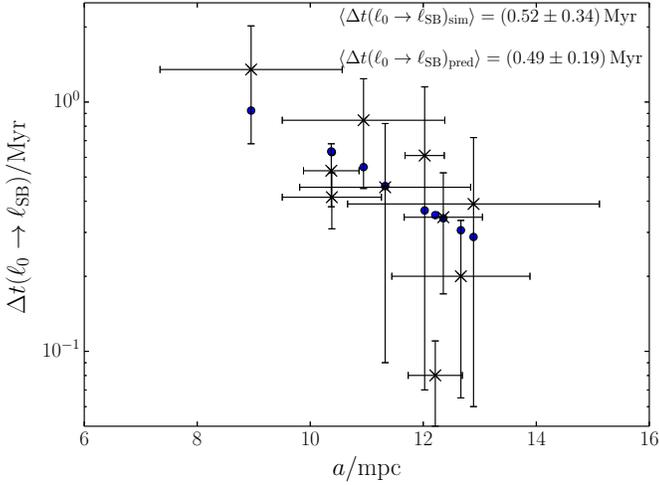}
\caption{\small The times $\Delta t(\ell_0 \rightarrow \ell_\mathrm{SB})$ required to diffuse from the initial angular momentum $\ell_0$ to the SB if $\ell_0 < \ell_\mathrm{SB}$. Crosses with error bars show diffusion times obtained from the S-star simulations. Horizontal error bars show the standard deviation of the values of $a$ from $t=0$ until reaching the SB for the first time for all realizations of the S-stars with $\ell_0 < \ell_\mathrm{SB}$. Vertical error bars show the median absolute value of the diffusion times based on the latter realizations. Blue bullets show the predicted times computed from equation~(\ref{eq:calc_time_scale}), where the mean semimajor axis from the simulations at $\ell=\ell_\mathrm{SB}$ was used. }
\label{fig:S_stars_dif_time_scales_test}
\end{figure}

Using equation~(\ref{eq:dif_coef_approx}) it is possible to estimate the time $\Delta t(\ell_1 \rightarrow \ell_2)$ to diffuse in angular momentum for any specified interval in $\ell$. We are most interested here in the time $\Delta t(\ell_0 \rightarrow \ell_\mathrm{SB})$ to diffuse from an initial value $\ell_0<\ell_\mathrm{SB}$ to $\ell_\mathrm{SB}$. Assuming -- as is appropriate for these nuclear models -- that the diffusion time is dominated by the AR regime, equation~(\ref{eq:dfc_AR}) implies:
\begin{subequations}\label{eq:calc_time_scale}
\begin{align}
\frac{\mathrm{d} \ell}{\mathrm{d} t} & \approx \langle \Delta \ell \rangle \approx \frac{C_1 \ell^3}{\tau} \Rightarrow \\
\Delta t (\ell_0\rightarrow \ell_\mathrm{SB}) &\approx \frac{\tau}{2C_1} \left ( \ell_0^{-2} - \ell_\mathrm{SB}^{-2} \right) \\
&\approx \frac{\tau}{2C_1 \ell_0^2}.
\end{align}
\label{eq:delta_t_AR}
\end{subequations}
The last step applies if $\ell_0 \ll \ell_\mathrm{SB}$. We tested equation~(\ref{eq:calc_time_scale}), and hence equation~(\ref{eq:dfc_AR}), by using the former to compute the time scales to diffuse from the initial value of $\ell$, $\ell_0$, to $\ell_\mathrm{SB}$, in the context of our simulations of the S-stars. In the latter simulations the initial values of $\ell$ are $0.14 \lesssim \ell_0 \lesssim 0.37$ with $\mathrm{d}N/\mathrm{d} \ell_0 = 2\,\ell_0$ (cf. \S\,\ref{sect:S-stars:initial_cond}). From the simulations we selected the S-stars with $\ell_0<\ell_\mathrm{SB}$ and we recorded the time $\Delta t(\ell_0 \rightarrow \ell_\mathrm{SB})_\mathrm{sim}$ it takes for $\ell$ to increase from $\ell_0$ to $\ell_\mathrm{SB}$. We also recorded the mean semimajor axis in this time interval. The latter value was used to compute the predicted time $\Delta t(\ell_0 \rightarrow \ell_\mathrm{SB})_\mathrm{pred}$ based on 1000 realizations of the initial value of $\ell$ (sampled similarly as in the simulations). For each realization we evaluated equation~(\ref{eq:calc_time_scale}) if $\ell_0<\ell_\mathrm{SB}$, with $C_{A_\mathrm{D}} = 0.5$ and $C_1=C_2=2.6$. From these realizations the mean was adopted as $\Delta t(\ell_0 \rightarrow \ell_\mathrm{SB})_\mathrm{pred}$. 

The predicted times are shown as function of semimajor axis with blue bullets in Figure \ref{fig:S_stars_dif_time_scales_test}; the times extracted from the simulations are shown as crosses with error bars. According to the prediction the time $\Delta t(\ell_0 \rightarrow \ell_\mathrm{SB})$ decreases with increasing $a$ which is borne out by the simulations, although there is considerable scatter. The mean values of $\Delta t(\ell_0 \rightarrow \ell_\mathrm{SB})$, averaged over the semimajor axes, are $\langle \Delta t(\ell_0 \rightarrow \ell_\mathrm{SB})_\mathrm{sim} \rangle \approx (0.5 \pm 0.3) \, \mathrm{Myr}$ and $\langle \Delta t(\ell_0 \rightarrow \ell_\mathrm{SB})_\mathrm{pred} \rangle \approx (0.5 \pm 0.2) \, \mathrm{Myr}$ for the simulations and predictions, respectively, and are consistent with each other.

\subsection{A new criterion for the location of the barrier}
\label{sect:SBcriterion}
In MAMW11, the SB was first observed as a locus in the $\log a$ vs. $\log (1-e)$ plane where the $N$-body trajectories ``bounced'' in the course of their RR-driven random walk in $L$. Equation (\ref{eq:ell_SB}), which was derived from a simple timescale argument, was found to reproduce the ``bounce'' location $e_\mathrm{SB}(a)$ with acceptable accuracy in those simulations.

The location of the barrier in the MAMW11 simulations was determined by eye from the $\log a$ vs. $\log(1-e)$ plane. Figures \ref{fig:introduction_figure}, \ref{fig:diffusion_coefficients} and \ref{fig:diffusion_coefficients_S_stars} from this paper suggest a new, more robust criterion for $e_\mathrm{SB}(a)$ in terms of the diffusion time scales or diffusion coefficients.

Under the influence of RR, the diffusion coefficient in $\ell$, $\langle(\Delta\ell)^2\rangle$, first increases toward smaller $\ell$, then sharply drops when $\ell$ is small enough that GR precession suppresses the effects of the torques. A natural definition for the angular momentum associated with the barrier at radius $a$ is the value $\ell=\ell_p(a)$ at which $\langle(\Delta\ell)^2\rangle$ peaks.

We can implement this criterion in two ways: using our analytic expressions for $\langle\left(\Delta\ell\right)^2\rangle$, or using the numerically-computed diffusion coefficients. To the extent that the analytic expressions correctly predict the numerical results, the two approaches should yield similar answers.

The analytic expressions for $\langle\left(\Delta\ell\right)^2\rangle$, equations (\ref{eq:dfc_lRR2}) and (\ref{eq:dfc_AR}), imply a maximum at $\ell=\ell_{\mathrm{b},2}(a)$, the latter given by equation (\ref{eq:labn}b). From that expression, the dependence of $\ell_{\mathrm{b},2}$ on $C_{\mathrm{b},2}$ in the limits of large- and small $a$ is easily shown to be
\begin{subequations}\label{eq:asymptote}
\begin{eqnarray}
\ell_{\mathrm{b},2} &\rightarrow& C_{\mathrm{b},2}^{1/4}, \ \ \ \ \ \ a\rightarrow\infty \ (C_{\mathrm{b},2}\ll 1) \\
&\rightarrow& 1 - \frac{2}{C_{\mathrm{b},2}}, \ \ a\rightarrow 0 \ \ (C_{\mathrm{b},2}\gg 1) .
\end{eqnarray}
\end{subequations}
In the models considered here, $C_{\mathrm{b},2}<1$ at the radii of interest. Equation (\ref{eq:asymptote}a) implies
\begin{subequations}\label{eq:newSB}
\begin{eqnarray}\label{eq:newSBa}
\ell_{\mathrm{b},2}^2(a) &\approx& \frac{2\alpha_s}{\sqrt{C_2}C_{A_\mathrm{D}}}
\left(\frac{r_g}{a}\right) \left[\frac{t_\mathrm{coh}}{P(a)}\right] \\
\label{eq:newSBb}
&\approx& 4.0\left(\frac{r_g}{a}\right) \left[\frac{t_\mathrm{coh}}{P(a)}\right] .
\end{eqnarray}
\end{subequations}
Unlike equation (\ref{eq:ell_SB}), the new expression (\ref{eq:newSB}) for the barrier location depends explicitly on the coherence time; in fact, $\ell_{\mathrm{b},2}$ is roughly the angular momentum for which the GR precession time equals $t_\mathrm{coh}$.

We can also estimate $\ell_p(a)$ directly from the numerically-computed diffusion coefficients. Since the numerical data are noisy, we fit smoothing splines to the measured ($X,Y$) values in Figure \ref{fig:diffusion_coefficients_S_stars}, where $X=\log\ell$ and $Y=\log\langle(\Delta\ell)^2\rangle$. The optimal choice of smoothing parameter for each data set was determined via the standard technique of generalized cross validation \citep{wahba90}. An estimate of the uncertainty associated with the location of the peak at each $a$ was then made via the bootstrap, by resampling at random from the measured points and repeating the spline fits, recalculating the smoothing parameter with each new bootstrap sample.

\begin{figure}
\center
\includegraphics[scale = 0.475, trim = 0mm 0mm 0mm 0mm]{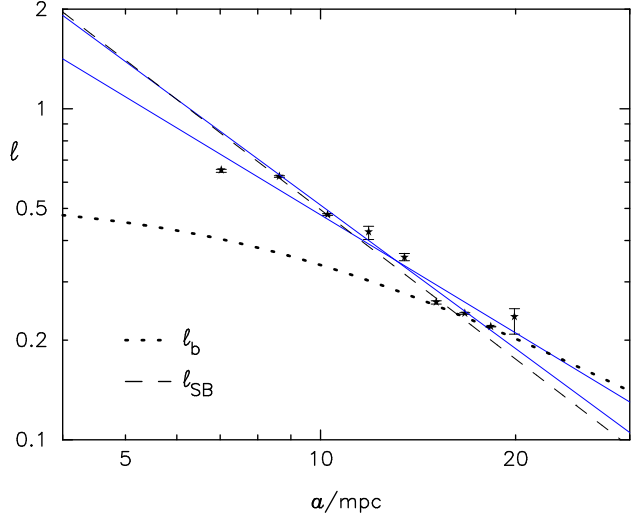}
\caption{\small Points with error bars are the values of $\ell$ at which the numerically-computed diffusion coefficients $\langle(\Delta\ell)^2\rangle$ peak in the S-star simulations (cf. Fig. \ref{fig:diffusion_coefficients_S_stars}). Thick (blue) lines are best-fit power laws: to all of the points (shallow slope) and to a subset that excludes the two data 
points at largest and smallest $a$ (steep slope). The dashed (black) line is equation (\ref{eq:ell_SB}) or (\ref{eq:lSBSstar}), and the dotted (black) curve is $\ell_{\mathrm{b},2}(a)$, equation (\ref{eq:labn}b).}
\label{fig:relation}
\end{figure}

Figure \ref{fig:relation} shows the results, for data having $7 \lesssim a/\mathrm{mpc} \lesssim 20$. Values of $\ell_p$ derived from data both at large and small $a$ are problematic:
the former because the data are noisy, the latter because there tends not to be a well-defined maximum. Excluding the two data points at largest and smallest $a$ in Figure \ref{fig:relation} results in a set of points that define a good power law; least-squares fit of a straight line to this subset of the data yields 
\begin{eqnarray}
\ell_p(a) &=& \ell_{p,10} \left(\frac{a}{10\;\mathrm{mpc}}\right)^{\beta}, \nonumber \\
\ell_{p,10} &=& 0.51 \pm 0.016, \nonumber \\
\beta &=& -1.43 \pm 0.086 .
\label{eq:lsqSB}
\end{eqnarray}
This relation is statistically indistinguishable from equation~ (\ref{eq:lSBSstar}), the ``Schwarzschild barrier'' as defined in MAMW11. Interestingly, that relation is a {\it better} fit to the points than $\ell_{\mathrm{b},2}(a)$, which is also plotted in Figure \ref{fig:relation}. The departure of the measured peak-values from the analytic prediction can be understood by referring to Figure  \ref{fig:diffusion_coefficients_S_stars}, which shows that for $a\lesssim 10$ mpc, the peak of the measured diffusion coefficients occurs increasingly at $\ell>\ell_{\mathrm{b},2}$.

The good agreement which we find between the barrier location as defined in MAMW11, and by our new criterion based on the diffusion coefficient, may be partly fortuitous. Nevertheless the agreement is encouraging, since it suggests that the ``barrier'' that was identified in MAMW11, based on the short-term behavior of orbits, can be recovered in a robust and quantitative way from simulations. It is also interesting to note that a single relation appears to define the barrier location both in these simulations and those of MAMW11, verifying that equation (\ref{eq:ell_SB}) holds true in systems with very different particle numbers and particle masses. At the same time, given the uncertainties in the numerical coefficients, we do not feel confident that we have necessarily ruled out our alternate expression (\ref{eq:newSB}) for the barrier location and we suggest that future work should compare both that expression and the one given in MAMW11 with the results of numerical simulations.

\section{Steady-state distribution}
\label{sect:steady-state}
In a nucleus where evolution in angular momentum was dominated by NRR, the steady-state phase-space density would be isotropic, $f=f(E)$, and the eccentricity distribution at any energy would be $\mathrm{d}N/\mathrm{d}e = 2e$, a ``thermal'' distribution. The steady-state eccentricity distribution under the influence of RR has not been well established. The semi-empirical model of \citet{madigan11} (hereafter MHL11) predicts an eccentricity distribution that is bimodal with peaks at both low ($\sim 0.2$) and high ($\sim 0.9$) eccentricities at small semimajor axes. Our $N$-body simulations include the effects of both NRR and RR on the orbital angular momenta, and relativistic corrections to the equations of motion are also taken into account (the latter were not included by MHL11). Using the diffusion coefficients that we obtained in \S\,\ref{sect:S-stars:dfc} it is therefore possible to investigate, for the first time, the expected steady-state distribution in angular momentum near a SBH under the joint influence of RR, NRR and general relativity.

Let $N(E,R,t)\, \mathrm{d} R \, \mathrm{d}E$ be the number of stars at time $t$ in angular momentum interval $\mathrm{d}R$, where $R \equiv L^2/L_c^2(E) \approx 1-e^2 \equiv\ell^2$, and energy interval $\mathrm{d}E$. The orbit-averaged Fokker-Planck equation is \citep[][5.5.1]{bookmerritt13}:
\begin{align}
\nonumber &\frac{\partial N(E,R,t)}{\partial t} \\
&= - \frac{\partial}{\partial R} \left [N(E,R,t) \langle \Delta R \rangle \right] + \frac{1}{2} \frac{\partial^2}{\partial R^2} \left [ N(E,R,t) \left \langle \left (\Delta R \right )^2 \right \rangle \right ].
\label{eq:FPsteady_state}
\end{align}
Here $\langle \Delta R \rangle = \langle \Delta R \rangle(E,R)$ and $\langle (\Delta R)^2 \rangle = \langle (\Delta R)^2 \rangle(E,R)$ are the first- and second-order, orbit-averaged diffusion coefficients in $R$; of course, the diffusion coefficients that we extract numerically from the $N$-body integrations are also orbit-averaged. Our motivation for expressing the Fokker-Planck equation in terms of the variable $R$, rather than $L$ or $\ell$, is that the first-order NRR diffusion coefficient in the limit $\ell \rightarrow 0$ diverges as $1/\ell$ (cf. equation~(\ref{eq:dfc_NRR_l0})), whereas this divergence in the equivalent limit $R\rightarrow0$ does not occur if expressed in terms of $R$. The diffusion coefficients in $R$ can be related, without approximation, to diffusion coefficients in $\ell$, i.e. $\langle \Delta \ell \rangle$ and $\langle (\Delta \ell)^2 \rangle$ \citep[][eq. 5.167]{bookmerritt13}. 

Before proceeding, we note the following caveats.

(1) We are finding the steady-state distribution of a set of test stars as they respond dynamically to a {\it specified} field-star distribution. In reality, the distribution of field stars would also evolve toward a steady state, both with respect to angular momentum (on the RR time scale) and energy (on the longer NRR time scale). It is often argued \citep[e.g.][]{cohnk78} that calculating diffusion coefficients from a non-self-consistent $L$-distribution is an adequate approximation, and in fact this was done in almost all studies prior to ours, including that of MHL11.

(2) Orbit averaging is a way of removing the short time scale (the radial orbital period) from the problem, by assuming that integrals like $L$ are fixed over this time scale. In the Newtonian problem, angular momentum is conserved (in a spherical cluster) in the absence of gravitational encounters. In the problem we are solving, there is a second short time scale when $\ell\lesssim\ell_\mathrm{SB}$: the time for GR precession. As noted above, $L$ is not precisely conserved over a GR precessional cycle: it oscillates in response to the (nearly) fixed torques from the field stars. One way to deal with this additional short time scale would be to express the Fokker-Planck equation in terms of a new quantity that is conserved during the precession; for instance, the ``secular Hamiltonian'' mentioned in \S \ref{sect:S-stars:dfc}. Instead, when applying the Fokker-Planck equation to the $\ell\lesssim\ell_\mathrm{SB}$ regime, we interpret $\ell$ as $\ell_\mathrm{av}$, its average value over a GR precessional cycle.
This interpretation is fully consistent with the manner in which the diffusion coefficients were extracted from the simulations. Furthermore, as noted in \S\,\ref{sect:S-stars:dfc}, the ``secular Hamiltonian'' is essentially $\ell_\mathrm{av}$.

(3) The Fokker-Planck equation assumes that the diffusion coefficients of third and higher order are negligible. In the case of diffusion driven by NRR, this approximation can be justified for intermediate and long time scales as compared to the relaxation time scale (e.g. \citealt{bookspitzer87}); at short time scales this is likely not the case \citep{bka13}. We are not aware of a justification of the neglect of higher-order diffusion coefficients in the case of RR and AR. In fact, extraction of the angular momentum transition probabilities from $N$-body simulations (D. Merritt, unpublished) reveals that the probability distributions are often extremely skewed near the SB, implying non-negligible third-order coefficients. The skewness is related to the ``bounce'' phenomenon near the SB, and by neglecting it in what follows, our results for the steady-state solutions are likely to have systematic errors near the SB.

(4) We are assuming either zero or constant flux $C$ of orbits in the $L$-direction (cf. equation~(\ref{eq:FPsteady_state_C})). This assumption cannot be strictly correct because $C$ must be zero at $L=L_c$, the angular momentum of a circular orbit, whereas it is nonzero near the loss cone $L_\mathrm{lc}$. In reality there must therefore also be a flux in the energy direction which supplies the loss of stars near the loss cone. In order to relax our assumption of constant flux it would be necessary to solve the 2D Fokker-Planck equation for $f(E,L)$, which is beyond the scope of the current paper. We expect, however, that the functional dependence of the steady-state distribution on $L$ is not strongly affected by assuming a constant flux in the $L$-direction. 

(5) As inner boundary condition, we set $f=0$ for orbits that satisfy the capture criterion that was defined in \S\,\ref{sect:S-stars:initial_cond}. In some contexts, a more appropriate condition would be to set $f=0$ at the smaller $L$ corresponding to orbits that intersect the sphere for capture of compact remnants; or at the larger $L$ for which the angular momentum diffusion time equals the time for gravitational-wave energy loss (cf. Figure \ref{fig:introduction_figure}). Our inner boundary condition is only strictly correct for test stars that have zero mass and radius, and this choice will affect both the steady-state solutions and the implied flux.

With these caveats in mind, we return to equation~(\ref{eq:FPsteady_state}) and set $\partial N/\partial t=0$. The result is:
\begin{align}
- N(R) \langle \Delta R \rangle + \frac{1}{2} \frac{\partial}{\partial R} \left ( N(R) \left \langle \left ( \Delta R \right )^2 \right \rangle \right ) = C.
\label{eq:FPsteady_state_C}
\end{align}
Here $C$ is an ``angular momentum flux''. The dependence of both $N$ and $C$ on $E$ (i.e. $a$) is understood. Equation~(\ref{eq:FPsteady_state_C}) has two types of solutions: those with $C=0$ (homogeneous; zero flux) and those with $C\neq0$ (inhomogeneous; constant flux). Exact solutions exist for both cases and are derived in Appendix \ref{app:steady-state:gen_sol}. These solutions require knowledge of the diffusion coefficients at arbitrary values of the angular momentum. 

\subsection{Analytic solutions}
\label{sect:steady-state:analytic}
In equations~(\ref{eq:dif_coef_approx})-(\ref{eq:labn}) we presented approximate analytic expressions for the diffusion coefficients. The second-order coefficients from our $N$-body simulations are well described in terms of equation~(\ref{eq:dif_coef_approx}), as was demonstrated in Figure \ref{fig:diffusion_coefficients_S_stars}. In the case of the first-order coefficients the agreement of equation~(\ref{eq:dif_coef_approx}) with the data is good for $\ell \lesssim \ell_{\mathrm{b},1}$. For $\ell \gg \ell_{\mathrm{b},1}$, however, this agreement is poor: $\langle \Delta \ell \rangle$ is expected and observed to be negative at $\ell \approx 1$; the latter feature is not described by equation~(\ref{eq:dfc_lRR1}). In Figure \ref{fig:diffusion_coefficients_S_stars} the value of $\ell$ for which $\langle \Delta \ell \rangle$ becomes negative, $\ell_\mathrm{c}$, is weakly dependent on semimajor axis. In addition, the results in Figure \ref{fig:diffusion_coefficients} suggest that $\ell_\mathrm{c}$ also only weakly depends on $N_\mathrm{max}$. We therefore assume that $\ell_\mathrm{c}$ is constant for our present purposes, and adopt the value $\ell_\mathrm{c} \approx 0.7$. Furthermore, we adopt $C_1 = C_2 = 2.6$, hence $\ell_{\mathrm{a},2} = 2^{1/4} \ell_{\mathrm{a},1}$ and $\ell_{\mathrm{b},1} = \ell_{\mathrm{b},2}$ (cf. equation~(\ref{eq:labn})).

To take into account the sign change of $\langle \Delta \ell \rangle$ at $\ell \approx \ell_\mathrm{c}$ observed in our $N$-body simulations, we make the following two changes to equation~(\ref{eq:dfc_lRR1}). (1) Instead of letting $\langle \Delta \ell \rangle \rightarrow 0$ as $\ell\rightarrow1$, we let $\langle \Delta \ell \rangle \rightarrow 0$ as $\ell \rightarrow \ell_\mathrm{c}$. As $\ell$ increases to $\ell>\ell_\mathrm{c}$, then $\langle \Delta \ell \rangle < 0$. (2) We multiply the resulting expression by a constant factor to ensure that $\langle \Delta \ell \rangle$ is continuous at $\ell=\ell_{\mathrm{b},1}$. The explicit form of $\langle \Delta \ell \rangle$ in the range $\ell_{\mathrm{b},1} < \ell < 1$ is included in Appendix \ref{app:steady-state:dfc_analytic}. For completeness, we have also included there explicit expressions of the diffusion coefficients in $\ell$ in the other regimes, based on equation~(\ref{eq:dif_coef_approx}). 

In the top panel of Figure \ref{fig:steady_state_analytic} we show the analytic functions for the diffusion coefficients described in Appendix \ref{app:steady-state:dfc_analytic} and we compare these to the coefficients obtained from the S-star simulations (cf. \S\,\ref{sect:S-stars:dfc}), for a single semimajor axis bin. The analytic relations capture the basic features of the coefficients obtained from the simulations. The boundary values given by equation~(\ref{eq:labn}) have been indicated in all of the panels of Figure \ref{fig:steady_state_analytic}. The quantity $R_\mathrm{loss}$ is the value of $R$ that corresponds to disruption of the star by the SBH; $R_\mathrm{loss}(a) = r_\mathrm{capt}/a \, (2-r_\mathrm{capt}/a)$. In the simulations $r_\mathrm{capt} = 2 R \, (M_\bullet/m)^{1/3} \approx 3 \times 10^{-2} \, \mathrm{mpc}$ with $R = 8 \, \mathrm{R}_\odot$ and $m=10\, \mathrm{M}_\odot$, giving $R_\mathrm{loss} = \mathcal{O}(10^{-2})$ for the semimajor axes of interest.

In the second panel of Figure \ref{fig:steady_state_analytic} we show the analytic coefficients transformed to $R$ (cf. equation~(\ref{eq:dfc_trans_R_l})). We note that we defined $R_\mathrm{c}$ as the value of $R$ for which $\langle \Delta \ell \rangle$ changes sign from positive to negative values. In general, the latter is different from the value of $R$ for which $\langle \Delta R \rangle$ changes sign from positive to negative values, as illustrated in the first and second panels of Figure \ref{fig:steady_state_analytic}. 

In the third panel of Figure \ref{fig:steady_state_analytic} we show the analytic solution of equation~(\ref{eq:FPsteady_state_C}) assuming $C\neq0$, computed from equation~(\ref{eq:solfull}) (black dashed line). The latter solution is given explicitly in Appendix \ref{app:steady-state:steady_state_analytic}. For verification of the analytical results we also include results from numerical integrations using the analytic diffusion coefficients (black solid line). The eccentricity distribution $N(e)$ that follows from the solution $N(R)$ is shown in the fourth panel of Figure \ref{fig:steady_state_analytic}. In this figure and those that follow in this section, all probability density functions are normalized to unit total number. 

\begin{figure}
\center
\includegraphics[scale = 0.435, trim = 0mm 0mm 0mm 0mm]{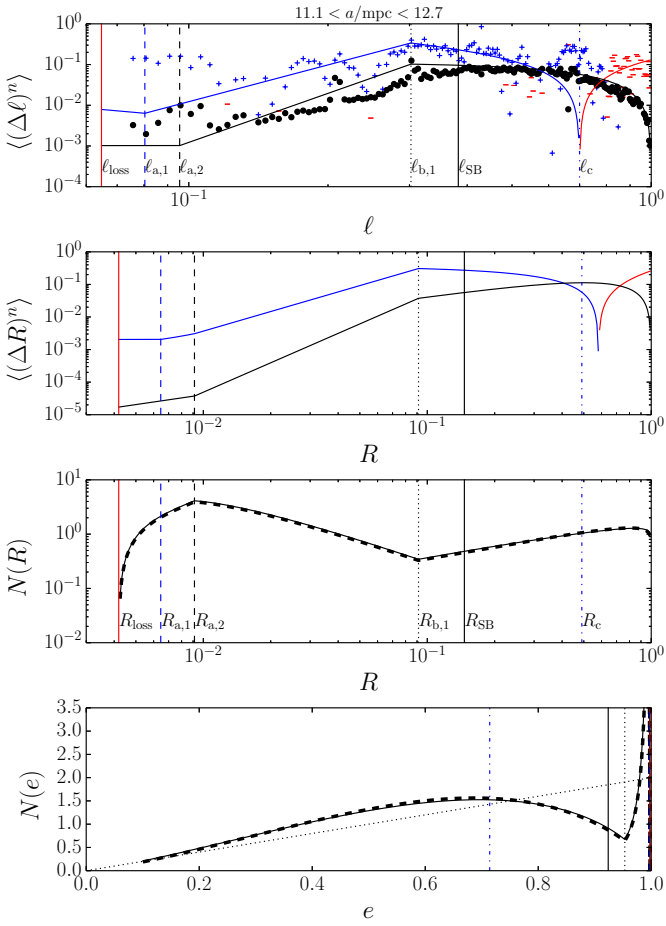}
\caption{\small Analytic solutions of the steady-state distributions in $R$ and $e$. Top panel: bullets, minusses and plusses show the diffusion coefficients in $\ell$ obtained from the S-star simulations (cf. \S\,\ref{sect:S-stars:dfc}). The continuous lines show our adopted analytic model, which is described explicitly in Appendix \ref{app:steady-state:dfc_analytic}. Here we set $a$ equal to the mean of the semimajor axis bin from the simulations. Second-order quantities are shown in black; positive (negative) first-order quantities are shown in blue (red). Various boundaries in terms of $\ell$ (and in terms of $R\equiv \ell^2$ and $e=\sqrt{1-\ell^2}$ in the other panels) are indicated with vertical lines (cf. equation~(\ref{eq:labn})). Second panel: the analytic coefficients transformed to $R$ using equation~(\ref{eq:dfc_trans_R_l}). Third panel: the solution to equation~(\ref{eq:FPsteady_state_C}) in terms of $R$. Thick dashed lines: analytic expressions, given in Appendix \ref{app:steady-state:steady_state_analytic}; solid lines: numerical solutions. Fourth panel: the corresponding solutions in terms of $e$. The black dotted line shows a thermal distribution. }
\label{fig:steady_state_analytic}
\end{figure}

\begin{figure}
\center
\includegraphics[scale = 0.435, trim = 0mm 0mm 0mm 0mm]{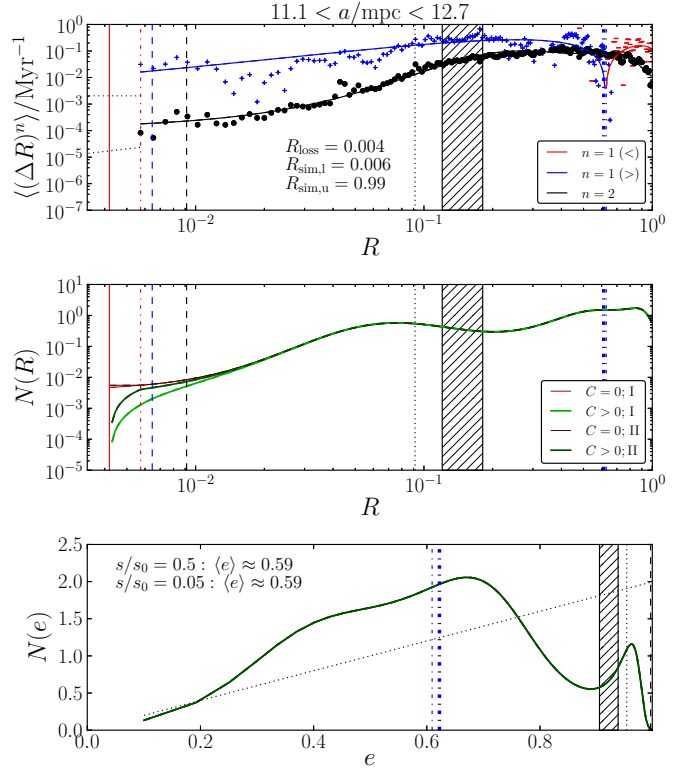}
\caption{\small Numerical solutions of the steady-state distributions in $R$ and $e$ based on interpolations of the diffusion coefficients extracted from the $N$-body simulations, for one semimajor axis bin. Top panel: bullets, minusses and plusses show the transformed diffusion coefficients in $R$ derived from the S-star simulations (cf. \S\,\ref{sect:S-stars:dfc}). The continuous lines show fifth-order spline interpolations to the data. Two different values of the smoothing parameter $s$ are adopted: $s=0.5\,s_0$ (solid lines) and $s=0.05\,s_0$ (dashed lines); in this case, however, the results for both values of $s$ are identical. Various boundaries in terms of $\ell$ and $R\equiv \ell^2$ and $e=\sqrt{1-\ell^2}$ are indicated with vertical lines (cf. Figure \ref{fig:steady_state_analytic}). The red vertical dot-dashed line indicates $R_\mathrm{sim,l}$, the smallest value of $R$ for which the diffusion coefficients were determined. The blue vertical thick dot-dashed line shows the value of angular momenta for which the interpolated $\langle \Delta \ell \rangle$ changes sign; the blue vertical thin dot-dashed line shows the value of angular momentum for which the interpolated $\langle \Delta R \rangle$ changes sign. The SB, which has a range in angular momentum because there is a range of semimajor axes, is indicated with the black hatched region. Second panel: the numerical solution to equation~(\ref{eq:FPsteady_state_C}) in terms of $R$ based on the interpolations. Two methods were used to extrapolate to the region $R_\mathrm{loss} < R < R_\mathrm{sim,l}$ that is missing in the data, cf. equation~(\ref{eq:dfc_numerical_extrapolation_methods}). Method I: light colour; method II: darker colour. We include solutions with $C=0$ (red lines) and $C\neq0$ (green lines). Third panel: the corresponding solution in terms of $e$. The black dotted line shows a thermal distribution. }
\label{fig:steady_state_numerical}
\end{figure}

We note the following features in the analytic solutions based on our analytic approximations of the diffusion coefficients obtained from the $N$-body simulations:
\begin{enumerate}
\item For $R_\mathrm{loss} < R < R_{\mathrm{a},1}$, $N(R)$ increases logarithmically with $R$, i.e. $N(R) \propto \log(R/R_\mathrm{loss})$. This is the well-known NRR ``empty loss cone'' result \citep{cohnk78} and reflects our assumed form of the diffusion coefficients in this regime. For $R_{\mathrm{a},1} < R < R_{\mathrm{a},2}$ the trend of increasing $N(R)$ continues, although the dependence on $R$ is no longer strictly logarithmic.
\item For $R_{\mathrm{a},2} < R_{\mathrm{b},1}$, $N(R)$ decreases with $R$. Approximately, $N \propto R^{-2}$ for $R\approx R_{\mathrm{b},1}$, independent of $C_1$ or $C_2$ if $C_1=C_2$ (cf. Appendix \ref{app:steady-state:steady_state_analytic}). 
\item For $R_{\mathrm{b},1} < R_{\mathrm{c}}$, $N(R)$ once again increases with $R$. As $R$ increases to $R>R_\mathrm{c}$, $N(R)$ drops. The latter reflects the rapid drop of $\langle (\Delta R)^2 \rangle$ as $R \rightarrow 1$, which can be interpreted as arising from the strongly reduced efficiency of RR as $R \rightarrow 1$. 
\end{enumerate}

The above features imply that there are {\it two} local maxima and {\it three} local minima in $N(R)$ (and, similarly, in the eccentricity distribution $N(e)$): two maxima at $R_{\mathrm{a},2}$ and near $R_\mathrm{c}$, and three minima at $R_\mathrm{loss}$, $R_{\mathrm{b},1}$ and $R=1$. The local minimum at $R_{\mathrm{b},1}$ is near the ``knee'' feature that was observed in the diffusion coefficients and, furthermore, $R_{\mathrm{b},1}$ is comparable to $R_\mathrm{SB}$. This suggests that the SB can be associated with a {\it deficit} of orbits in the steady-state angular momentum distribution. 

An interesting feature of the steady-state solution $N(R)$ is a local maximum in $N$ at $R < R_\mathrm{SB}$. We suggest that this can be explained by the inefficiency of AR, which is the dominant form of relaxation in the angular-momentum regime $R_{\mathrm{a},2} < R < R_{\mathrm{b},1}$. If we imagine that the region below the SB was initially unpopulated,  stars would diffuse to $R<R_\mathrm{SB}$ at some rate determined by $\langle(\Delta R)^2\rangle_\mathrm{SB}$. Once ``below the barrier,'' stars would experience diffusion at much lower rates, causing them to ``pile up'' until reaching a high enough density that the fluxes in the AR and RR regions are equalized. Apparently, achieving this equality can result in higher values of $N$ below the SB than above -- a non-intuitive result given the difficulty of crossing the SB from above. At even lower $R$, $N$ drops again because of losses to the SBH. 

In the next section we show that, although still clearly present, the increase in the value of $N$ below the SB is probably less extreme than suggested by these analytic solutions.

\begin{figure}
\center
\includegraphics[scale = 0.435, trim = 0mm 0mm 0mm 0mm]{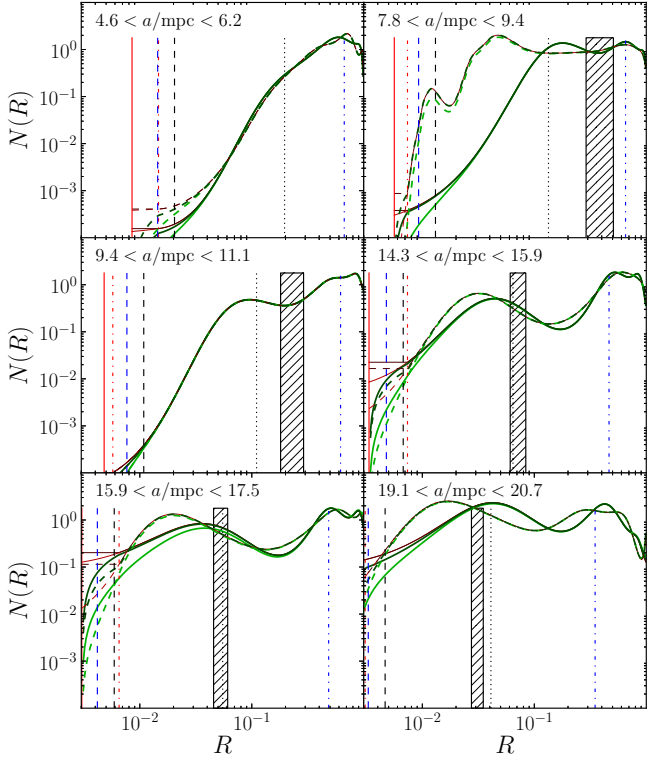}
\caption{\small Numerical steady-state distributions in $R$ based on interpolated diffusion coefficients extracted from the $N$-body simulations as in Figure \ref{fig:steady_state_numerical}, for different semimajor axis bins. Line colors and styles are the same as in Figure \ref{fig:steady_state_numerical}. The blue vertical dot-dashed line shows the value of $R$ for which the interpolated $\langle \Delta \ell \rangle$ changes sign. The time lags for the diffusion coefficients are chosen according to the criterion discussed in \S\,\ref{sect:belowSB:diffusion}. }
\label{fig:steady_state_sma_R}
\end{figure}

\begin{figure}
\center
\includegraphics[scale = 0.435, trim = 0mm 0mm 0mm 0mm]{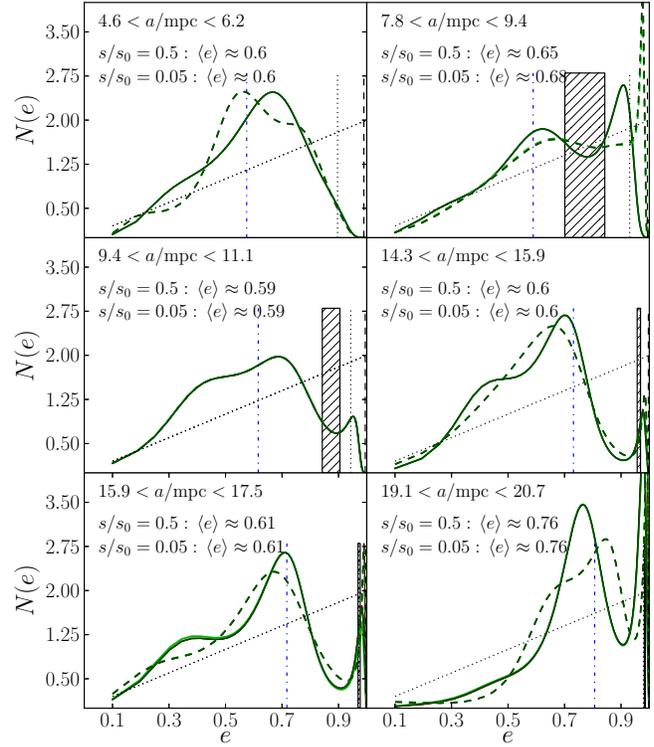}
\caption{\small Steady-state eccentricity distributions based on the solutions shown in Figure \ref{fig:steady_state_sma_R}. }
\label{fig:steady_state_sma_e}
\end{figure}

\subsection{Numerical solutions}
\label{sect:steady-state:numerical}
In \S\,\ref{sect:steady-state:analytic} we presented analytic functions that approximate the diffusion coefficients obtained from our $N$-body simulations, and we obtained analytic solutions for the steady-state angular momentum distribution. This method facilitates insight into the steady-state solutions, but it turns out to be inaccurate insofar as the relative heights of the peaks in $N(R)$ are concerned. We also obtained numerical solutions by fitting splines to the diffusion coefficients obtained from the simulations. Although we find the same basic features in $N(R)$ discussed above, the analytic method fails to accurately describe the relative importance of the two local maxima in $N$. This is likely due to the sensitivity of the solution $N(R)$ to $\langle \Delta \ell \rangle$ in the regime $\ell_{\mathrm{b},1} < \ell < 1$. For example, by multiplying $\langle \Delta \ell \rangle$ by factors of a few in the analytic prescription (this does not make the fit to the data much worse), we find that the peak near $R_\mathrm{c}$ becomes much more dominant compared to the peak near $R_{\mathrm{a,2}}$. 

In this section we present numerical steady-state solutions based on fifth-order spline fitting\footnote{We used the \textsc{splprep} routine implemented in \textsc{SciPy}, a \textsc{Python} library. } of the diffusion coefficients in $R$. The latter were derived from transformation of the measured coefficients in $\ell$ to $R$. For all the results shown in this section we adopted the same criterion for the time lags as in \S\,\ref{sect:S-stars:dfc}. We show an example of the spline fitting in the top panel of Figure \ref{fig:steady_state_numerical}. To obtain better fit results for a large range in $R$ we fitted the logarithm of $\langle (\Delta R)^2 \rangle$; this is not the case for $\langle \Delta R \rangle$, which changes sign at $R \approx 0.6$. A parameter that affects the result of the interpolation is the smoothness $s$ of the interpolated spline. For a data set $(x_i,y_i)$ this parameter is defined via the condition that $\sum_i[y_i - h(x_i)]^2 \leq s$, where $h(x)$ is the interpolation function. In order to obtain a measure of uncertainty associated with the choice of $s$ we adopted two values, $s = 0.5 \, s_0$ and $s=0.05\,s_0$, where $s_0 = N - \sqrt{2N}$ and $N$ is the number of data points. Generally, the former value yields a smooth interpolation, whereas the latter yields a more detailed, but less smooth interpolation, which is more sensitive to scatter in the data. 

The interpolated diffusion coefficients have a range $R_\mathrm{sim,l} \leq R \leq R_\mathrm{sim,u}$; the boundaries vary per semimajor axis bin. Typically $R_\mathrm{sim,l} \sim 10^{-2}$ and $R_\mathrm{sim,u} \sim 1-10^{-2}$. The lower limit $R_\mathrm{sim,l}$ is comparable to, but slightly larger than the value of $R$ that corresponds to the assumed tidal disruption radius in the simulations, $R_\mathrm{loss}(a) = r_\mathrm{capt}/a \, (2-r_\mathrm{capt}/a) = \mathcal{O}(10^{-2})$. For the semimajor axis range shown in the top panel of Figure \ref{fig:steady_state_numerical} $R_\mathrm{loss} \approx 0.004$, whereas $R_\mathrm{sim,l} \approx 0.006$.

As mentioned above, in the solutions with $C\neq0$, $N(R)$ is set to zero at $R_\mathrm{loss}$. This constraint is physically desirable since close to the SBH the distribution function should be zero at $R<R_\mathrm{loss}$ (``empty loss cone''). Implementing $N(R_\mathrm{loss}) = 0$ requires knowledge of the diffusion coefficients in the range $R_\mathrm{loss} < R < R_\mathrm{sim,l}$, which is not available in our data. Therefore, we imposed two different extrapolations for the diffusion coefficients in this regime:
\begin{align}
\left \{ \begin{array}{ll}
\displaystyle \left \langle \left (\Delta R \right )^n \right \rangle (R) = \left \langle \left (\Delta R \right )^n \right \rangle (R_\mathrm{sim,l}); & (\mathrm{method \, I}) \\
  \left \{ \begin{array}{ll}
  \displaystyle \left \langle \Delta R \right \rangle (R) &= A(E); \\
  \displaystyle \left \langle \left (\Delta R \right )^2 \right \rangle (R) &= 2 R A(E). \\
\end{array} \right. & (\mathrm{method \, II})
\end{array} \right.
\label{eq:dfc_numerical_extrapolation_methods}
\end{align}
Here $A(E)$ is given by equation~(\ref{eq:AE}). Method I amounts to imposing the constant values at $R_\mathrm{sim,l}$, whereas method II adopts the NRR diffusion coefficients in the limit $R\rightarrow0$. The latter coefficients are shown with dotted lines in the first panel of Figure \ref{fig:steady_state_numerical} in the range $R_\mathrm{loss} < R < R_\mathrm{sim,l}$. 

In the middle panel of Figure \ref{fig:steady_state_numerical} we show the steady-state solution $N(R)$ for a single semimajor axis range, computed for the cases $C=0$ (red lines) and $C\neq0$ (green lines). In both cases we employ the two different extrapolation methods in the regime $R_\mathrm{loss} < R < R_\mathrm{sim,l}$ (light color: method I; dark color: method II). Dashed (solid) lines correspond to $s=0.05 \, s_0$ ($s=0.5\,s_0$). The corresponding eccentricity distribution is shown in the bottom panel of Figure \ref{fig:steady_state_numerical}. While qualitatively similar to the corresponding plots in Figure~\ref{fig:steady_state_analytic}, there are important differences. Most notably, the peak in $N(R)$ below the SB is much less dominant compared to the peak above the SB. 

We show similar results for different semimajor axis bins in Figures \ref{fig:steady_state_sma_R} (in terms of $R$) and \ref{fig:steady_state_sma_e} (in terms of $e$). In these plots, at low $R$, $R_\mathrm{loss} < R < R_\mathrm{sim,l}$, the solutions with $C=0$ and $C\neq0$ deviate from each other. However, for $R\gtrsim R_\mathrm{sim,l} \sim 10^{-2}$ the solutions are indistinguishable for the same smoothness $s$. Likewise, the choice of extrapolation in the regime $R_\mathrm{loss} < R < R_\mathrm{sim,l}$ (i.e. method I or II) does not noticably affect the solution for $R\gtrsim R_\mathrm{sim,l}$. The latter value of $R$ corresponds to a very high eccentricity, $e \sim 0.995$. Consequently, the eccentricity distributions (cf. Figure \ref{fig:steady_state_sma_e}) are visually unaffected by the flux constraints nor by the choice of extrapolation. The only parameter that does noticeably affect the solutions, is the interpolation smoothness parameter $s$ (i.e. compare the solid and dashed lines). Nevertheless, the solutions are qualitatively similar for both values of $s$. 

Except for $4.6 < a/\mathrm{mpc} < 6.2$, the SB is present in the semimajor axis bins shown in Figures \ref{fig:steady_state_sma_R} and \ref{fig:steady_state_sma_e}, i.e. $a>C_\mathrm{SB}$. Here $C_\mathrm{SB} = [r_g a_\mathrm{max}^{1/2} (M_\bullet/m_\star) (1/N_\mathrm{max}^{1/2}) ]^{3/2} \approx 6.25 \, \mathrm{mpc}$ is the smallest value of $a$ for which the SB exists (MAMW11). Although the solutions for these larger $a$ values do depend somewhat on the degree of smoothing, two local maxima and three local minima can always be observed in the distributions in $R$. These extrema have the following locations: 
\begin{enumerate}
\item A minimum near $R\approx 1$.
\item A maximum near $R_\mathrm{c}$, which we determine from the value of $R$ for which the (interpolated) $\langle \Delta \ell \rangle$ changes sign (blue vertical dot-dashed line).
\item A minimum near or slightly above $R_\mathrm{SB}$.
\item A maximum between $R_{\mathrm{a},2}$ (black dashed line) and $R_\mathrm{SB}$. 
\item A minimum near $R_\mathrm{loss}$ (red solid line).
\end{enumerate}
These locations are generally consistent with those found using our analytic expressions for the diffusion coefficients (cf. \S\,\ref{sect:steady-state:analytic}). An exception is the maximum between $R_{\mathrm{a},2}$ and $R_\mathrm{SB}$ which, according to the analytic solutions, should occur near $R_{\mathrm{a},2}$. In the solutions based on the interpolations, this maximum occurs at a somewhat larger value of $R$. 

For the smallest values of semimajor axis, shown in the top left panel of Figures \ref{fig:steady_state_sma_R} and \ref{fig:steady_state_sma_e}, the SB does not exist, i.e. $a<C_\mathrm{SB}$. It is not surprising that the steady-state solutions at these small radii are systematically different compared to those farther out. In this regime, orbits at all $R$ are strongly affected by GR precession and RR is not effective at any $R$. Our analytic prescription of the diffusion coefficients breaks down in this regime, as illustrated by the first panel of Figure \ref{fig:diffusion_coefficients_S_stars}, where the measured diffusion coefficients are systematically lower than our predictions. A detailed description of diffusion in this regime is beyond the scope of this paper. Nevertheless, the numerical solutions indicate that the local minimum that was observed near $R_\mathrm{SB}$ for larger semimajor axes, disappears. This is not surprising, considering that the SB does not exist in this radial range. The maximum near $R_\mathrm{c}$, on the other hand, becomes more pronounced.

\section{Discussion}
\label{sect:discussion}
\subsection{Limits on the typical S-star age from the $N$-body simulations}
\label{sect:discussion:S_star_implications}
In our simulations of the S-stars, we assumed that their orbits about the SBH were initially very eccentric, $0.93 < e_0 < 0.99$. We  considered two possibilities for the nature of the rate of supply of S-stars to the GC: formation in a burst or continuous formation. The former assumption is consistent with the infall of a young stellar cluster, possibly with a central intermediate mass black hole, into the GC that subsequently dissolves and leaves massive stars tightly bound to the SBH \citep{hm03,berukoffhansen06,fujii10}. There are numerous problems with this scenario, however (see e.g. \citealt{peretsgual10}). An alternative possibility is binary disruption, in which case the rate of supply of S-stars to the GC is expected to be continuous, if averaged over a sufficiently long time. Massive perturbers like giant molecular clouds (GMCs) are a promising candidate for strongly perturbing the orbits of stellar binaries outside the central parsec into loss-cone orbits at a rate that is high enough to account for the current number of S-stars and high-velocity stars \citep{perets07,peretsgual10}. 

In the case of burst formation we have found in our simulations that the cumulative eccentricity distribution rapidly evolves to a distribution that is consistent with observations in $\sim 7 \, \mathrm{Myr}$ (cf. Figure \ref{fig:S_stars_p_evolution}). These results have also been extrapolated to include continuous formation and we have found that in this case the minimum time to evolve to the observed distribution is $\sim 25 \, \mathrm{Myr}$ (cf. Figure \ref{fig:S_stars_p_evolution_continuous}). If our assumptions of the formation process (high initial eccentricities) and the field star distribution (a cusp of stellar black holes) are correct, then the consistency of the eccentricity distribution with observations after a certain time implies a lower limit on the typical S-star lifetime and hence an upper limit on the typical S-star mass. We emphasize that only conclusions can be drawn for the typical age, because $p(t)$ applies to the S-stars as a whole population. Assuming solar metallicity the lower limit of the typical age of $\sim 7 \, \mathrm{Myr}$ in the burst scenario corresponds to an upper limit of the typical mass of $\sim 24 \, \mathrm{M_\odot}$. The lower limit of the typical age of $\sim 25 \, \mathrm{Myr}$ in case of continuous S-star formation corresponds to an upper limit of the typical mass of $\sim 10 \, \mathrm{M_\odot}$. The observed spectral types of the S-stars range from B0 V to B9 V \citep{eis05}, or $3 \lesssim m/\mathrm{M}_\odot \lesssim 20$. Furthermore, the initial mass function (IMF) of the S-stars is consistent with a Salpeter IMF, $\mathrm{d}N/\mathrm{d}m \propto m^{-2.15\pm0.3}$ \citep{bartko10}, which implies a mean mass of $\langle m \rangle = (6.4 \pm 0.5) \, \mathrm{M}_\odot$. The latter mass is consistent with our upper limits of the typical mass for both burst and continuous formation. 

\subsection{S-star relaxation times for different field star models}
\label{sect:discussion:S_star_relaxation_times}
In the $N$-body simulations of the S-stars a cusp of stellar black holes was assumed. Although predicted by theory \citep{bahcallwolf76}, so far no direct evidence for the presence of such a cusp in the GC has been found. Observations of late-type stars in the GC \citep{bse09,do_ea09,bartko10} indicate that there is a core of size $\sim0.5\,\mathrm{pc}$ in the distribution of these stars, which is well outside the radial extent of the S-star cluster. Such a core can be represented by a density slope $\gamma=1/2$ \citep{merritt10}, which is the lowest possible value consistent with an isotropic velocity distribution.

In order to estimate the effect of a core of late-type stars on the typical time scale for the orbits of the S-stars to evolve to eccentricities consistent with observations (as opposed to a cusp of stellar black holes), we applied equation~(\ref{eq:delta_t_AR}) using a similar method as in \S\,\ref{sect:S-stars:dfc}. Here we adopted $\gamma=1/2$, $N_\mathrm{max} = 8\times10^4$ and $m_\star=1.0\,\mathrm{M_\odot}$ from the stellar core model that was assumed in AM13. 

For a range of semimajor axes an initial value of $\ell$ consistent with binary disruption ($0.93 < e_0 < 0.99$), $\ell_0$, was sampled in $10^4$ Monte-Carlo realizations. In each of these a value $\ell_p>\ell_0$ was sampled from the cumulative distribution $\mathrm{CDF}(\ell) = 1-(1-\ell^2)^{p/2}$ which corresponds to a cumulative eccentricity distribution $\mathrm{CDF}(e) = e^p$, where $p=2.6$ was adopted to be consistent with observations \citep{gil09}. The time scale for $\ell$ to increase from $\ell_0$ to $\ell_p$ was then computed as follows. For $\ell\leq\ell_\mathrm{SB}$ equation~(\ref{eq:delta_t_AR}) was applied assuming $C_\mathrm{A_D} = 0.5$ and $C_1 = 2.6$; the mass precession time scale $\langle t_\mathrm{MP}\rangle$ was approximated by $\langle t_\mathrm{MP} \rangle \approx (1/1.2) [M_\bullet/M_\star(a)] P(a)$ \citep[][4.4.1]{bookmerritt13}. For $\ell>\ell_\mathrm{SB}$ the estimate $\Delta t_\mathrm{RR} = t_{\mathrm{RR},0} \Delta \ell^2$ was applied, where $t_{\mathrm{RR},0} = [M_\bullet/M_\star(a)] N_\star(a) P(a)^2/t_\mathrm{coh}(a)$. In the latter estimate the dependence of the RR time scale on $\ell$ was neglected for simplicity (cf. equation~(\ref{eq:dfc_lRR2})). Subsequently, by averaging over the Monte-Carlo realizations we obtained $\langle \Delta t(\ell_0\rightarrow \ell_p) \rangle$, the approximate time scale for $\ell$ to increase from a value consistent with a highly eccentric orbit (e.g. as a result of binary disruption) to a value consistent with the ``super-thermal'' eccentricities of the S-stars. In this method the semimajor axes were assumed to be constant during the relaxation process. 

\begin{figure}
\center
\includegraphics[scale = 0.435, trim = 0mm 0mm 0mm 0mm]{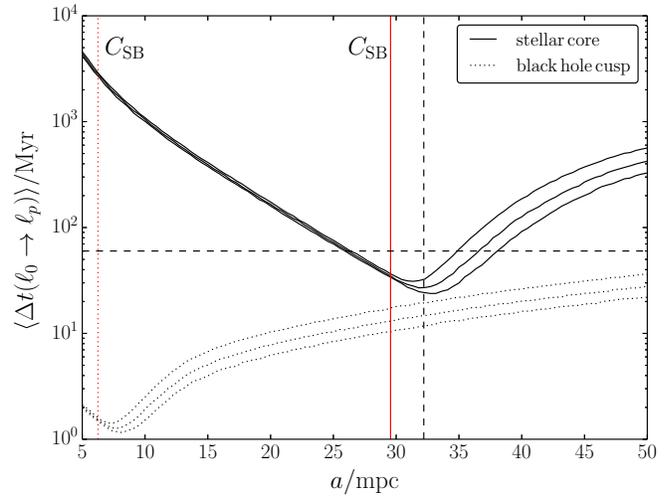}
\caption{\small Evolution time scales for the eccentricity of the S-stars to reach a value consistent with observations $(\ell=\ell_p$) starting from high eccentricities consistent with binary disruption, computed using Monte-Carlo realizations of equations (\ref{eq:dfc_lRR2}) and (\ref{eq:delta_t_AR}). Black solid lines: asssuming a stellar core model; black dotted lines: assuming a black hole cusp model (see text). For each model the time scales were computed for three different values of $p$ to take into account the uncertainty of the observed value of $p_\mathrm{obs}=2.6\pm0.9$ \citep{gil09}: $p=2.6-0.9$, $p=2.6$ and $p=2.6+0.9$. The red solid and dotted lines indicate the value of $C_\mathrm{SB}$ for the stellar core and black hole cusp models, respectively, where $C_\mathrm{SB}$ is the minimum semimajor axis at which the SB is predicted to exist (cf. \S\,\ref{sect:steady-state:numerical}). The black vertical dashed line indicates $32.2 \, \mathrm{mpc}$, approximately the outer extent of the S-star cluster. The black horizontal dashed line indicates $60 \, \mathrm{Myr}$, an estimate of the mean age of the S-stars. }
\label{fig:evolution_times_stellar_core_and_bh_cusp}
\end{figure}

The resulting time scales are plotted as function of semimajor axis in Figure \ref{fig:evolution_times_stellar_core_and_bh_cusp} (solid lines). In that figure we also included similar calculations for a cusp of stellar black holes as was assumed in \S\,\ref{sect:S-stars} (dotted lines). For $a<C_\mathrm{SB}$, where $C_\mathrm{SB}$ is the smallest value of $a$ for which the SB exists (MAMW11), the time scale decreases with increasing $a$. This can be understood from equation~(\ref{eq:delta_t_AR}): neglecting $\ell_\mathrm{SB}$ in that equation it can be shown that $\Delta t \propto a^{\gamma-5/2}$ and $\propto a^{2\gamma-13/2}$ assuming that orbital precession is dominated by relativity and mass precession, respectively. Therefore the time scales decrease with $a$ for both $\gamma=1/2$ and $\gamma=2$. Conversely, for $a>C_\mathrm{SB}$ the time scale increases with increasing $a$, which can be understood from the scaling of the RR time scale with $a$ (cf. equation~(\ref{eq:dfc_lRR2})): assuming mass precession the scaling is $\Delta t \propto a^{3/2}$, independent of $\gamma$. Note that in the stellar core model most of the orbits of the S-stars lie below the SB for any value of $\ell$ (cf. the left panel of Fig. 1 of AM13). 

For the stellar core model the evolution time scale exceeds 60 Myr, an estimate of the mean S-star life time assuming a mean mass of $\sim 6 \, \mathrm{M_\odot}$ \citep{eis05}, for a large range of semimajor axes. This is consistent with the result of AM13 that the eccentricity distribution of the S-stars cannot evolve to the observed distribution over the life time of the S-stars in the case of a stellar core (cf. the top left panel of Fig. 3 of AM13). The long evolution time scales in the case of a stellar core would suggest that either (1) the assumption of the formation mechanism of the S-stars (i.e. high initial eccentricities) is incorrect, or that (2) a stellar core of is not the dominant cause of relaxation of the S-stars. Interestingly, the evolution time scales are consistent with (i.e. shorter than) the ages of the S-stars when assuming a cusp of stellar black holes (cf. the dotted lines in Figure \ref{fig:evolution_times_stellar_core_and_bh_cusp}).

\begin{figure*}
\center
\includegraphics[scale = 0.455, trim = 0mm 0mm 0mm 0mm]{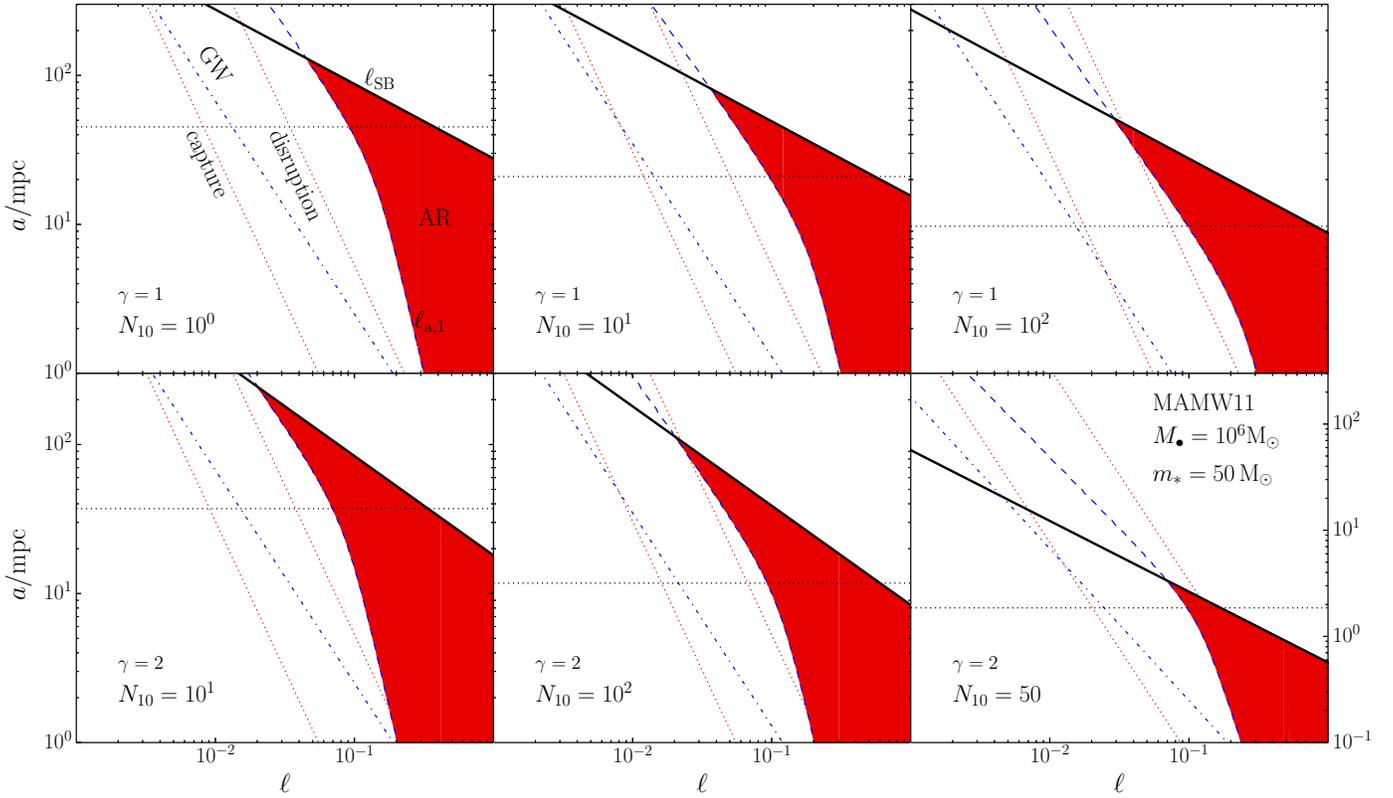}
\caption{\small Several quantities of importance for AR shown in the $(a,\ell)$-plane for nuclear models with different $\gamma$ and $N_\mathrm{10}$, the number of stars within 10 mpc. Black solid line: the SB (equation~(\ref{eq:ell_SB})). Blue dashed lines: $\ell_{\mathrm{a},1}$ (equation~(\ref{eq:labn})). The black horizontal dotted line shows an estimate of the transition between GR and mass precession, $a_\mathrm{trans}$ (equation~(\ref{eq:a_trans})). Two red dotted lines: radii for capture of compact objects (left) and tidal disruption of stars (right) by the SBH. Blue dot-dashed line: an estimate of where changes in orbital eccentricity due to gravitational-wave enery loss occur at the same rate as changes due to two-body relaxation (MAMW11, Eq. 62). We have indicated with red shading the region in which we expect AR to dominate angular momentum relaxation. In all panels $M_\bullet = 4 \times 10^6 \, \mathrm{M}_\odot$ and $m_\star = 10 \, \mathrm{M}_\odot$ is assumed, with the exception of the bottom right panel, where we adopted the model of MAMW11: $M_\bullet = 10^6 \, \mathrm{M}_\odot$, $m_\star = 50 \, \mathrm{M}_\odot$, $\gamma=2$ and $N_{10} = 50$. Note that there is a different range of the vertical axis in the latter panel. } 
\label{fig:discussion_implications}
\end{figure*}

\subsection{Generalizations to other galactic nuclei}
\label{sect:discussion:gen}
In \S\,\ref{sect:S-stars:dfc} we presented analytic expressions for the diffusion coefficients which were calibrated using $N$-body simulations with assumed parameters $\rho_\star(r) \propto r^{-2}$, $m_\star = 10 \, \mathrm{M}_\odot$ and $N_\mathrm{max} = 4800$. It is of interest to investigate whether these relations also apply to nuclear star clusters with different properties. We have also carried out a set of simulations with $\rho_\star(r) \propto r^{-1}$, $m_\star = 10 \, \mathrm{M}_\odot$ and $N_\mathrm{max} = 2500$. The results of the latter simulations, i.e. with $\gamma=1$, are presented in Appendix \ref{app:Nbody_gamma_1}. The main conclusions that can be drawn from these additional simulations is that the features in the diffusion coefficients that are associated with AR and that were observed in the simulations with $\gamma=2$, are also present in the simulations with $\gamma=1$. In particular, the analytic approximation of the diffusion coefficients that was presented in equation~(\ref{eq:dif_coef_approx}) also describes the data well for $\gamma=1$. There is an exception for larger semimajor axes, for which it appears that $C_1$ and $C_2$ (cf. equation~(\ref{eq:dfc_AR})) increase with semimajor axis.

These results indicate that it is justified to extrapolate the relations presented in \S\,\ref{sect:S-stars:dfc} to nuclear star clusters with different properties. We adopt the reference values $M_\bullet = 4.0 \times 10^6 \, \mathrm{M}_\odot$ and $m_\star = 10 \, \mathrm{M}_\odot$, hence $\log(\Lambda) = \log[M_\bullet/(2m_\star)] \approx 12.2$. We consider two values of $\gamma$, $\gamma=1$ and $\gamma=2$, for which $C_\mathrm{NRR}(1) \approx 0.07$ and $C_\mathrm{NRR}(2) \approx 0.18$ (cf. equation~(\ref{eq:AE}) and Appendix \ref{app:C_gamma}). Furthermore, for both values of $\gamma$ we adopt $C_{A_\mathrm{D}} = 0.5$ (cf. Figure \ref{fig:e_amplitude_S_stars}) and $C_1=C_2=2.6$ (cf. Figure \ref{fig:diffusion_coefficients_S_stars}). For the coherence time $t_\mathrm{coh}$ we assume $t_\mathrm{coh}^{-1} = \langle t_\mathrm{GR} \rangle^{-1} + \langle t_\mathrm{MP} \rangle^{-1}$ as before, with $\langle t_\mathrm{GR} \rangle = (1/12) (a/r_g) P(a)$ and $\langle t_\mathrm{MP} \rangle = C_\mathrm{MP}(\gamma) [M_\bullet/M_\star(a)] P(a)$, where $C_\mathrm{MP}(1) = 1$ and $C_\mathrm{MP}(2) = 3/2$ \citep[][4.4.1]{bookmerritt13}. 

We show the main relations in the $(a,\ell)$-plane in Figure \ref{fig:discussion_implications}. Models are included with the two values of $\gamma$ and various values of $N_{10}$, the number of stars within 10 mpc. The SB (equation~(\ref{eq:ell_SB})) is shown with the black solid line. We show two relevant periapsis distances that are associated with losses to the SBH: the stellar tidal disruption radius, $r_\mathrm{dis} = 2 R \, (M_\bullet/m)^{1/3} \approx 2.7 \times 10^{-3} \, \mathrm{mpc}$ (assuming $R = 8 \, \mathrm{R}_\odot$ and $m=10\, \mathrm{M}_\odot$) \citep{antlomb11} and the radius for capture of compact remnants, $r_\mathrm{capt} = 8 \, r_g \approx 1.5 \times 10^{-3} \, \mathrm{mpc}$ \citep{will12}. 

In \S\,\ref{sect:S-stars:dfc} an expression was presented for $\ell_{\mathrm{a},n}$, the lower boundary in $\ell$ for which we expect AR to dominate diffusion in angular momentum. Note that $\ell_{\mathrm{a},2} = 2^{1/4} \ell_{\mathrm{a},1}$ if $C_1=C_2$, which we find is the case in our $N$-body simulations and which we adopt here. If precession of the field star orbits is dominated by GR precession, then $\ell_{\mathrm{a},n}$ can be written as:
\begin{subequations}\label{eq:lan_phys_GR}
\begin{align}
\ell_{\mathrm{a},n} &= n^{1/4} \tilde{C}_\mathrm{N}(\gamma) \left ( \frac{1}{12} \frac{r_g}{a} \right )^{1/4} \\
&\approx 0.11 \, n^{1/4} \left ( \frac{M_\bullet}{4 \times 10^6 \, \mathrm{M}_\odot} \right )^{1/4} \left ( \frac{a}{10 \, \mathrm{mpc}} \right )^{-1/4},
\end{align}
\end{subequations}
the latter assuming $\gamma=2$. Here we defined
\begin{align}
\tilde{C}_\mathrm{N} = \tilde{C}_\mathrm{N}(\gamma) = \left [ \frac{\log(\Lambda)}{C_\mathrm{NRR}(\gamma) C_n C^2_{A_\mathrm{D}}} \right ]^{1/4}.
\label{eq:Ctilde}
\end{align}

On the other hand, if $t_\mathrm{MP} \ll t_\mathrm{GR}$, then precession is dominated by mass precession. In this case:
\begin{subequations}\label{eq:lan_phys_MP}
\begin{align}
\ell_{\mathrm{a},n} &= n^{1/4} \tilde{C}_\mathrm{N}(\gamma) \, C^{1/4}_\mathrm{MP}(\gamma) \left ( \frac{r_g}{a} \right )^{1/2} \left (\frac{M_\bullet}{m_\star} \right )^{1/4} N_\star(a)^{-1/4} \\
\nonumber &\approx 0.12 \, n^{1/4} \left ( \frac{M_\bullet}{4 \times 10^6 \, \mathrm{M}_\odot} \right )^{3/4} \left ( \frac{m_\star}{10 \, \mathrm{M}_\odot} \right )^{-1/4}  \\
&\quad \times \left ( \frac{N_\star(a)}{10^2} \right )^{-1/4} \left ( \frac{a}{10 \, \mathrm{mpc}} \right )^{-1/2},
\end{align}
\end{subequations}
the latter assuming $\gamma=2$.

The transition between the two regimes of field star precession occurs near $a_\mathrm{trans}$, which we define as the value of $a$ for which $\langle t_\mathrm{GR}\rangle  = \langle t_\mathrm{MP} \rangle$. With our assumptions, $a_\mathrm{trans}$ is given by:
\begin{subequations}\label{eq:a_trans}
\begin{align}
a_\mathrm{trans} &= \left [ 12 \, C_\mathrm{MP}(\gamma) \, r_g a_\mathrm{max}^{3-\gamma} N_\mathrm{max}^{-1}  \left ( \frac{M_\bullet}{m_\star} \right ) \right]^{1/(4-\gamma)} \\
&\approx 11.7 \, \mathrm{mpc} \left ( \frac{M_\bullet}{4 \times 10^6 \, \mathrm{M}_\odot} \right ) \left ( \frac{m_\star}{10 \, \mathrm{M}_\odot} \right )^{-1/2} \left ( \frac{N_{10}}{10^2} \right )^{-1/2},
\end{align}
\end{subequations}
the latter assuming $\gamma=2$. In Figure \ref{fig:discussion_implications} $a_\mathrm{trans}$ is indicated with the horizontal black dotted line. Furthermore we show in that figure $\ell_{\mathrm{a},1}$ (blue dashed line), with the coherence time computed from $t_\mathrm{coh}^{-1} = \langle t_\mathrm{GR} \rangle^{-1} + \langle t_\mathrm{MP} \rangle^{-1}$. 

In Figure \ref{fig:discussion_implications} we have indicated with red shading the approximate region in which we expect that AR dominates evolution in angular momentum. In general, AR is expected to be important in the region $\ell_{\mathrm{a},1} \lesssim \ell \lesssim \ell_\mathrm{SB}$. It can be seen in Figure \ref{fig:discussion_implications} that there is a critical value of $a$, $a_\mathrm{ AR,max}$, where $\ell_{\mathrm{a},1} = \ell_\mathrm{SB}$. For $a>a_\mathrm{ AR,max}$, $\ell_{\mathrm{a},1} > \ell_\mathrm{SB}$, and we expect AR not to be active at any $\ell$. Instead, we expect that NRR dominates angular momentum relaxation below the SB and that RR dominates above the SB. Hence we expect that the AR regime disappears for $a>a_\mathrm{ AR,max}$. 

As shown in Figure \ref{fig:discussion_implications} the value of $a_\mathrm{ AR,max}$ is large for the nuclear models considered here. Assuming that near $a_\mathrm{ AR,max}$ precession is dominated by mass precession, which is borne out by Figure \ref{fig:discussion_implications}, and combining equations~(\ref{eq:ell_SB}) and (\ref{eq:labn}), we find:
\begin{subequations}\label{eq:AR_crit}
\begin{align}
a_\mathrm{ AR,max} &\approx \left [ \tilde{C}^{-1}_\mathrm{N}(\gamma) C^{-\frac{1}{4}}_\mathrm{MP}(\gamma) r_g^\frac{1}{2} a_\mathrm{max}^\frac{3-\gamma}{4} N_\mathrm{max}^{-\frac{1}{4}} \left ( \frac{M_\bullet}{m_\star} \right )^\frac{3}{4} \right ]^\frac{4}{5-\gamma} \\
\nonumber &\approx 114 \, \mathrm{mpc} \left ( \frac{M_\bullet}{4 \times 10^6 \, \mathrm{M}_\odot} \right )^{5/3} \left ( \frac{m_\star}{10 \, \mathrm{M}_\odot} \right )^{-1} \\
&\quad \times \left ( \frac{N_{10}}{10^2} \right )^{-1/3},
\end{align}
\end{subequations}
the latter assuming $\gamma=2$. Unless $N_{10}$ is very large, $N_{10} \gtrsim 10^3$, $a_\mathrm{ AR,max} \sim 10^2 \, \mathrm{mpc}$ is large compared to other values of $a$ of interest in Figure \ref{fig:discussion_implications}. This shows that for many models of galactic nuclei there is a large regime in the energy and angular momentum space in which AR is important. We note that the quantity $a_\mathrm{ AR,max}$ derived above is the same as another critical semimajor axis that was defined in \S\,VC of MAMW11. The latter quantity was argued to be the minimum value of $a$ for which NRR would allow orbits to ``penetrate'' the SB. The equivalence of these two quantities is shown explicitly in Appendix \ref{app:aARcrit}. 

For comparison purposes have also included in the bottom right panel of Figure \ref{fig:discussion_implications} the $N$-body model that was adopted in MAMW11. In that model, $M_\bullet = 10^6 \, \mathrm{M}_\odot$, $m_\star = 50 \, \mathrm{M}_\odot$ and $N_{10} = 50$. Semimajor axes were sampled from a distribution consistent with $\gamma=2$ with $0.1<a/\mathrm{mpc} < 10$. Equations~(\ref{eq:a_trans}) and (\ref{eq:AR_crit}) applied to this model give $a_\mathrm{trans} \approx 1.9 \, \mathrm{mpc}$ and $a_\mathrm{ AR,max} \approx 3.1 \, \mathrm{mpc}$. Unlike the other models considered above, in the MAMW11 model $a_\mathrm{trans}$ and $a_\mathrm{ AR,max}$ are comparable, implying that in the AR regime field star precession is driven mainly by relativistic precession. Moreover, in the latter model, $a_\mathrm{AR,max} \approx 2-3$ mpc, while the stellar orbits had 0.1 mpc $\leq a \leq $ 10 mpc . It follows that for the stars with the larger $a$-values in MAMW11, NRR was the dominant diffusion mechanism acting on stars after they had crossed the SB; only for $a\lesssim 3$ mpc was AR effective. Indeed it was shown in that paper that essentially all of the stars that were captured by the SBH had $a\gtrsim 2$ mpc, and it was argued that ``penetration'' of the SB was probably driven by NRR for these stars.

The trend seen in the panels in Figure \ref{fig:diffusion_coefficients} with different $N_\mathrm{max}$ can similarly be explained by the scaling of $a_\mathrm{ AR,max}$ with $N_\mathrm{max}$: as $N_\mathrm{max}$ increases, $a_\mathrm{ AR,max}$ increases, thereby increasing the importance of AR. More quantitatively, for fixed stellar mass $M_\star(a) = m_\star N_\star(a)$, as was assumed in Figure \ref{fig:diffusion_coefficients}, equation~(\ref{eq:AR_crit}) implies $a_\mathrm{ AR,max} \propto N_\star(a)^{2/3}$. The values of $a_\mathrm{ AR,max}$ in the models shown in the different panels in Figure \ref{fig:diffusion_coefficients} are $\approx 3.1, 4.8,7.5$ and $11.7 \, \mathrm{mpc}$ for $N_\mathrm{max} = 50, 100, 200$ and $400$, respectively. In the latter model $a_\mathrm{ AR,max} > a_\mathrm{max} = 10 \, \mathrm{mpc}$. 

\subsection{Caveats of \textsc{TPI}}
\label{sect:discussion:TPIcaveats}
The code presented in \S\,\ref{sect:methods} has the advantage of linear scaling with the number of field stars (for a fixed number of test stars), enabling simulations with much larger numbers of stars ($\gtrsim 10^3$) than are currently feasible using fully general  $N$-body codes. The disadvantage is that the motion of the field stars is not reproduced precisely. By allowing the field star orbits to precess, we do reproduce in an approximate way the dynamical effects of the smoothly-distributed field-star mass (``mass precession'') and of the 1PN relativistic corrections (``Schwarzschild precession''). But the \textsc{TPI} algorithm does not reproduce either (i) interactions between field stars due to discreteness of the mass distribution, or (ii) the dynamical influence of the test stars on the field stars. 

In the time- and spatial domains of interest here, discrete interactions between field stars can change both the magnitude and the direction of the field-star ${\boldsymbol L}$-vectors (changes in energy occur on longer time scales). Changes in the magnitude of ${\boldsymbol L}$, i.e. in orbital eccentricity, would cause $N(e)$ for the field stars to evolve with a characteristic time $\sim t_\mathrm{RR}$ toward some steady-state distribution. Insofar as the steady-state $N(e)$ which we infer for the {\it test} stars is not hugely different from a ``thermal'' distribution -- the same distribution which we {\it assumed} for the field stars -- we do not expect this evolution to be of much consequence for any of our results. Changes in the {\it direction} of the field-star ${\boldsymbol L}$ vectors on the other hand, constitute an additional form of precession and as such would play a role in determining the coherence time -- which we recall is defined as the time for a typical (field) star orbit to precess and so is a function only of $r$ or $a$. Changes in orbital orientation due to $\sqrt{N}$ torques occur on the coherent RR time scale, $t_\mathrm{RR,coh} \approx [M_\bullet/M_\star(a)] N_\star(a)^{1/2} P(a)$ \citep[][p. 275]{bookmerritt13}; comparing this time scale to the mass precession time scale (cf. equation~(\ref{eq:MP})) one finds $t_\mathrm{MP}/t_\mathrm{RR,coh} \propto N_\star(a)^{-1/2}$. Therefore, for sufficiently large $N_\star(a)$, precession of orbital planes can be neglected compared to mass precession. The consistency between the different codes in \S\,\ref{sect:belowSB} suggests that $N_\mathrm{max} = 50$ is already sufficiently large for this to be the case.

With regard to (ii), i.e., neglect of test star - field star perturbations, the consequences are less certain. Discussions in the literature of RR almost always ignore the dynamical influence of the test star on the stars producing the $\sqrt{N}$ torques. In the limit of small test star mass, that influence tends to zero, and so a test particle code like \textsc{TPI} is correct.

\subsection{Location of the sign change of $\langle \Delta \ell \rangle$ at high $\ell$}
\label{sect:discussion:sign_change}
\begin{figure}
\center
\includegraphics[scale = 0.435, trim = 0mm 0mm 0mm 0mm]{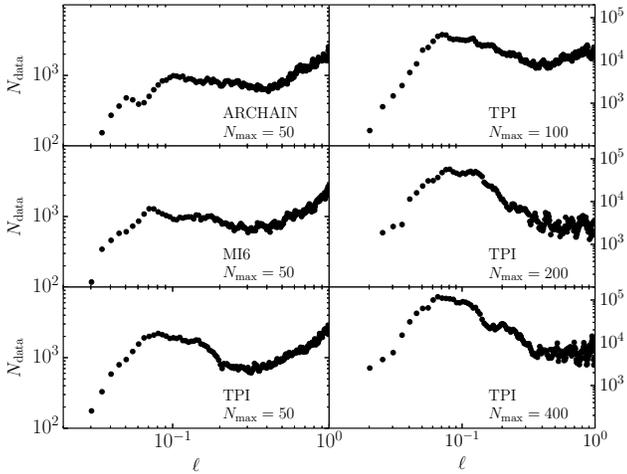}
\caption{\small The number of data points $N_\mathrm{data}$ in each bin of $\ell$ and the time lag bin that was adopted in this paper, i.e. corresponding to the coherence time (cf. Figure \ref{fig:diffusion_coefficients}). Left column: simulations with \textsc{ARCHAIN}, \textsc{MI6} and \textsc{TPI} with $N_\mathrm{max}=50$. Right column: simulations with \textsc{TPI} with $N_\mathrm{max}=100,200$ and $400$. }
\label{fig:diffusion_coefficients_data_length}
\end{figure}

In the simulations presented in \S\,\ref{sect:belowSB} and \S\,\ref{sect:S-stars} the first-order diffusion coefficient $\langle \Delta \ell \rangle$ was found to change sign from positive to negative values as $\ell$ increases towards unity. This is to be expected, considering that $\ell$ cannot exceed unity. The value of $\ell$ where this sign change occurs, $\ell_\mathrm{c}$, is well-defined in the simulations with $N_\mathrm{max} = 50$, but becomes less well-defined as $N_\mathrm{max}$ increases (cf. Figure \ref{fig:diffusion_coefficients}). In the S-star simulations, where $N_\mathrm{max}$ is greater still, $\ell_\mathrm{c}$ is likewise not well-defined. A possible explanation for this trend with $N_\mathrm{max}$ is that for low $N_\mathrm{max}$ the number of data points in each bin at high $\ell$ is larger compared to this number at low $\ell$, whereas for larger $N_\mathrm{max}$, the relative number of bins at high $\ell$ decreases. This is demonstrated in Figure \ref{fig:diffusion_coefficients_data_length}, where the number of data points in each bin of $\ell$ is plotted for the simulations of \S\,\ref{sect:belowSB}. The trend of the number of data points with $N_\mathrm{max}$ can be explained by the increase of the RR time scale as $N_\mathrm{max}$ increases (cf. equation~(\ref{eq:dfc_lRR2}); note that here $M_\star$ is kept fixed): as $N_\mathrm{max}$ increases, the relative time spent at high $\ell$ in the simulations decreases, thereby decreasing the number of available data points. 

\begin{figure}
\center
\includegraphics[scale = 0.435, trim = 0mm 0mm 0mm 0mm]{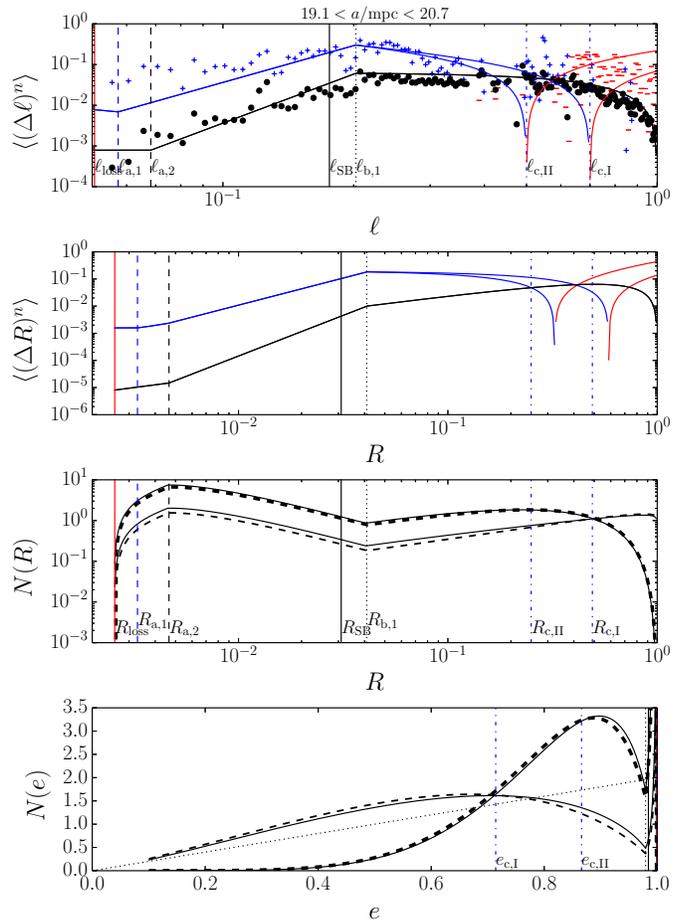}
\caption{\small Steady-state solutions of the Fokker-Planck equation in angular momentum space based on our analytic functions of the diffusion coefficients (cf. equation~(\ref{eq:dfc_analytic_modified})) similar to Figure \ref{fig:steady_state_analytic}, but now assuming two values of $\ell_\mathrm{c}$: $\ell_\mathrm{c,I}=0.7$ (as before; thin black lines) and $\ell_\mathrm{c,II} = 0.5$ (thick black lines), and for a larger semimajor axis bin. } 
\label{fig:steady_state_analytic_different_l_c}
\end{figure}

The uncertainty of $\ell_\mathrm{c}$ in simulations with large $N_\mathrm{max}$ is a caveat for our approximate analytic functions of the angular momentum diffusion coefficients which depend on $\ell_\mathrm{c}$ (cf. equation \ref{eq:dfc_analytic_modified}), and therefore for the analytic steady-state solutions (cf. \S\,\ref{sect:steady-state:analytic}). To explore the implications of this uncertainty we show in Figure \ref{fig:steady_state_analytic_different_l_c} a figure similar to Figure \ref{fig:steady_state_analytic} for the steady-state solutions based on the analytic functions for the coefficients, but now also including a lower value of $\ell_\mathrm{c,II}=0.5$ and assuming the largest semimajor axis bin shown in Figure \ref{fig:diffusion_coefficients_S_stars}. The latter bin is associated with large uncertainty in $\ell_\mathrm{c}$ and we adopt $\ell_\mathrm{c,II}=0.5$ as an alternative value for $\ell_\mathrm{c}$ for this semimajor axis. From Figure \ref{fig:steady_state_analytic_different_l_c} we conclude that the steady-state solutions for small $\ell$ ($\ell \ll \ell_\mathrm{c}$) are not strongly affected by the uncertainty in $\ell_\mathrm{c}$. For larger $\ell$ the steady-state solution is sensitive to the value of $\ell_\mathrm{c}$, however. Nevertheless, our result that the maximum in the steady-state eccentricity distribution occurs near $e_\mathrm{c}\equiv \sqrt{1-\ell_\mathrm{c}^2}$ is robust (cf. the third panel of Figure \ref{fig:steady_state_analytic_different_l_c}).

\subsection{Comparison of steady-state solutions}
\label{sect:discussion:steady_state_comparison}
As mentioned in \S\,\ref{sect:steady-state}, MHL11 have previously investigated the effect of RR on the steady-state eccentricity distribution of stars near a SBH. MHL11 used a semi-empirical model and found a bimodal eccentricity distribution with two peaks at small semimajor axes (cf. Fig. 18 of MHL11). Although we have also found a bimodal distribution with two peaks, the positions of these peaks are quite different in our work (cf. Figure \ref{fig:steady_state_sma_e}). In MHL11 the lower peak occurs at $e\sim 0.2$, whereas in our work the lower peak occurs at much higher eccentricity, $e\sim 0.7-0.8$. Furthermore, in MHL11 the upper peak occurs at $e\sim 0.9$, whereas in our work the upper peak occurs at even higher eccentricity, $e\sim 0.98$. 

An important difference between our work and that of MHL11 is that in the latter general relativistic corrections in the equations of motion for the test stars were not taken into account, whereas these corrections were included here (cf. equation~(\ref{eq:a_pert})). This would suggest that these terms in the equations of motion tend to increase eccentricities in the steady-state distribution. 

\begin{figure}
\center
\includegraphics[scale = 0.435, trim = 0mm 0mm 0mm 0mm]{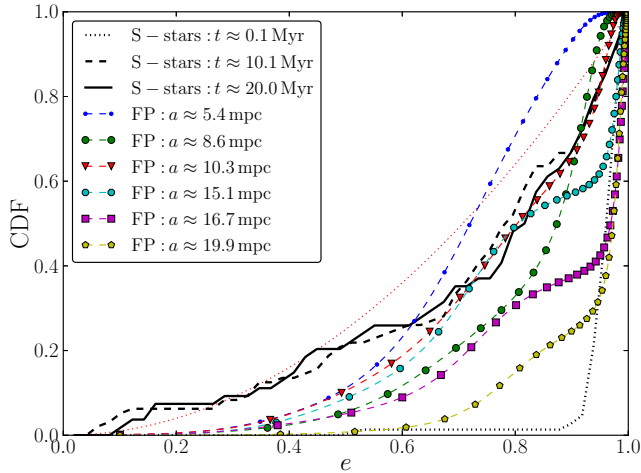}
\caption{\small Cumulative eccentricity distributions obtained directly from the S-star simulations at three different times (black dotted, dashed and solid lines for $t \approx 0, 10$ and $20 \, \mathrm{Myr}$, respectively), and according to the steady-state solutions of the Fokker-Planck equation ($C>0$; smoothness parameter $s = 0.5 \, s_0$ and method II, cf. \S\,\ref{sect:steady-state:numerical}). For the latter the same semimajor axis bins are shown as in Figure \ref{fig:steady_state_sma_e}. The red dotted line shows a ``thermal distribution''. }
\label{fig:fokker_planck_comparison_S}
\end{figure}

Finally, we briefly compare our distributions $N(e)$ obtained from solving the steady-state Fokker-Planck equation, equation~(\ref{eq:FPsteady_state_C}), to the eccentricity distributions that we obtained directly from the $N$-body simulations of the S-stars in \S\,\ref{sect:S-stars:orb_ev}. One expects that the former apply in the limit $t\rightarrow \infty$. The latter are limited by the simulation time, in our case $t < 20 \, \mathrm{Myr}$. We show both distributions in Figure \ref{fig:fokker_planck_comparison_S}. In the case of the steady-state Fokker-Planck solutions, the semimajor axis bins are shown that were included in Figures \ref{fig:steady_state_sma_R} and \ref{fig:steady_state_sma_e}; in the case of the distributions obtained directly from the $N$-body simulations, three times are shown. For semimajor axes that are comparable to the typical S-star semimajor axes, $a\sim 10 \, \mathrm{mpc}$, the Fokker-Planck solutions are consistent with the direct S-star distributions for $t \gtrsim 10 \, \mathrm{Myr}$ at high eccentricities, $e \gtrsim 0.7$. For smaller eccentricities the direct distributions are on average less eccentric than suggested by our steady-state solutions. The latter may be due to the following reasons. (1) The simulated time of 20 Myr is too short for low-eccentricity orbits to each a steady-state. This would be consistent with the RR diffusion time scale, which approaches $\sim 10^2 \, \mathrm{Myr}$ for $e \rightarrow 0$ in the assumed nuclear model (cf. Figure \ref{fig:introduction_figure}). (2) At low eccentricities the diffusion coefficients that were obtained from the simulations suffer from large scatter. In particular, large scatter is present in the first-order diffusion coefficients shown in Figure \ref{fig:diffusion_coefficients_S_stars} at high $\ell$, and this may produce bias in our results at low eccentricities.

\section{Conclusions}
\label{sect:conclusions}
We have presented a new $N$-body algorithm, \textsc{Test Particle Integrator} (\textsc{TPI}), that allows us to efficiently integrate orbits of test stars around a supermassive black hole (SBH) including post-Newtonian corrections to the equations of motion and interactions with a large ($\gtrsim 10^3$) number of field stars. We compared results obtained with this code to results obtained with two more accurate, but slower, $N$-body codes \textsc{ARCHAIN} and \textsc{MI6}; we focussed on the behavior of orbits above and below the ``Schwarzschild barrier'' (SB), the region in (energy, angular momentum) space where relativistic precession vitiates torques from $\sqrt{N}$ asymmetries (resonant relaxation; RR). In addition we have performed simulations of the Galactic center (GC) to test models for the origin of the S-stars; these simulations used 4800 field stars close to the SBH, a number that is not currently feasible with other $N$-body algorithms. Our main conclusions are as follows.

\medskip
\noindent 1. The behavior of test-particle orbits in \textsc{TPI} is consistent with what is found using the codes \textsc{ARCHAIN} and \textsc{MI6}, which do not make our simplifying assumptions.  

\medskip
\noindent 2. We analysed several aspects of eccentricity oscillations below the SB ($e>e_\mathrm{SB}$) that are associated with rapid GR precession in the presence of Newtonian torques from the field stars. Using power spectra of the eccentricity time series we found evidence for enhanced power at higher integer frequencies than the relativistic frequency $f_\mathrm{GR}$ (Figure \ref{fig:power_spectrum_all}). The peak at the latter frequency can be interpreted as implying that the torquing potential (due to the $\mathcal{O}(N^{-1/2})$ asymmetry in the field star distribution) is basically lopsided, or $m=1$, in character (MAMW11). Higher-order terms in the multipole expansion of the field star potential would give rise to eccentricity oscillations at higher integer frequencies of $f_\mathrm{GR}$. Our results indicate that these higher-order contributions are important, though typically not dominant.

In addition, we determined the amplitude of the eccentricity oscillations and we  verified the expected dependence $\Delta \ell \propto \langle \ell \rangle^2$, where $\ell\equiv L/L_c=\sqrt{1-e^2}$ is the dimensionless angular momentum, $\Delta \ell$ is the amplitude of angular momentum oscillations over a precessional cycle, and $\langle \ell \rangle$ is its average value. By fitting our data to the model of MAMW11 we also determined the fitting constant $C_{A_\mathrm{D}}$ that captures unspecified uncertainties in this model (Figures \ref{fig:e_amplitude} and \ref{fig:e_amplitude_S_stars}). 

\medskip
\noindent 3. We applied the \textsc{TPI} algorithm to the evolution of the S-stars in the GC, assuming that they were deposited initially onto orbits of very high eccentricity. This is expected for the tidal disruption of a stellar binary. We adopted a distribution of field stars that is consistent with the steady-state distribution of stellar remnants at the GC. Assuming formation of S-stars in a burst, we  found that their cumulative eccentricity distribution evolves to $N(e) \propto e^{2.6}$ on a time scale of $7 \pm 0.1 \, \mathrm{Myr}$. The latter distribution is consistent with observations. We also extrapolated our results to a continuous-formation model. Our results suggest a lower limit on the typical age of the S-stars of $\sim 7 \, \mathrm{Myr}$ in the case of burst formation and $\sim 25 \, \mathrm{Myr}$ in the case of continuous formation. 

\medskip
\noindent 4. From our simulations we extracted first- and second-order diffusion coefficients in the normalized angular momentum variable $\ell$. We identified three angular momentum regimes, in which the diffusion coefficients depend in functionally different ways on $\ell$. Regimes of lowest and highest $\ell$ are well described in terms of non-resonant relaxation (NRR) and resonant relaxation (RR), respectively. Near and below the SB, a third regime exists, ``anomalous relaxation'' (AR), which is not well described in terms of either NRR or RR. In this regime, the time scale for angular momentum diffusion increases rapidly with increasing eccentricity. We found that the features associated with the new AR regime are only clearly present in simulations with larger numbers of field stars than considered previously. We presented analytic expressions, in terms of physical parameters, that describe the diffusion coefficients in all three angular momentum regimes. 

\medskip
\noindent 5. We proposed a new, empirical criterion for the location of the barrier, based on the $L$- dependence of the diffusion coefficients. This criterion was found to predict essentially the same $\ell_\mathrm{SB}(a)$ relation as equation (1) which was derived in MAMW11 from simple timescale arguments. Our results also demonstrate the validity of that relation in systems that differ greatly in terms of particle number and mass. 

\medskip
\noindent 6. We derived a simple expression for the typical time scale of angular momentum diffusion in the ``anomalous'' (AR) regime (equation~(\ref{eq:delta_t_AR})) and verified its correctness by applying it to the $N$-body simulations (cf. Figure \ref{fig:S_stars_dif_time_scales_test}). We applied this relation assuming both a core of late-type stars and a cusp of stellar black holes in the GC, and confirmed the earlier result (AM13) that in the case of a core of late-type stars the time scales for the S-stars to reach the observed ``super-thermal'' distribution of eccentricities is much longer than the typical age of the S-stars (cf. Figure \ref{fig:evolution_times_stellar_core_and_bh_cusp}).

\medskip
\noindent 7. Using our expressions for the angular-momentum diffusion coefficients, we derived the steady-state distribution of orbital angular momenta implied by the Fokker-Planck equation for stars near a SBH. This distribution differs significantly from the distribution predicted by NRR, $f(E,L)\propto f(E)$. There is a deficit of orbits near the SB and an excess just above it (i.e. $e < e_\mathrm{SB}$). Furthermore, we found evidence for a local excess of orbits below the SB ($e>e_\mathrm{SB}$) in a steady state, which can be attributed to the slow nature of diffusion in the AR regime, causing orbits to accumulate in this region.

\medskip
\noindent 8. Using our analytic expressions we derived an approximate relation for the maximum semimajor axis for which we expect AR to be important (cf. equation~(\ref{eq:AR_crit})). This relation implies that AR is important in a large radial range for physically realistic nuclear star clusters.

\section*{Acknowledgements}
We would like to thank M. Atakan G\"{u}rkan for making his Kepler solver freely available\footnote{The code can be downloaded from the web page \href{http://home.strw.leidenuniv.nl/~gurkan/kepler/sol_kep/sol_kep.html}{http://home.strw.leidenuniv.nl/$\sim$gurkan/kepler/sol\_kep/sol\_kep.html}.}, Jeroen B\'{e}dorf for invaluable help with implementing GPU acceleration in \textsc{TPI} using the \textsc{Sapporo} library and Fabio Antonini for useful comments on the manuscript. We also thank the anonymous referee for providing comments that helped to improve the paper. We are grateful for the hospitality of the Institut Henri Poincar\'{e} where parts of this work were carried out. We also thank the organizers of the ``Al\'{a}jar Meeting 2013: Stellar dynamics and growth of massive black holes'' for a stimulating venue for discussions of issues related to this work. This work was supported by the Netherlands Research Council NWO (grants \#639.073.803 [VICI],  \#614.061.608 [AMUSE] and \#612.071.305 [LGM]), the Netherlands Research School for Astronomy (NOVA), the National Science Foundation under grant no. AST 1211602 and the National Aeronautics and Space Administration under grant no. NNX13AG92G. 

\bibliographystyle{mn2e}
\bibliography{literature}

\appendix
\section{Simple tests of \textsc{TPI}}
\label{app:simple_tests}

\subsection{Test stars orbiting the SBH}
\label{sect:validation:basic:teststarSBH}
In the absence of post-Newtonian (PN) and field star perturbations, test stars should maintain fixed Kepler orbits about the supermassive black hole (SBH) indefinitely. A sensitive test of the \textsc{Test Particle Integrator} (\textsc{TPI}) algorithm is to check whether it conserves the Keplerian elements over many periods. For example, assuming constant $a$ implies $t/P \approx 3.4 \times 10^5 \, (t/\mathrm{Myr}) (a/\mathrm{mpc})^{-3/2} (M_\bullet/10^6 \, \mathrm{M}_\odot)^{1/2}$ orbital revolutions after time $t$. 

\begin{figure}
\center
\includegraphics[scale = 0.435, trim = 0mm 0mm 0mm 0mm]{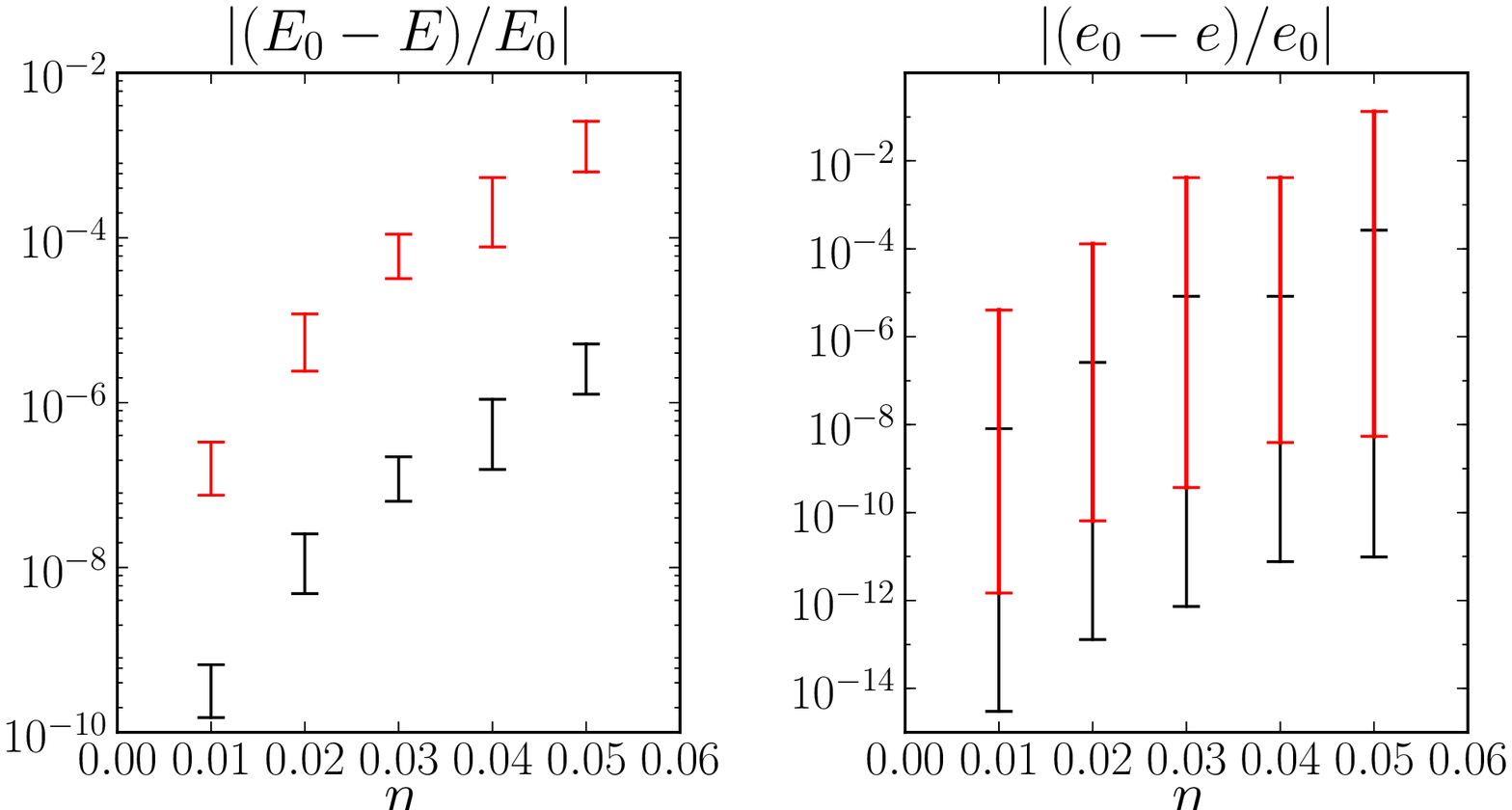}
\caption{\small Accuracy of the \textsc{TPI} integrator for a single star orbiting a SBH; PN terms are excluded. For eight initial eccentricities $e_0 \in \{0.01,0.1,0.5,0.9,0.99,0.999,0.9999,0.99999\}$ we show the range in the absolute values of the relative energy errors $(E_0-E)/E_0$ (left panel) and the relative eccentricity errors $(e_0-e)/e_0$ (right panel) as function of the time step parameter $\eta$. Black ranges are after 1000 orbits; red ranges are after $3.4 \times 10^5$ orbits. }
\label{fig:bh_energy_eccentricity_errors}
\end{figure}

We initialized eight test stars in Kepler orbits around a SBH with $M_\bullet = 1.0 \times 10^6 \, \mathrm{M}_\odot$ and initial semimajor axis $a_0 = 1 \, \mathrm{mpc}$ (this corresponds to an orbital period $P_0\approx 2.96 \, \mathrm{yr}$) and eight initial eccentricities $e_0 \in \{0.01,0.1,0.5,0.9,0.99,0.999,0.9999,0.99999\}$. We integrated these stars with \textsc{TPI} for $3.4 \times 10^5$ orbital periods in the absence of PN terms and field stars. All orbital elements except the orbital phase should therefore remain constant. We show in Figure \ref{fig:bh_energy_eccentricity_errors} the relative energy errors $(E_0-E)/E_0$ and the relative eccentricity errors $(e_0-e)/e_0$ of the orbit around the SBH after 1000 orbits (black lines) and after $3.4 \times 10^5$ orbits (red lines). Energy and eccentricity errors are included for five values of the time step parameter $\eta$ (cf. equation~(\ref{eq:timestepfunction})). The relative energy errors do not exceed $10^{-9}$ ($10^{-5}$) for $\eta = 0.01$ ($0.05$) after 1000 orbits and $10^{-6}$ ($10^{-2}$) for $\eta = 0.01$ ($0.05$) after $3.4 \times 10^5$ orbits. The relative eccentricity errors are larger than the relative energy errors but still do not exceed $10^{-3}$ for $\eta = 0.02$ after $3.4 \times 10^5$ orbits. We have made similar plots for the orbital angles $i$, $\omega$ and $\Omega$ and the relative errors of their cosines are $<10^{-11}$ after 1000 orbits for $\eta \leq 0.05$ and $<10^{-5}$ after $3.4 \times 10^5$ orbits for $\eta \leq 0.04$. The high precision in this test can be attributed to the use of regularization in the equations of motion (cf. \S\,\ref{sect:methods}). 

Because we are interested in the regime where precession due to general relativity is important, we also tested the ability of the code to reproduce relativistic (Schwarzschild) precession of test stars. In the implementation of the 1PN terms we assume that the SBH is fixed at the origin. To test whether this fixing of the SBH systematically affects the magnitude of secular precession expected from the theoretical expectation, equation~(\ref{eq:1PNprecession}), we integrated three different orbits with $a_0 = 0.1 \, \mathrm{mpc}$ and eccentricities $e_0 \in \{0.5,0.9,0.99\}$ with the 1PN terms included. The magnitude of pericenter shift during one orbit was computed by numerically determining the moments of two consecutive apocenter passages. By varying the numerically determined moments of apocenter by one output time we obtained a measure of the error in the pericenter shift. We show the results in Table \ref{table:precessiontest}. The pericenter shift calculated with \textsc{TPI} is in very good agreement with the expected 1PN shift. For example, in the test with the smallest pericenter distance (corresponding to $r_p \approx 20.9 \, r_g$, where $r_g = G M_\bullet/c^2$ is the gravitational radius) the relative error is smaller than 0.005. 

We also include in Table \ref{table:precessiontest} results for the same test of the implementation of the 1PN terms carried out with the direct summation code \textsc{MI6} (\citealt{nitmak08,iwasawa11}; see also \S\,\ref{sect:belowSB:init}). In the latter code the SBH is assumed to be fixed as in \textsc{TPI}. The errors made in \textsc{MI6} should therefore be comparable to those of \textsc{TPI}. The close similarity of the errors made by the two codes suggests that this is indeed the case (cf. the last column of Table \ref{table:precessiontest}).

\begin{table*}
\begin{tabular}{cccccccc}
\toprule
& & \multicolumn{2}{c}{$\Delta \omega|_\mathrm{sim}/{}^\circ$} & \multicolumn{2}{c}{$\Delta \omega|_\mathrm{pred}/{}^\circ$} & \multicolumn{2}{c}{1PN error} \\
$a/\mathrm{mpc}$ & $e$ & MI6 & TPI & MI6 & TPI & MI6 & TPI \\
\midrule
0.1 & 0.5 & $0.69017 \pm 0.00008$ & $0.69010 \pm 0.00010$ & 0.68986 & 0.68979 & -0.00046 & -0.00045 \\
0.1 & 0.9 & $2.75946 \pm 0.00003$ & $2.75913 \pm 0.00002$ & 2.75578 & 2.75548 & -0.00135 & -0.00133 \\
0.1 & 0.99 & $30.141905 \pm 0.000008$ & $30.138717 \pm 0.000005$ & 30.28069 & 30.27620 & 0.00457 & 0.00454 \\
\bottomrule
\end{tabular}
\caption{Pericenter shift per radial period for three different initial orbital configurations as computed by \textsc{MI6} and \textsc{TPI} including 1PN terms. Shown are the values determined from the simulations ($\Delta \omega|_\mathrm{sim}$), the predicted values computed from equation~(\ref{eq:1PNprecession}) based on $a$ and $e$ as determined in the simulations and the errors of the simulated values with respect to the predicted values, $(\Delta \omega|_\mathrm{pred} - \Delta \omega|_\mathrm{sim})/\Delta \omega|_\mathrm{pred}$. }
\label{table:precessiontest}
\end{table*}

\begin{figure}
\center
\includegraphics[scale = 0.435, trim = 0mm 0mm 0mm 0mm]{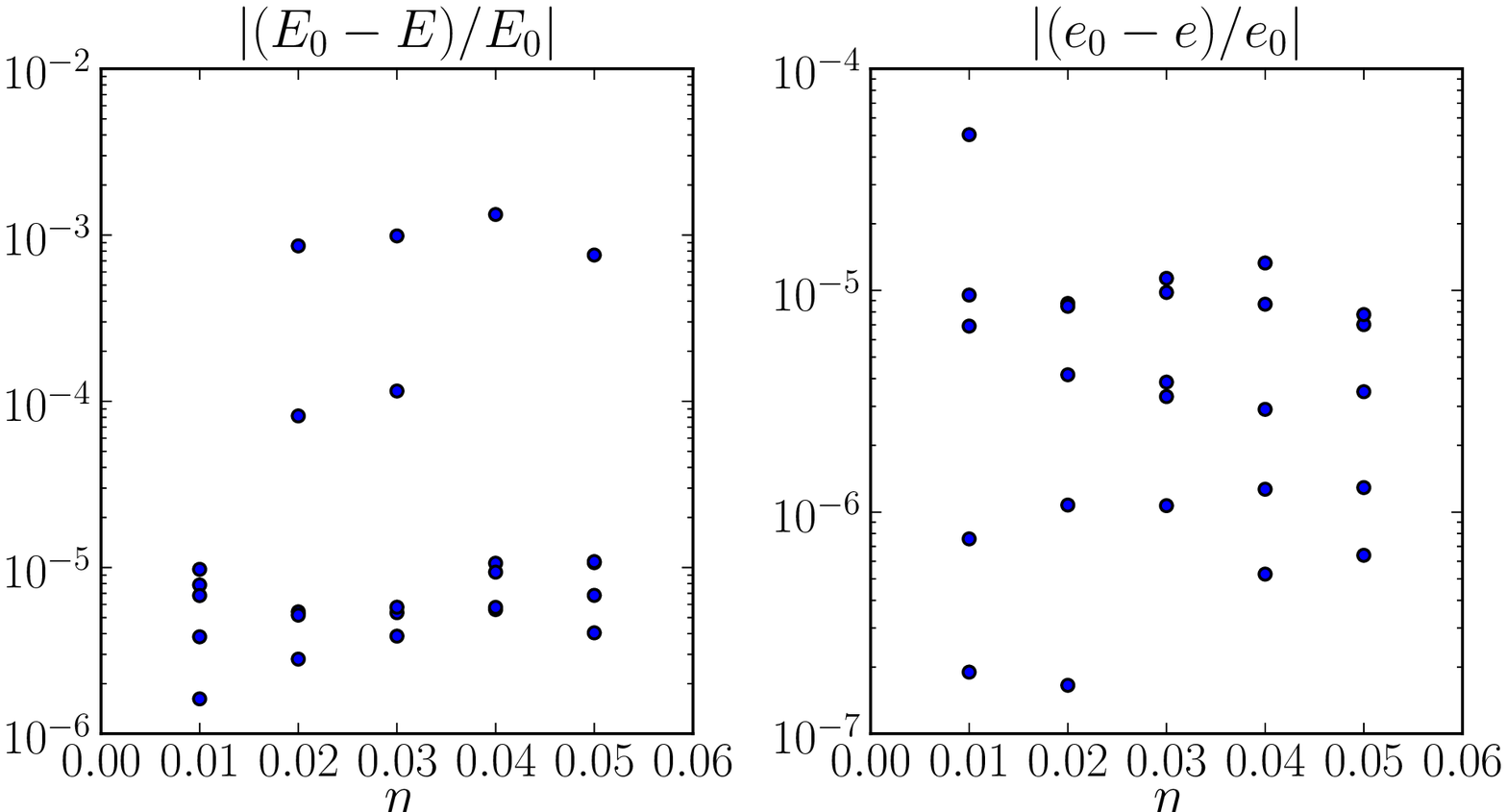}
\caption{\small Similar to Figure \ref{fig:bh_energy_eccentricity_errors}, now for test stars orbiting a single $1 \, \mathrm{M_\odot}$ field star that orbits the SBH (without PN terms). The points correspond to the absolute values of the relative energy errors $(E_0-E)/E_0$ (left panel) and relative eccentricity errors $(e_0-e)/e_0$ (right panel) for five initial eccentricities $e_0 \in \{0.01,0.1,0.5,0.9,0.99\}$. Errors are shown after 100 orbits of the test stars around the field star. }
\label{fig:per_energy_eccentricity_errors}
\end{figure}

\subsection{Test stars orbiting a field star that orbits the SBH}
\label{sect:validation:basic:teststarstarSBH}
In \textsc{TPI}, we tested interactions between test and field stars by placing test stars in Kepler orbits around a single field star that orbits the SBH. The field star has a mass $m_\star = 1.0 \, \mathrm{M}_\odot$ and its orbital parameters around the SBH ($M_\bullet = 1.0 \times 10^6 \, \mathrm{M}_\odot$) are $a_\star = 10^6 \, \mathrm{AU} \approx 4.85 \, \mathrm{pc}$ and $e_\star = 0.01$. For the latter orbit the star is situated far from the SBH at all times, hence the orbital elements except the orbital phases of test stars orbiting the star should remain constant. The initial orbital elements of the test stars orbiting the field star are set to $a_0 = 1 \, \mathrm{AU}$ and $e_0 \in \{0.01,0.1,0.5,0.9,0.99\}$. The test stars are integrated for 100 orbital periods around the field star (i.e. 100 yr). We show in Figure \ref{fig:per_energy_eccentricity_errors} the relative energy and eccentricity errors of the motion of the test stars around the field star. These errors are larger compared to those for the motion of test stars around the SBH, cf. Figure \ref{fig:bh_energy_eccentricity_errors}. This is not surprising considering that in \textsc{TPI} the motion around the SBH is regularized, whereas the motion around the field stars is not. Nevertheless, after 100 orbital periods the relative energy error remains less than $\approx 10^{-3}$ even for rather eccentric ($e_0 = 0.99$) orbits.

We also tested interactions between test and field stars by initiating a test star and a field star in nearly circular and nearly intersecting orbits around the SBH. We set $\eta = 0.02$, $m_\star = 50 \, \mathrm{M}_\odot$ and $M_\bullet = 1.0 \times 10^6 \, \mathrm{M}_\odot$. The closest approach between the test and field star in the resulting interaction is $\approx 0.05 \, \mathrm{AU}$ and the orbit of the test star is strongly perturbed, i.e. $\Delta a/a \approx 0.85$ and $\Delta e/e \approx 8 \times 10^3$. We also computed the same encounter with the direct $N$-body code \textsc{Hermite} implemented in \textsc{AMUSE} \citep{zwart13}. Here the test star mass was set to $1.0 \times 10^{-30} \, \mathrm{M}_\odot$. The discrepancies between the integrations with \textsc{Hermite} and \textsc{TPI} are very small: the differences between the two codes in the test star semimajor axis and eccentricity after the strong encounter, are $(a_{\mathrm{final},\textsc{TPI}} - a_{\mathrm{final},\textsc{Hermite}})/a_{\mathrm{final},\textsc{TPI}} \approx 0.007$ and $(e_{\mathrm{final},\textsc{TPI}} - e_{\mathrm{final},\textsc{Hermite}})/e_{\mathrm{final},\textsc{TPI}} \approx 0.002$. 

In \textsc{TPI} decreasing $\eta$ increases the number of integration steps and this decreases performance. Based on the above tests we have chosen the value $\eta = 0.02$ for the simulations presented in this paper, which we believe is a good compromise between accuracy and performance.

\section{Dependence of the NRR diffusion coefficients on $\gamma$}
\label{app:C_gamma}

\begin{figure}
\center
\includegraphics[scale = 0.435, trim = 0mm 0mm 0mm 0mm]{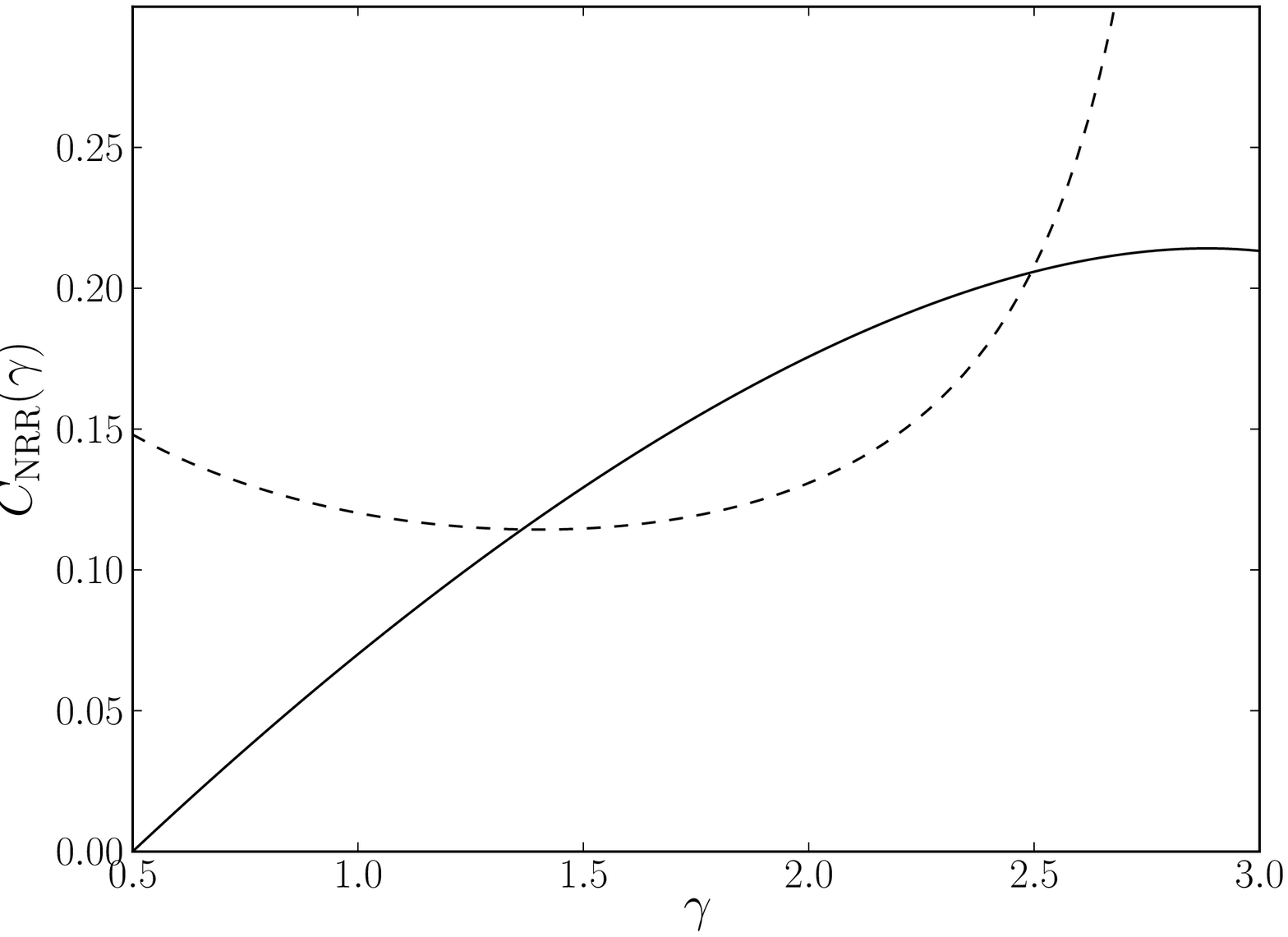}
\caption{\small The quantity $C_\mathrm{NRR}(\gamma)$ as funtion of $\gamma$. Solid line: equation~(\ref{eq:C_gamma}). Dashed line: a common approximation, $0.68/[(3-\gamma)(1+\gamma)^{3/2}]$. }
\label{fig:C_gamma}
\end{figure}

Here we include an expression for the quantity $C_\mathrm{NRR}(\gamma)$ for arbitrary $\gamma$. This expression appears in the NRR diffusion coefficients in which the limit $\ell\rightarrow0$  was taken (cf. equations~(\ref{eq:dfc_NRR_l0}) and (\ref{eq:AE})). It can be computed using the procedure described in Appendix B of MAMW11 in which the potential of the stars was neglected. In that appendix an explicit expression was derived for $\gamma=2$:
\begin{align}
C_\mathrm{NRR}(2) = \frac{9}{7} \frac{1}{12 \log(2) - 1} \approx 0.175698.
\end{align}
We have derived an expression that is valid for arbitrary $\gamma$ in the range $1/2 < \gamma < 3$:
\begin{align}
C_\mathrm{NRR}(\gamma) = \frac{3\pi}{64} \left[ K_{1/2}(\gamma) - \frac{1}{5} K_{3/2}(\gamma) + \frac{5\pi}{8} \frac{1}{2\gamma-1} \right ]^{-1}.
\label{eq:C_gamma}
\end{align}
Here $K_{1/2}(\gamma)$ and $K_{3/2}(\gamma)$ are integral functions defined as:
\begin{subequations}
\begin{align}
K_{1/2}(\gamma) &= \int_0^1 x^{3-\gamma} \sqrt{1-x} \, {}_2F_1 \left ( \frac{3}{2},\frac{3}{2}-\gamma,\frac{5}{2},1-x \right) \, \mathrm{d} x; \\
K_{3/2}(\gamma) &= \int_0^1 x^{3-\gamma} \sqrt{1-x} \, {}_2F_1 \left ( \frac{5}{2},\frac{3}{2}-\gamma,\frac{7}{2},1-x \right) \, \mathrm{d} x,
\end{align}
\end{subequations}
where ${}_2F_1(a,b,c;x)$ is the Gauss hypergeometric function. For $\gamma=1$, equation~(\ref{eq:C_gamma}) yields:
\begin{align}
C_\mathrm{NRR}(1) = \frac{1}{48 \log(2) - 19} \approx 0.0700719.
\end{align}

We show $C_\mathrm{NRR}(\gamma)$ as function of $\gamma$ in Figure \ref{fig:C_gamma} (solid line). For reference we have also plotted in that figure with the dashed line a less accurate, but more common approximation, $0.68/[(3-\gamma)(1+\gamma)^{3/2}]$ \citep[][p. 276]{bookmerritt13} (both relations neglect the potential of the stars).

\section{Extrapolating the shape of the cumulative eccentricity distribution for the S-star simulations}
\label{app:cont}
\begin{figure}
\center
\includegraphics[scale = 0.435, trim = 0mm 0mm 0mm 0mm]{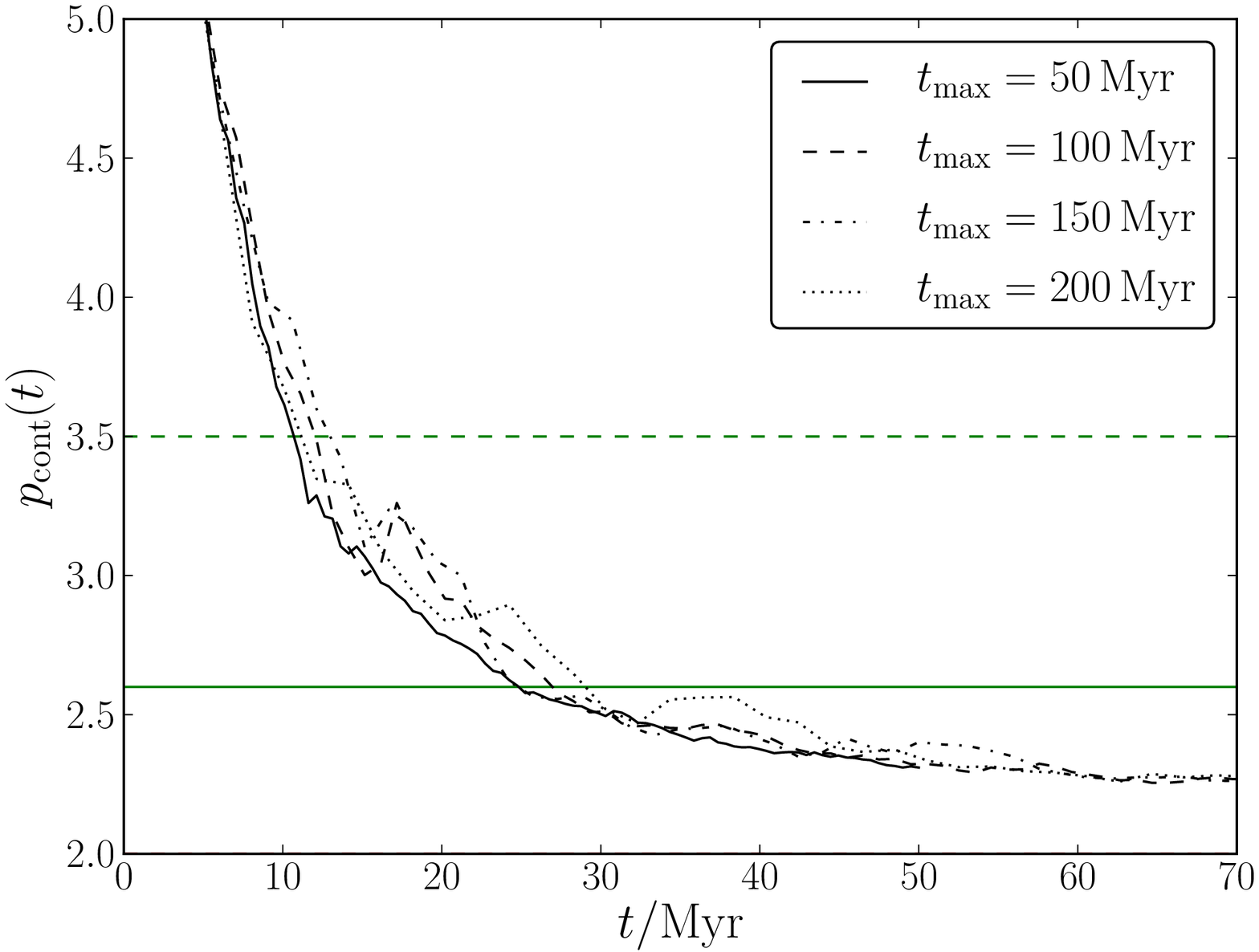}
\caption{\small The evolution of the slope $p$ with time extrapolated to the case of continuous formation of S-stars, determined using the fitted form $p_\mathrm{fit}(t)$ shown in Figure \ref{fig:S_stars_p_evolution}). Four different values are adopted for $t_\mathrm{max}$, the maximum time of deposition. The green solid and dashed lines indicate the observed value $p=2.6 \pm 0.9$  \citep{gil09}. }
\label{fig:S_stars_p_evolution_continuous}
\end{figure}
Here we present a method to extrapolate our results of $p(t)$ from the S-star simulations assuming burst formation, to the case of continuous formation (cf. \S\,\ref{sect:S-stars:orb_ev}). We assume that the probability density function (PDF) for a single S-star $i$ is of the form $\mathrm{d}N_i/\mathrm{d}e \propto e^{p_i(t-t_i)-1}$ for $t>t_i$ and $\mathrm{d}N_i/\mathrm{d}e = 0$ for $t<t_i$. Here $t_i$ is the time at which star $i$ is deposited. We assume that $t_i = x t_\mathrm{max}$, where $x\in[0,1]$ is a random number and $t_\mathrm{max}$ is a time scale for which the upper limit is set by the MS lifetime of the S-star. Normalization of $\mathrm{d}N_i/\mathrm{d}e$ with $0\leq e<1$ yields $\mathrm{d}N_i/\mathrm{d}e = p_i(t-t_i) e^{p_i(t-t_i)-1}$ for $t>t_i$. The PDF $\mathrm{d} N/\mathrm{d}e$ for the ensemble of $N_S=19$ S-stars is composed of the PDFs for the individual S-stars and it is therefore given by the sum of the latter PDFs, i.e. $\mathrm{d} N/\mathrm{d}e \propto \sum_{i=1}^{N_S} H(t-t_i) \, \mathrm{d}N_i/\mathrm{d}e = \sum_{i=1}^{N_S} H(t-t_i) p_i(t-t_i) e^{p_i(t-t_i)-1}$, where $H$ is the Heaviside step function. Normalization of the latter PDF with $0\leq e < 1$ gives $\mathrm{d} N/\mathrm{d}e = [\sum_{i=1}^{N_S} H(t-t_i)]^{-1} \sum_{i=1}^{N_S} [ H(t-t_i) p_i(t-t_i) e^{p_i(t-t_i)-1} ]$. This yields the following cumulative density function $N(e,t)$ for the ensemble of the S-stars at time $0 \leq t \leq t_\mathrm{max}$:
\begin{align}
N(e,t) = \left [ \sum_{i=1}^{N_S} H(t-t_i) \right ]^{-1} \sum_{i=1}^{N_S} \left [ H(t-t_i) e^{p_i(t-t_i)} \right ].
\label{eq:S_star_cum_cont}
\end{align}
It is assumed that $p_i(t-t_i) = p_\mathrm{fit}(t-t_i)$, i.e. that the fitted curve $p_\mathrm{fit}(t)$ in the burst scenario is representative for each individual S-star. For four different values of $t_\mathrm{max}$, $t_\mathrm{max}/\mathrm{Myr} \in \{50,100,150,200\}$, we take 100 different random realizations of $t_i$ and for each realization we fit the CDF equation~(\ref{eq:S_star_cum_cont}) to $N(e,t) = e^{p(t)}$. Subsequently we average the curves $p(t)$ over the 100 different random realizations and we adopt the averaged curve as $p_\mathrm{cont}(t)$, the value of $p$ in case of continuous formation. We show the results in Figure \ref{fig:S_stars_p_evolution_continuous} for the different values of $t_\mathrm{max}$. As expected, $p(t)$ evolves more slowly for continuous formation of S-stars compared to formation in a burst (cf. Figure \ref{fig:S_stars_p_evolution}); the speed of evolution is reduced by a factor of $\sim 4$. The peaks are due to depositions of individual S-stars at the random times $t_i$. There is no strong dependence of $p_\mathrm{cont}(t)$ on $t_\mathrm{max}$; after 50 Myr $p_\mathrm{cont}(t)$ for the four values of $t_\mathrm{max}$ is nearly identical.

\section{Dependence of derived diffusion coefficients on time lag in the simulations}
\label{sect:app:deptau}

\begin{figure*}
\center
\includegraphics[scale = 0.46, trim = 0mm 0mm 0mm 0mm]{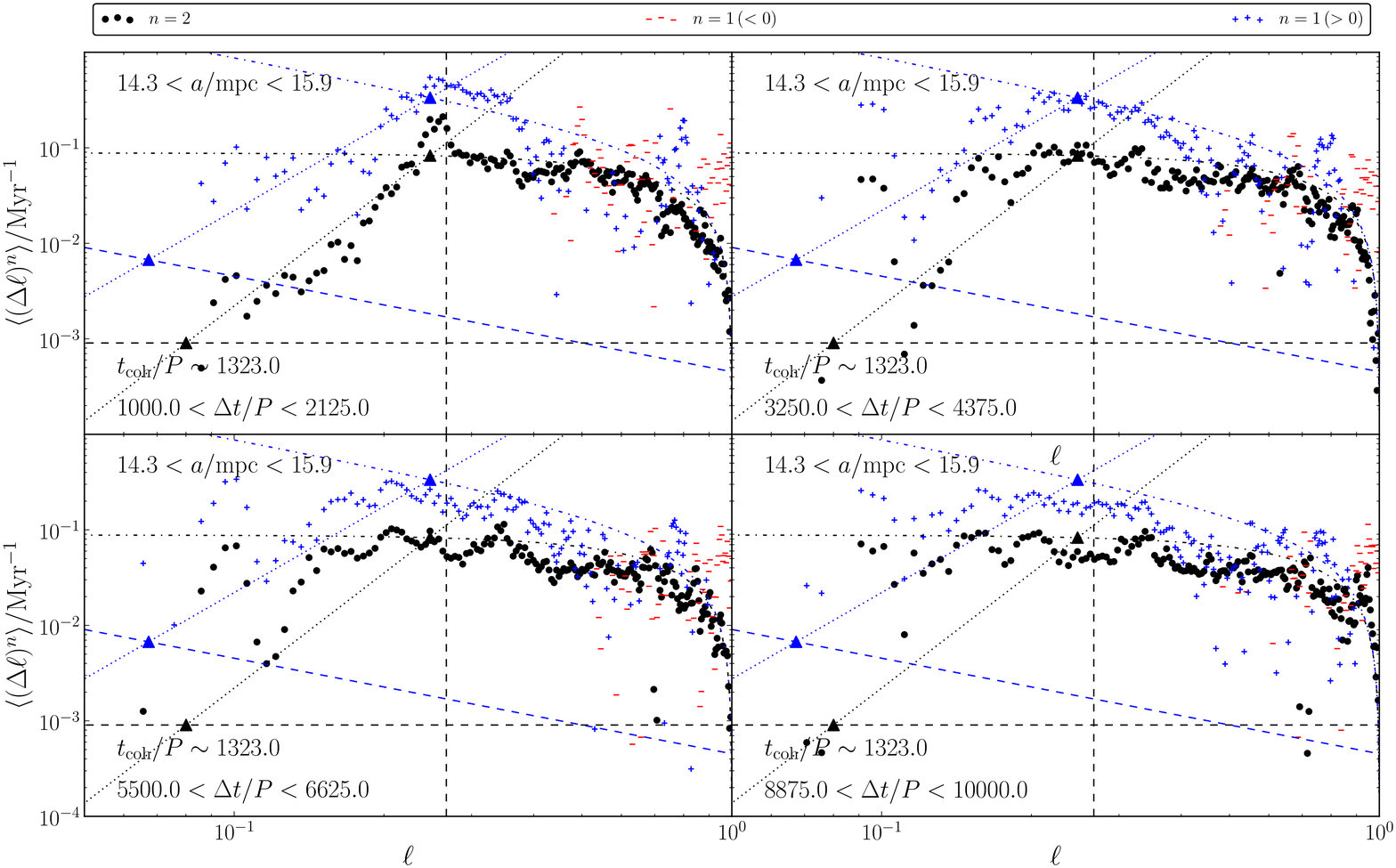}
\caption{\small Diffusion coefficients obtained from the S-star simulations, similar to Figure \ref{fig:diffusion_coefficients_S_stars}. Results are shown for a single semimajor axis bin and for different time lags. }
\label{fig:diffusion_coefficients_S_stars_tau_2}
\end{figure*}

In Figure \ref{fig:diffusion_coefficients_S_stars} the time lags were chosen to match the coherence time $t_\mathrm{coh}$. Here we illustrate the importance of choosing the appropriate time lag, by showing in Figure \ref{fig:diffusion_coefficients_S_stars_tau_2} an example of the dependence of the diffusion coefficients on time lag. We computed the coefficients for much longer time lags than in \S\,\ref{sect:S-stars:dfc}: $10^3 < \Delta t/P < 10^4$. For $\Delta t \sim t_\mathrm{coh}$ the ``knee" feature (cf. \S\,\ref{sect:S-stars:dfc}) below the SB is clearly present; for $\Delta t \gg t_\mathrm{coh}$ this feature gradually disappears. This can be understood from the argument that was presented in \S\,\ref{sect:belowSB:diffusion}: below the SB, the time lag should not be much longer than $t_\mathrm{coh}$, since for longer time lags the changes in $\ell$ become comparable to $\ell$ itself.

\section{Steady-state solutions to the Fokker-Planck equation}
\label{app:steady-state}
\subsection{Solution of the steady-state equation}
\label{app:steady-state:gen_sol}
The aim is to solve equation~(\ref{eq:FPsteady_state_C}), which we write as:
\begin{align}
- N(R) \mathcal{D}_1(R) + \frac{1}{2} \frac{\partial}{\partial R} [N(R)\mathcal{D}_2(R)] = C.
\label{eq:STC}
\end{align}
Here we have used the notation $\langle \Delta R \rangle = \mathcal{D}_1$ and $\langle (\Delta R)^2 \rangle = \mathcal{D}_2$.  We are interested in solutions $N(R)$ in the range $R_\mathrm{loss}<R<R_\mathrm{up}$. Here $R_\mathrm{loss} = r_\mathrm{capt}/a \, (2-r_\mathrm{capt}/a) = \mathcal{O}(10^{-2})$ is the loss boundary and $R_\mathrm{up}$ is the largest value of $R$; if the diffusion coefficients are completely known then $R_\mathrm{up} = 1$. If $C=0$ then equation~(\ref{eq:STC}) can readily be integrated, with solution:
\begin{align}
\nonumber N_H(R) &= N_H(R_\mathrm{loss}) \exp \left [ \int_{R_\mathrm{loss}}^R \frac{2\mathcal{D}_1(R') - \mathcal{D}'_2(R')}{\mathcal{D}_2(R')} \, \mathrm{d} R' \right ] \\
&\equiv N_H(R_\mathrm{loss}) \, g(R).
\label{eq:solh}
\end{align}
Here $\mathcal{D}'_n(R) \equiv (\partial/\partial R) \mathcal{D}_n(R)$ and we have implicitly defined the function $g(R)$; note that $g(R_\mathrm{loss}) = 1$. The function $N_H(R)$ is a homogeneous solution to equation~(\ref{eq:STC}). To find the inhomogeneous solution, we apply the method of variation of constants and write $N(R) = N_H(R) N_I(R)$. Substituting the latter into equation~(\ref{eq:STC}), we find:
\begin{subequations}\label{eq:STin}
\begin{align}
C &= -N_H N_I \mathcal{D}_1 + \frac{1}{2} \frac{\partial}{\partial R} (N_H N_I \mathcal{D}_2 ) \\
&= N_I \left [ -N_H \mathcal{D}_1 + \frac{1}{2} \frac{\partial}{\partial R} (N_H \mathcal{D}_2) \right ] + \frac{1}{2} N_H \mathcal{D}_2 \frac{\partial N_I}{\partial R} \\
&= \frac{1}{2} N_H \mathcal{D}_2 \frac{\partial N_I}{\partial R}.
\end{align}
\end{subequations}
The last step is by virtue of equation~(\ref{eq:STC}) with $N =N_H$ and $C=0$. Equation~(\ref{eq:STin}) is readily integrated:
\begin{align}
N_I(R) = \int_{R_\mathrm{loss}}^R \frac{2C}{N_H(R') \mathcal{D}_2(R')} \, \mathrm{d} R' + C_I.
\label{eq:soli}
\end{align}
Here $C_I$ is an integration constant. By imposing $N(R_\mathrm{loss}) = 0$ and substituting the solutions equations~(\ref{eq:solh}) and (\ref{eq:soli}) we find $C_I = 0$. The general solution is therefore given by:
\begin{subequations}\label{eq:solfull}
\begin{align}
N(R) &= 2\,C g(R) I(R), \\
g(R) &= \exp \left [ \int_{R_\mathrm{loss}}^R \frac{2\mathcal{D}_1(R') - \mathcal{D}'_2(R')}{\mathcal{D}_2(R')} \, \mathrm{d} R' \right ], \\
I(R) &= \int_{R_\mathrm{loss}}^R \frac{\mathrm{d} R'}{\mathcal{D}_2(R') g(R')}. 
\end{align}
\end{subequations}
As expected for a second-order differential equation, the solution to equation~(\ref{eq:solfull}) contains two parameters, $R_\mathrm{loss}$ and $C$ (we do not consider $R_\mathrm{up}$ to be a free parameter). By imposing an additional constraint on the solution, the number of parameters is reduced by one. For example, requiring that $N(R)$ is normalized to unit total number, i.e. $\int N(R) \, \mathrm{d} R = 1$, we find for the flux in terms of $R_\mathrm{loss}$:
\begin{align}
C = \frac{1}{2} \left [ \int_{R_\mathrm{loss}}^{R_\mathrm{up}} g(R) I(R) \, \mathrm{d} R  \right ]^{-1}.
\label{eq:solfull_flux}
\end{align}

\subsection{Analytic expressions for the diffusion coefficients}
\label{app:steady-state:dfc_analytic}
For completeness we give the explicit functional expressions for our approximation of the diffusion coeffcients, equation~(\ref{eq:dif_coef_approx}):
\begin{subequations}\label{eq:dfc_analytic_modified}
\begin{align}
\langle \Delta \ell \rangle &= \left \{ \begin{array}{lc}
\displaystyle 1/( 4\ell t_\mathrm{N1}), & \ell_\mathrm{loss} < \ell < \ell_{\mathrm{a},1}; \\
\displaystyle C_1 \ell^3/\tau, & \ell_{\mathrm{a,1}} \leq \ell < \ell_{\mathrm{b},1}; \\
\displaystyle C_1 \ell_{\mathrm{b},1}^4/(\tau \ell) \frac{ \ell_\mathrm{c}^2 - \ell^2 }{\ell_\mathrm{c}^2 - \ell_{\mathrm{b},1}^2}, & \ell_{\mathrm{b},1} \leq \ell \leq 1;
\end{array} \right. \\
\left \langle \left ( \Delta \ell \right )^2 \right \rangle &= \left \{ \begin{array}{lc}
\displaystyle 1/( t_\mathrm{N1}), & \ell_\mathrm{loss} < \ell < \ell_{\mathrm{a},2}; \\
\displaystyle C_2 \ell^4 / \tau, & \ell_{\mathrm{a,2}} \leq \ell < \ell_{\mathrm{b},2}; \\
\displaystyle \left (1 - \ell^2\right) \alpha_s^2 / t_\mathrm{R1}, & \ell_{\mathrm{b},2} \leq \ell \leq 1. \\
\end{array} \right. 
\end{align}
\end{subequations}
Here $t_\mathrm{N1} \equiv A(E)^{-1}$ (cf. equation~(\ref{eq:AE})), $\tau \equiv t_\mathrm{coh}/A^2_\mathrm{D}$ (cf. equation~(\ref{eq:dfc_AR})) and $t_\mathrm{R1} \equiv [M_\bullet/M_\star(a)]^2 N_\star(a) P(a)^2/t_\mathrm{coh}$ (cf. equation~(\ref{eq:dfc_lRR2})). In equation~(\ref{eq:dfc_analytic_modified}) the first-order diffusion coeffcient in the range $\ell_{\mathrm{b},1} < \ell < 1$ has been modified to account for negative $\langle \Delta \ell \rangle$ for $\ell>\ell_\mathrm{c}$, as described in \S\,\ref{sect:steady-state:analytic}.
A comparison of equation~(\ref{eq:dfc_analytic_modified}) to $N$-body data is given in Figure \ref{fig:steady_state_analytic}. 

\subsection{Explicit analytic steady-state solutions}
\label{app:steady-state:steady_state_analytic}
We derive explicit expressions for the steady-state distribution function $N(R)$ for the analytic functions of the diffusion coefficients presented in equation~(\ref{eq:dfc_analytic_modified}). First we transform $\langle (\Delta \ell)^n\rangle$ to $\langle (\Delta R)^n\rangle$ using the transformations (\citealt{cohn79}; \citealt[][eq. 5.167]{bookmerritt13}):
\begin{align}
\left \{ \begin{array}{cc} 
\displaystyle \langle \Delta R \rangle = 2 \ell \langle \Delta \ell \rangle + \left \langle \left (\Delta \ell \right )^2 \right \rangle; \\
\displaystyle \left \langle \left ( \Delta R \right )^2 \right \rangle = 4 \ell^2 \left \langle \left (\Delta \ell \right )^2 \right \rangle. 
\end{array} \right.
\label{eq:dfc_trans_R_l}
\end{align}
We subsequently substitute $\langle (\Delta R)^n\rangle$ into equation~(\ref{eq:solfull}). Here we assume that  $\ell_{\mathrm{b,1}} = \ell_{\mathrm{b},2}$, which is the case if $C_1=C_2$. The result is:

\begin{subequations}\label{eq:steady_state_analytic_solution}
\begin{align}
\bar{N}(R) &= 2\,C g(R) I(R); \\
g(R) &= \left \{ \begin{array}{lr}
\displaystyle g_0, & R_\mathrm{loss} < R < R_{\mathrm{a},1}; \\
\displaystyle g_1(R), & R_{\mathrm{a},1} \leq R < R_{\mathrm{a},2}; \\
\displaystyle g_2(R), & R_{\mathrm{a},2} \leq R < R_{\mathrm{b},1}; \\
\displaystyle g_3(R), & R_{\mathrm{b},1} \leq R < 1; \\
\end{array} \right.
\end{align}
\begin{align}
I(R) &= \left \{ \begin{array}{ll}
\displaystyle I_0(R), & R_\mathrm{loss} < R < R_{\mathrm{a},1}; \\
\displaystyle I_0(R_{\mathrm{a,1}}) + I_1(R), & R_{\mathrm{a},1} \leq R < R_{\mathrm{a},2}; \\
\displaystyle I_0(R_{\mathrm{a,1}}) + I_1(R_{\mathrm{a,2}}) + I_2(R), & R_{\mathrm{a},2} \leq R < R_{\mathrm{b},1}; \\
\displaystyle I_0(R_{\mathrm{a,1}}) + I_1(R_{\mathrm{a,2}}) + I_2(R_{\mathrm{b,1}}) \\
\displaystyle \quad + I_3(R), & R_{\mathrm{b},1} \leq R < 1. \\
\end{array} \right.
\end{align}
\end{subequations}
The auxiliary functions are given by:
\begin{align}
\nonumber g_0 &= 1; \\
\nonumber g_1(R) &= g_0 \left ( \frac{R}{R_{\mathrm{a},1}} \right )^{-\frac{1}{2}} \exp \left [ c_\mathrm{N} \left ( R^2 - R_{\mathrm{a},1}^2 \right ) \right ]; \\
\nonumber g_2(R) &= g_1(R_{\mathrm{a},2}) \left (\frac{R}{R_{\mathrm{a,2}}} \right )^\frac{2C_1-5C_2}{2C_2}; \\
\nonumber g_3(R) &= g_2(R_{\mathrm{b},1}) \left (\frac{R}{R_{\mathrm{b,1}}} \right )^{c_\mathrm{R} c_\mathrm{B} R_\mathrm{c} - \frac{1}{2}} \left ( \frac{1-R_{\mathrm{b,1}}}{1-R} \right )^{c_\mathrm{R} c_\mathrm{B} (R_\mathrm{c}-1) + 1} \\
\end{align}
and
\begin{align}
\nonumber I_0(R) &= \frac{t_\mathrm{N1}}{2} \log \left ( \frac{R}{R_\mathrm{loss}} \right ); \\
\nonumber I_1(R) &= \frac{t_\mathrm{N1}}{4} R_{\mathrm{a},1}^{-\frac{1}{2}} c_\mathrm{N}^{-\frac{1}{4}} \exp \left (c_\mathrm{N} R_{\mathrm{a},1}^2 \right ) \\
\nonumber &\times \left [ \Gamma \left ( \frac{1}{4}, c_\mathrm{N} R_{\mathrm{a},1}^2 \right ) - \Gamma \left ( \frac{1}{4}, c_\mathrm{N} R^2 \right ) \right ]; \\
\nonumber I_2(R) &= \frac{\tau}{2C_2} \left ( \frac{R_{\mathrm{a},2}}{R_{\mathrm{a},1}} \right)^\frac{1}{2} \exp \left [ c_\mathrm{N} \left ( R_{\mathrm{a},1}^2 - R_{\mathrm{a},2}^2 \right ) \right ] \\
\nonumber &\times \frac{C_2}{2C_1-C_2} R_{\mathrm{a},2}^{-2} \left [ 1 - \left ( \frac{R}{R_{\mathrm{a},2}} \right )^{\frac{C_2-2C_1}{2C_2}} \right ]; \\
\nonumber I_3(R) &= \frac{t_\mathrm{R1}}{2\alpha_s^2} \left ( \frac{R_{\mathrm{a},2}}{R_{\mathrm{a},1}} \right)^\frac{1}{2} \exp \left [ c_\mathrm{N} \left ( R_{\mathrm{a},1}^2 - R_{\mathrm{a},2}^2 \right ) \right ] \left ( \frac{R_{\mathrm{b},1}}{R_{\mathrm{a},2}} \right )^\frac{5C_2-2C_1}{2C_2} \\
\nonumber &\times \left (1-R_{\mathrm{b},1} \right )^{-c_\mathrm{R} c_\mathrm{B} (R_\mathrm{c}-1) - 1} \left (2 c_\mathrm{R} c_\mathrm{B} R_\mathrm{c} - 1 \right)^{-1}   \\
\nonumber &\times \left [ \left ( \frac{R}{R_{\mathrm{b},1}} \right )^{c_\mathrm{R} c_\mathrm{B} R_\mathrm{c} - \frac{1}{2}} \times {}_2F_1 \left (d_1,d_2,d_3; R \right ) \right. \\
&\quad - \left. {}_2F_1 \left (d_1,d_2,d_3; R_{\mathrm{b},1} \right ) \right ].
\end{align}

Here $c_\mathrm{N} \equiv (t_\mathrm{N1}/\tau) C_1$, $c_\mathrm{R} \equiv (t_\mathrm{R1}/\tau) C_1$, $c_\mathrm{B} \equiv (R^2_{\mathrm{b},1}/\alpha_s^2)[1/(R_\mathrm{c} - R_{\mathrm{b},1})]$, $\Gamma(s,x) = \int_x^\infty t^{s-1} \exp(-t) \, \mathrm{d} t$ is the upper incomplete Gauss function and ${}_2F_1(d_1,d_2,d_3;R)$ is the Gauss hypergeometric function, with $d_1 \equiv c_\mathrm{R} c_\mathrm{B} (1-R_\mathrm{c})$, $d_2 \equiv c_\mathrm{R} c_\mathrm{B} R_\mathrm{c} - \frac{1}{2}$ and $d_3 = c_\mathrm{R} c_\mathrm{B} R_\mathrm{c} + \frac{1}{2}$.

\section{$N$-body simulations with $\gamma=1$}
\label{app:Nbody_gamma_1}

\begin{figure*}
\center
\includegraphics[scale = 0.46, trim = 0mm 0mm 0mm 0mm]{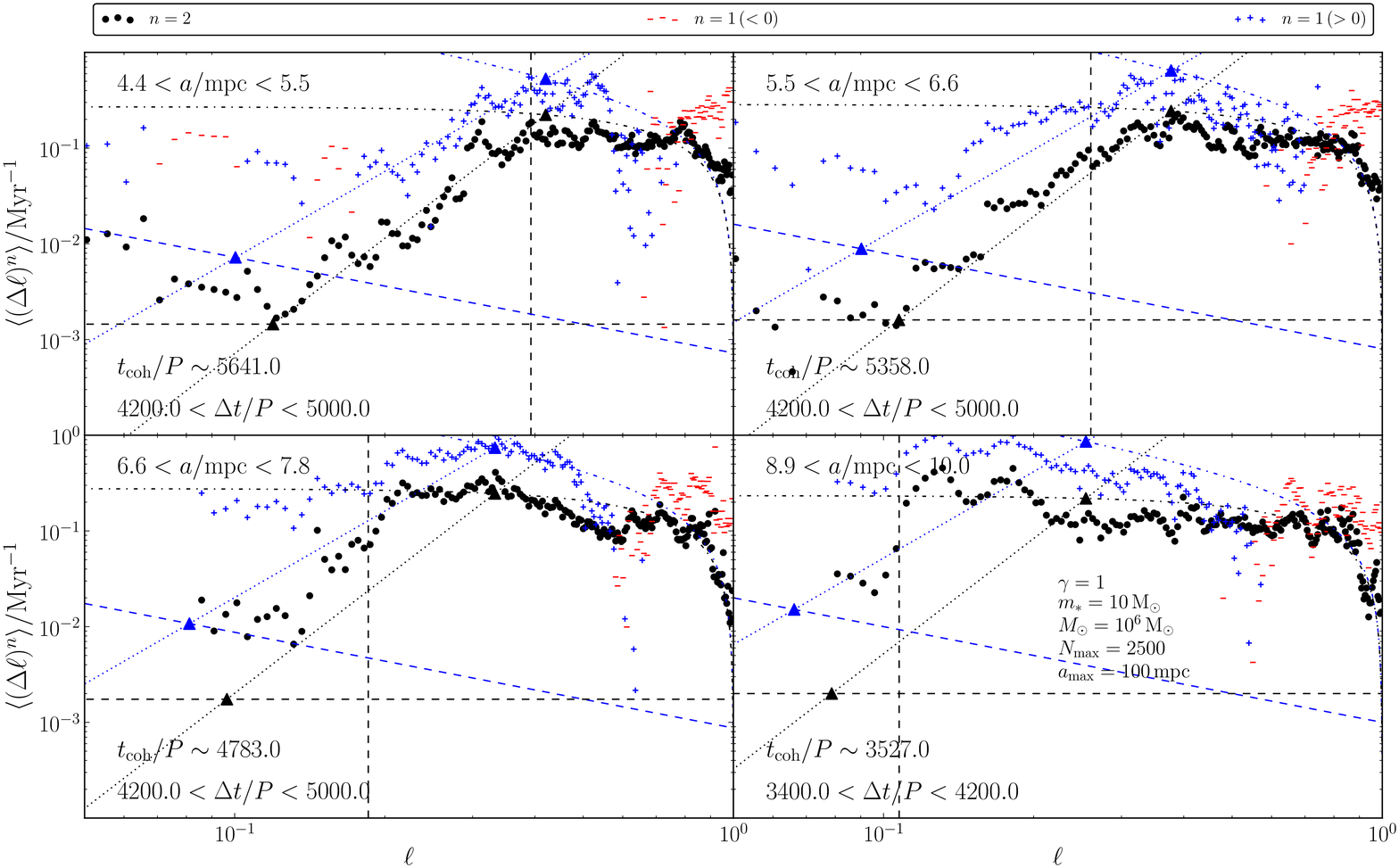}
\caption{\small Diffusion coefficients obtained from an additional set of simulations with $\gamma=1$; parameters are indicated in the bottom right panel. The lines show the analytic model of equation~(\ref{eq:dif_coef_approx}) as in Figure \ref{fig:diffusion_coefficients_S_stars}, but now evaluated for the corresponding nuclear model with $\gamma=1$ (see \S\,\ref{sect:discussion:gen} for details). } 
\label{fig:diffusion_coefficients_gamma_1}
\end{figure*}

In the simulations presented in \S\,\ref{sect:S-stars} a field star density profile $\rho_\star(r) \propto r^{-2}$ was assumed. In order to establish whether the ``knee'' feature in the diffusion coefficients that can be associated with AR is also present in simulations with different $\gamma$, we have carried out an additional set of simulations with \textsc{TPI} with $\gamma=1$. These additional simulations provide verifcation of some of our expectations for the regime in which AR is important, as discussed in \S\,\ref{sect:discussion:gen}.

The parameters of the additional set of simulations were as follows. The field star mass was $m_\star = 10 \, \mathrm{M}_\odot$ and the SBH mass was $M_\bullet = 10^6 \, \mathrm{M}_\odot$. Field stars were distributed according to $N(a) = N_\mathrm{max} (a/a_\mathrm{max})^{3-\gamma}$, with $N_\mathrm{max} = 2500$, $a_\mathrm{max} = 100 \, \mathrm{mpc}$ and $\gamma=1$, and their eccentricities were sampled from a ``thermal'' distribution $\mathrm{d}N/\mathrm{d}e = 2\,e$. In total 200 test particles were included, with initial semimajor axes sampled from $N(a) \propto a^{3-\gamma}$ with $3 \lesssim a/\mathrm{mpc} \lesssim 14$ and $\mathrm{d}N/\mathrm{d}e = 2\,e$. The orbits of the test and field stars were initially randomly oriented. The capture radius was $r_\mathrm{capt} = 8 \, r_g$ and the integration time was $10 \, \mathrm{Myr}$. Only the 1PN terms were included. 

The diffusion coefficients obtained from these simulations are shown for different semimajor axes in Figure \ref{fig:diffusion_coefficients_gamma_1}. In that figure we have included the same analytical functions for the coefficients that were also included in Figure \ref{fig:diffusion_coefficients_S_stars} (cf. equation~(\ref{eq:dif_coef_approx})), but now evaluated for the model with $\gamma=1$. The results for $\gamma=1$ are consistent with those for $\gamma=2$, which were presented in Figure \ref{fig:diffusion_coefficients_S_stars}. In particular, the ``knee'' feature is clearly present which, as we argued, can be associated with the rapid quenching of RR below the SB. The position of this ``knee'' agrees well with the predicted position of the SB, equation~(\ref{eq:ell_SB}), suggesting that this relation is also valid for nuclear models with $\gamma=1$. 

For small semimajor axes our predictions for the AR diffusion coeffients with $C_1=C_2 \approx 2.6$ (cf. equation~(\ref{eq:dfc_AR})) are in good agreement with the data obtained from the simulations with $\gamma=1$. At larger semimajor axes the slopes predicted by these relations are still consistent with the data, but the normalization is not: it appears that in order to remain consistent with the data, both $C_1$ and $C_2$ must {\it increase} with increasing semimajor axis. We note that this trend can also be observed in Figure \ref{fig:diffusion_coefficients_S_stars}, although the dependence of $C_1$ and $C_2$ on semimajor axis appears to be weaker in the latter figure.

\section{Equivalence of two critical radii}
\label{app:aARcrit}

Here we show that the quantity $a_\mathrm{ AR,max}$ defined in \S\,\ref{sect:discussion:gen} is the same as the critical semimajor axis that was defined in \S\,VC of MAMW11. The latter quantity, which we here denote by $a_\mathrm{MAMW}$, was argued to be the minimum value of $a$ for which NRR would allow orbits to ``penetrate'' the SB.

The criterion in MAMW11 was that -- for orbits near the SB --
\begin{align}
\left(\Delta\ell\right)_\mathrm{NRR} \equiv \left(\frac{t_\mathrm{coh}}{t_\mathrm{NRR}}\right)^{1/2} \gtrsim \ell_\mathrm{max}-\ell_\mathrm{min} \approx 2 \ell_\mathrm{av}^2 A_\mathrm{D}
\end{align}
(MAMW11, equations 66, 67). Thus $a_\mathrm{MAMW}$ is the value of $a$ for which:
\begin{align}
\frac{t_\mathrm{coh}}{t_\mathrm{NRR}} \approx 4 A_\mathrm{D}^2\; \ell_\mathrm{SB}^4 .
\label{eq:cond1}
\end{align}
The quantity defined as $t_\mathrm{NRR}$ in MAMW11 is essentially the inverse of the quantity $A(E)$ defined in this paper (cf. equation~(\ref{eq:AE})).

In \ref{sect:discussion:gen} of this paper, $a_\mathrm{ AR,max}$ was defined as the value of $a$ for which $\ell_{\mathrm{a},1} = \ell_\mathrm{SB}$. The quantity $\ell_{\mathrm{a},1}$ was defined, in turn, as the angular momentum for which
\begin{align}
\langle\Delta\ell\rangle_\mathrm{NRR} \equiv \frac{1}{4\ell}A(E)
= \langle\Delta\ell\rangle_\mathrm{AR} \approx \frac{A_\mathrm{D}^2\ell^3}{t_\mathrm{coh}}
\end{align}
(equation~(\ref{eq:dif_coef_approx})). Thus
\begin{align}
\ell_{\mathrm{a},1}^4 \approx \frac{t_\mathrm{coh}}{A_\mathrm{D}^2}\frac{A(E)}{4} 
\end{align}
and setting $\ell_{\mathrm{a},1}=\ell_\mathrm{SB}$ then yields:
\begin{align}
4 A_\mathrm{D}^2\; \ell_\mathrm{SB}^4 \approx A(E) t_\mathrm{coh} ,
\end{align}
the same as equation~(\ref{eq:cond1}).

\end{document}